\documentclass{emulateapj}

\usepackage{psfig}
\usepackage{amsmath}
\usepackage{amssymb}
\usepackage{graphicx}
\usepackage{appendix}
\usepackage{color}

\newcommand{\lSect}[1]{{\label{sec:#1}}}
\newcommand{\lFig}[1]{{\label{fig:#1}}}

\newcommand{\lTab}[1]{{\label{tab:#1}}}
\def\gtaprx {\lower .1ex\hbox{\rlap{\raise .6ex\hbox{\hskip .3ex
	{\ifmmode{\scriptscriptstyle >}\else
		{$\scriptscriptstyle >$}\fi}}}
	\kern -.4ex{\ifmmode{\scriptscriptstyle \sim}\else
		{$\scriptscriptstyle\sim$}\fi}}}
\def\ltaprx {\lower .1ex\hbox{\rlap{\raise .6ex\hbox{\hskip .3ex
	{\ifmmode{\scriptscriptstyle <}\else
		{$\scriptscriptstyle <$}\fi}}}
	\kern -.4ex{\ifmmode{\scriptscriptstyle \sim}\else
		{$\scriptscriptstyle\sim$}\fi}}}
\newcommand{\FIGFF}[2]{{\ref{fig:#2}{#1}}}

\newcommand{\FIG}[2]{{Fig.~\FIGFF{#1}{#2}}}
\newcommand{\Fig}[1]{{\FIG{}{#1}}}

\newcommand{\Sectff}[1]{{\ref{sec:#1}}}
\newcommand{\Sect}[1]{{\S~\Sectff{#1}}}

\newcommand{\Msun}{\ensuremath{\mathrm{M}_\odot}}
\newcommand{\Lsun}{\ensuremath{\mathrm{L}_\odot}}
\newcommand{\Rsun}{\ensuremath{\mathrm{R}_\odot}}
\newcommand{\Zsun}{\ensuremath{\mathrm{Z}_\odot}}
\newcommand{\Tab}[1]{{Table \ref{tab:#1}}}

\newcommand{\black}{\color{black}}

\begin{document}

%\shorttitle{Pulsational Pair-Instability Supernovae}
%\shortauthors{Woosley}

\title{Pulsational Pair-Instability Supernovae}

\author{S. E. Woosley\altaffilmark{1}}
%\author{S. E. Woosley\altaffilmark{1} and Alexander Heger\altaffilmark{2}}

\altaffiltext{1}{Department of Astronomy and Astrophysics, University
  of California, Santa Cruz, CA 95064; woosley@ucolick.org}

\begin{abstract} 
The final evolution of stars in the mass range 70 - 140 \Msun \ is
explored. Depending upon their mass loss history and rotation rates,
these stars will end their lives as pulsational pair-instability
supernovae producing a great variety of observational transients with
total durations ranging from weeks to millennia and luminosities from
10$^{41}$ to over 10$^{44}$ erg s$^{-1}$.  No non-rotating model
  radiates more than $5 \times 10^{50}$ erg of light or has a kinetic
  energy exceeding $5 \times 10^{51}$ erg, but greater energies are
  possible, in principle, in magnetar-powered explosions which are
  explored. Many events resemble Type Ibn, Icn, and IIn supernovae,
  and some potential observational counterparts are mentioned. Some
  PPISN can exist in a dormant state for extended periods, producing
  explosions millennia after their first violent pulse. These dormant
  supernovae contain bright Wolf-Rayet stars, possibly embedded in
  bright x-ray and radio sources. The relevance of PPISN to supernova
impostors like Eta Carinae, to super-luminous supernovae, and to
sources of gravitational radiation is discussed. No black holes
between 52 and 133 \Msun \ are expected from stellar evolution in
close binaries.
\end{abstract}


\section{INTRODUCTION}
\lSect{intro}

For helium cores more massive than about 30 \Msun, post-carbon burning
stages are, initially at least, unstable \citep{Woo07,Woo15}. The
production of electron-positron pairs at high entropy and temperatures
over about $7 \times 10^8$ K softens the equation of state, reducing
the structural adiabatic index below 4/3. Roughly speaking, the
creation of the rest mass of the pairs takes energy that might have
gone into providing pressure support. A contraction to a higher
temperature does not encounter as much resistance as it might have
otherwise, and the star becomes unstable. This is the
``pair-instability'' \citep{Fow64,Rak67,Bar67}.

This instability results in a dynamical implosion of the helium and
heavy element core which, provided the mass of that core does not
exceed 133 \Msun, is reversed by nuclear burning \citep{Heg02}. Within
this range of presupernova helium core masses, 30 - 133 \Msun, which
corresponds to a larger, less certain range of main sequence mass of
roughly 70 - 260 \Msun, ignoring rotation, a diverse range of outcomes
is expected. Helium cores above about 64 \Msun \ experience a single
violent pulse that disrupts the entire star as a ``pair-instability
supernova'' (PISN). These events have been well studied
\citep[e.g.][]{Obe83,Bon84,Gla85,Heg02,Ume02,Sca05,Kas11}, in part
because they are easy to simulate. Unlike iron core-collapse
supernovae, the explosion mechanism is well understood and easily
calculated in 1D. The major uncertainties lie instead with the
formation and evolution of the progenitor stars.

Less well studied are the ``pulsational pair-instability supernovae''
(PPISN) powered by the ``pulsational pair instability'' (PPI). Here,
the nuclear flashes are not sufficiently energetic to disrupt the
entire star. Instead a series of pulsations occurs. The core
contracts, ignites burning, typically of oxygen or silicon, expands
and cools, then contracts and ignites burning again, either on a
hydrodynamic time scale in low mass cores or on a Kelvin-Helmholtz
time scale in higher mass ones. In the Kelvin-Helmholtz case, pulses,
followed by cooling by radiation and neutrino emission, recur until
the mass and entropy of the helium and heavy element remnant are
reduced sufficiently to avoid the PPI. As a result, the final core
masses converge on a relatively narrow range of values in the range
roughly 35 - 50 \Msun. These remnants complete their lives, finishing
silicon burning in hydrostatic equilibrium with no further pulsing
activity and mass ejection. The duration of activity, from the onset
of pulsations until the iron core collapses, can span many orders of
magnitude, from a few hours to 10,000 years.

Though PPISN have also been extensively studied
\citep{Bar67,Woo86,Heg02,Woo07,Cha12,Che14,Woo15,Yos16}, these studies
have not been as thorough and systematic as for PISN, and there is
some confusion about how these explosions might appear. They certainly
are not all super-luminous. Very few light curves of pure PPISN (i.e.,
PPISN without an artificial core explosion) have been calculated
except for the 110 \Msun \ model of \citet{Woo07};
\citep[e.g.,][]{Bli10}, or parametrized representations thereof
\citep{Mor13}. Other studies of light curves have considered only bare
helium cores \citep{Woo15} or assumed parametrized core explosions to
calculate light curves \citep{Yos16}. The latter violates the common
assumption (which may be wrong; \Sect{superl}) that the cores of stars
that experience the PPI do not explode, but collapse into black holes.
The recent discovery of gravitational radiation from merging
intermediate mass black holes \citep{Abb16a} has also heightened
interest in the evolution of stars in this mass range \citep{Woo16},
and offers new insights into their deaths.

The present paper addresses these issues. The focus is on PPISN in
metal poor stars (10\% \Zsun \ = $1.6 \times 10^{-3}$), primarily as a
way of suppressing, but not eliminating mass loss in the presupernova
star. The key quantities of a PPISN progenitor are the mass of its
helium and heavy element core and the mass and radius of its hydrogen
envelope, if any, when the star first encounters the PPI at central
carbon depletion. Various choices of uncertain mass loss rates give
similar values for these quantities for progenitors with different
initial masses and compositions.  For example, solar metallicity stars
with greatly reduced mass loss rates also give similar
results. Rotation, including the extreme case of chemically
homogeneous evolution (CHE), increases the helium core mass for a
given main sequence mass and may affect the explosion mechanism, but
otherwise gives similar outcomes. The same is true for stars in
interacting binaries in which the envelope and part of the core may be
lost, or the envelope mass increased by accretion.

A great variety of light curves results from explosions of differing
pulsational power and interval in progenitors of different mass and
radius.  Some are ultra-luminous, others are quite faint, and many are
relatively normal Type IIp and IIn supernovae. The
nucleosynthesis is unique, however. Since it is usually assumed
(though see \Sect{superl}) that the elements deep in the core all end
up in a black hole, the new elements are restricted to lighter ones
ejected in the pulses. He, C, N, and O are abundant, and some Ne, Na,
and Mg may be ejected.

To begin our discussion, the physics of the PPI is briefly
described. This includes both the physics used in the code
(\Sect{physics}), especially for mass loss, as well as a brief
discussion of the physics of the PPI itself (\Sect{taukh}). Since the
outcome of the PPI depends critically upon the helium core mass, some
time is spent (\Sect{hecoreexp}) reviewing the outcome of instability
and explosion in bare helium cores of constant mass. This has the
advantage of removing some of the uncertainties in the mass loss rate,
convection theory, rotationally-induced mixing, and binary mass
exchange which affect the final helium core mass as a function of main
sequence mass. It also produces a set of models that are appropriate
for hydrogen-stripped supernovae or for the products of CHE. Surveys
of helium core evolution have been done before
\citep{Woo07,Woo15}. These differ in carrying a larger nuclear
reaction network, using improved stellar phsics, and providing more
detail of the observational outcomes.

The discussion then moves to full star models calculated for a
metallicity 10\% that of the sun. A grid of masses is treated that
spans the range in which the PPI is observed to occur in the stars
without rotation or binary interaction, 70 - 140 \Msun, and shows the
results of varying the uncertain mass loss rate
(\Sect{tmods}). Attention is paid to the bolometric light curves
expected for stars of different masses. These turn out to include
long, low luminosity red transients (\Sect{t70}), ordinary Type II
supernovae (\Sect{t90}), long, irregular, luminous supernovae
(\Sect{t100}), recurrent supernovae, some ultraluminous (\Sect{t110}),
and long transients that are not ordinary supernovae, but essentially
young supernova remnants with intense circumstellar interaction
(\Sect{t120}).

Subsequent sections explore a more limited grid of masses calculated
for solar metallicity (\Sect{solar}), for low-metallicity blue
progenitors (\Sect{lbv}), and for rotating stars (\Sect{rotate}). With
a dramatic reduction in mass loss, near solar metallicity stars are
capable of making PPISN virtually identical to those calculated for
the low-metallicity stars. Models where a small presupernova radius is
enforced show what might happen if the progenitor star is a blue
supergiant (BSG) or luminous blue variable (LBV). The light curves,
even for the same helium core masses, are appreciably different,
especially at early times. The rotating models show a shift downwards
in the main sequence mass necessary to produce PPISN, and also
demonstrate that the cores of the stars are rapidly rotating when they
die, which may have interesting implications for {\sl how} they die.

None of the conservative, ``first principles'' models considered here
produce supernovae as bright as the brightest ``superluminous
supernovae'' (SLSN).  Most stay below 10$^{44}$ erg s$^{-1}$ and none
emit more than $5 \times 10^{50}$ erg of light, with only a few models
briefly surpassing that luminosity. More speculative models are thus
considered (\Sect{superl}) in which rapid rotation launches at least a
partial explosion of the star when the iron core collapses. One
motivation is the observation in recent 2D and 3D simulations of MHD
core collapse of jet formation. It may be that leaving a large black
hole remnant and producing an energetic explosion are not incompatible
hypotheses. Given the freedom to invoke rotationally-powered
explosions {\sl and} the ejection of large masses by the PPI, more
luminous transients with smoother light curves are possible.

The next section (\Sect{etacar}) discusses the (highly speculative)
possibility that Eta Carinae is a PPISN in progress. The idea has
appeal, but requires that Eta Carinae have been a more luminous
supernova and a more energetic explosion some time in the past than
most people presently believe. It offers the tantalizing prospect,
however, that the main ``star'' in Eta Carinae is actually a
Wolf-Rayet remnant experiencing Kelvin-Helmholtz evolution on its way
to becoming a massive black hole.

The nucleosynthesis expected from PPISN is then briefly reviewed
(\Sect{nucleo}) and the relevance of stars in this mass range for
gravitational radiation briefly explored
(\Sect{gwave}). \Sect{conclude} summarizes the principle conclusions
of the paper and gives a number of possible observational counterparts
to PPISN in need of further study.

\section{Code Physics and Assumptions}
\lSect{physics}

\subsection{ Basic Code Physics}
\lSect{procedure}

All stars and explosions were modeled using the KEPLER code
\citep{Wea78,Wea93,Woo02}.  A value of 1.3 times the \citet{Buc96}
rate for $^{12}$C($\alpha,\gamma)^{16}$O was employed. Additional
description of the code physics is given in \citet{Woo07a} and
\citet{Suk16}. Rotationally-induced mixing was treated, for those
models that included rotation, as described by \citet{Heg00}, and
magnetic torques were included as described by \citet{Heg05}. The use
of an implicit hydrodynamics code was essential to the study of PPSN
which often required modeling cores that were still in tight
hydrostatic equilibrium, while simultaneously following shock waves in
tenuous, previously ejected matter. A typical calculation required
from several days to a week on a single desktop CPU, with most of the
time being spent in the large reaction network.  Calculations employed
1200 to 1900 zones, continuously redistributed so as to resolve
gradients in temperature, density, and composition. Typical runs took
from 20,000 time steps for simple PISN, to 60,000 or more steps for
PPISN with several pulses. Several modifications to the standard set
up were necessary to follow these events which often made extreme
excursions in density and temperature as they contracted, exploded,
and then contracted repeatedly.

A nuclear reaction network of at least several hundred nuclei was
directly coupled to the stellar model. Use of the
``quasi-equilibrium'' and ``nuclear statistical equilibrium''
approximations was avoided. Frequently, the cores would experience
oxygen and silicon burning in their centers, producing a central
region of iron, and then explode to low density and temperature, and
then contract back to ignite silicon burning again later. It was
important to follow the weak interactions during both the high
temperature burning and the long, cool phases where the temperature
was frequently less than 10$^9$ K and the quasiequilibrium network
would have failed to converge. A small network would not have sufficed
to follow weak interactions accurately. In all cases the adaptive
network approach proved stable and conserved mass to high accuracy.
The network used was complete up to germanium (Z = 32), which is
sufficient for following energy generation and electron capture. In
several cases, the network was extended to bismuth to accurately track
the weak s-process of nucleosynthesis (\Sect{nucleo}).

It was also important to follow convective mixing in the bound
remnants, but not in shock waves or in the ejected shells which were
being carried in the same simulation. Convective mixing during the
interpulse period affected the distribution of fuel for the next flash
and needed to be included, but tracking convective mixing in a shock
wave in the explosively ejected matter would have been unphysical
and unstable.  Convection extending all the way to the surface of any
bound remnant gave that remnant an unphysically large luminosity
during the interpulse period. The solution was simple. Convection was
turned off in all zones exterior to a few tenths of a solar mass
beneath the final remnant.

Shells ejected to very large radii, greater than 10$^{17}$ cm
sometimes needed to be manually removed, especially when they became
compressed and thin. Resolving fine structure in both the distant
shells and a collapsing iron core required a greater precision than
the code was set up to handle. The effect of these distant shells on
subsequent pulses was negligible, amounting to no more than a dense
interstellar medium surrounding the star. All mass that was ejected
had its composition and energy added to the totals.

The opacity in the ejecta posed a special problem. The ejecting shells
expanded to such low density that it was not on existing grids of
stellar opacities. Treatment of the bound-free and bound-bound
opacities in the presence of large velocity shear would have posed
special problems. The approximation made here was to employ only
electron scattering opacity everywhere after the pulsations began, and
to calculate the electron abundance with an accurate Saha solver that
included all ionization stages of 19 elements up to nickel. An opacity
floor of 0.001 to 0.01 cm$^2$ g$^{-1}$ was also assumed to account
very approximately for other low temperature sources of opacity
besides electron scattering. This was important, for example, during
the long interpulse periods when substantial matter fell back from an
ejected shell and accreted, essentially at the Eddington luminosity on
the core. The luminosity of the core from this accretion was
substantial and opacity dependent because zoning at the accreting
surface was coarse.

Particularly challenging were the (unrealistic) density spikes that
developed when fast moving shells snowplowed into ones moving slower.
In a 1D code, there was no way for mixing and overturn to occur, and
no resistance to compression in regions with small velocity gradients,
unless the density became so high that ideal gas pressure offered some
resistance. The pile up often included a large fraction of the entire
ejected mass in a thin dense shell, all moving with the same speed.
These density spikes could sometimes had a contrast $\Delta \rho/\rho$
with their surroundings of several orders of magnitude. In a 2D study
that followed mixing instabilities \citep{Che82}, the pile up would
still exist, but not with such great contrast
\citep{Che14,Che16}. Very tight convergence criteria on the radius (as
small as $\Delta r/r = 10^{-13}$) were necessary to keep the
calculation stable and, even then, often failed. Collisions of these
thin massive shells often produced unphysical spikes in the
luminosity.  Smearing out the spikes would broaden the peaks in the
light curve due to the collisions of individual shells, while roughly
preserving the total radiated energy. In cases where more than one
collision occurs, the medium through which subsequent shocks propagate
might be clumpy. See also \Sect{conclude}.

\subsection{Pre-Explosive Mass Loss and Opacity}
\lSect{mdot}

The most uncertain aspect of thermonuclear PPISN is not how they
explode, but how presupernova evolution produces the necessary helium
core masses. A proper treatment of mass loss is critical to
associating a given final behavior with a main sequence mass. If the
star loses all its hydrogen envelope, and enough of its helium core to
shrink below 30 \Msun, the pair-instability is avoided. This is
probably the case for all stars of solar metallicity.

Mass loss in very massive stars is a subject of great interest and
considerable uncertainty. Generally speaking, for single stars, the
mass loss is of three varieties: line-driven mass loss which dominates
on the main sequence and for other hot stars
\citep[e.g.][]{Vin01,SmiOw06,Vin11}; less well understood continuum
driven mass loss and envelope instabilities which may play an
important role in luminous blue variables
\citep[e.g.][]{Bes14,Owo15,Pet16}; and mass loss where grain formation
is important, as in red supergiants \citep[RSGs;][]{Voo00,Gro09}. Of
these, line-driven mass loss is most studied and best
understood. Analytic functions and routines are available to
facilitate the use of such rates in computer codes.

The mass loss formula used here for the stars with hydrogen-rich
envelopes is taken from \citet{Nie90}. Correcting a typo in the
abstract,
\begin{equation}
\dot M \ = \ 9.63 \times 10^{-15} \left(\frac{L}{\Lsun}\right)^{1.24}\left(\frac{M}{\Msun}\right)^{0.16} \left(\frac{R}{\Rsun}\right)^{0.81} \ {\rm \Msun \ y^{-1}}.
\end{equation}
This formula is dated and of questionable accuracy, especially for the
stars considered here, which lie far from the masses and metallicities
for which the fit was originally calibrated. The expression is simple,
however, and easily applied across the HR-diagram. It was adopted, but
multiplied by various constants less than one to explore the
sensitivity of outcomes. It was also multiplied by $(Z/\Zsun)^{1/2}$
to approximate its scaling with metallicity. For hot stars with
line-driven winds, a better scaling might be $(Z/\Zsun)^{0.64}$ or
$(Z/\Zsun)^{0.69}$ \citep{Vin01}.  Based upon an analysis of RSGs in
the Milky Way, LMC, and SMC, \citet{Mau11} suggest a scaling of
$(Z/\Zsun)^{0.7}$, but with an uncertain factor overall of at least a
factor of four. Given the limited metallicity range studied here, the
difference could be accounted for by a small shift in the overall mass
loss rate. Unfortunately, supergiant mass loss is important for many
of the stars and the scaling with metallicity there is unknown.

Using this formula and some simple approximations, one can estimate
the necessary conditions for the pair instability to occur in stars
that still retain some hydrogen envelope. Because such massive stars
all have luminosities near the Eddington limit, their lifetimes on the
main sequence are nearly constant at 3 million years, and their helium
burning lifetime is close to 300,000 y (i..e., the Eddington
luminosity divided by the energy release from hydrogen and helium
burning assuming that the whole star burns). These numbers are
validated later in the stellar models (\Sect{tmods}) and are good to a
factor of two.  There is some trade off in that the luminosity is not
quite Eddington and the fraction of the star's mass that burns to
helium is only about 3/7. From 70 to 140 \Msun, the main sequence
lifetime for non-rotating stars decreases from 3.3 My to 2.5 My and
the helium burning lifetime varies from 330,000 to 270,000 y.  The
luminosities both on the main sequence and during helium burning are,
at all times, within a factor of two of $7 \times 10^{39}$ erg
s$^{-1}$. For hydrogen burning, a more accurate approximation is
$L_{\rm ms} \approx 6 \times 10^{39} (M/100 \, \Msun)^{3/2}$ erg
s$^{-1}$; for helium burning, $L_{\rm He} \approx 9 \times 10^{39}
(M/100 \, \Msun)$ erg s$^{-1}$, where M is the zero age main sequence
mass of the star.  Furthermore, the radius does not vary greatly on
the main sequence, $R_{\rm ms} \approx 1.0 \times 10^{12} (M/100 \,
\Msun)$ cm. The radius during helium burning does vary greatly,
however, according to whether the star is a RSG, $R_{\rm He} \approx 1
- 2 \times 10^{14}$ cm, or a blue one, $R_{\rm He} \ltaprx {\rm few}
\times 10^{13}$ cm.  This variation introduces uncertainty into the
esimated mass loss.

Together these approximations for L and R, an assumed hydrogen burning
lifetime of 3 My and a helium burning lifetime of 0.3 My imply a total
mass lost in solar masses of
\begin{equation}
\Delta M \ = \ \left(8 \, M_{100}^{2.83} \ + 55\, R_{14}^{0.81}
M_{100}^{1.40}\right)\left(\frac{Z}{\Zsun}\right)^{1/2},
\end{equation}
where the first term is mass lost on the main sequence, the second
term during helium burning, and $R_{14}$ is the average radius during
helium burning in units of $10^{14}$ cm. Assuming the mass of the
envelope is 4/7 the mass of the star, the entire envelope will be lost
when
\begin{equation}
F_{\rm env} \ \approx \ \left(0.13 M_{100}^{1.83} \ + \ 0.96
M_{100}^{0.40} R_{\rm He,14}^{0.81}\right) (\frac{Z}{0.1
  \Zsun})^{1/2}.
\end{equation}
For solar composition $R_{\rm He,14} \approx 2$; for 10\% solar
metallicity $R_{\rm He,14}$ varies from 0.3 to 2, but is usually
closer to 2.  Since $F_{\rm env}$ must be less than one, this equation
implies that no solar metallicity star will end its life as a PPISN
(or a PISN), but an appreciable fraction of stars with metallicity
below 1/3 \Zsun \ might \citep[see also][]{Geo13,Spe15,Lan07,Yus13},
especially if the metallicity scaling of \citet{Vin01} is employed
instead of Z$^{0.5}$, or the mass loss rate of \citet{Nie90} is an
overestimate.  This motivates the choice Z = 0.1 \Zsun \ for emphasis
in the present survey.

This estimate is very uncertain. The mass loss rate implied for a 100
\Msun \ solar metallicity star with radius $2 \times 10^{14}$ cm and
luminosity 10$^{6.5}$ \Lsun \ using the analytic expression above
would be 10$^{-3}$ \Msun \ y$^{-1}$. For the same star, \citet{Vin01}
give a mass loss rate of $2.5 \times 10^{-4}$ \Msun \ y$^{-1}$ (though
the temperature here, 4500 K, may be too cool to apply the Vink
formula). \citet{SmiOw06} estimate an upper limit to line-driven mass
loss of $3 \times 10^{-4}$ \Msun \ y$^{-1}$ for a star of this
luminosity \citep[though see][for a different approach]{Mue08}, and
all of this theoretical work on line-driven winds neglects continuum
driven processes, instabilities, and grain formation. The default mass
loss rates employed here are probably uncertain to at least a factor
of a few and are likely overestimates. This affects the metallicity
range where PPISN might occur, and even {\sl if} they occur. To
compensate for this uncertainty, the mass loss rates in the models
were multiplied by various factors less than 1.

Opacity, semiconvection, and rotation also play important indirect
roles in determining the mass loss. Less efficient semiconvection
favors more time as a RSG and thus more mass
loss. Rotational mixing affects the composition of the hydrogen
envelope and its opacity as well as the luminosity of the helium
core. Higher opacities favor a larger stellar radius and hence greater
mass loss. The OPAL opacities used here \citep{Rog92,Igl96} have a
well-known ``iron bump'' at low temperature that can lead to
artificial density inversions in the outer envelopes of RSGs
\citep{Owo15}. This tends to overinflate the star and thus boost
its mass loss. All of these effects need further study. 

So long as the helium core is not uncovered prior to death, similar
core masses and similar explosions, including e.g., remnant masses,
result. The exact value of the mass lost, as well as the secondary
uncertainties in opacity, semiconvection and rotation, serve to define
the observational properties of the presupernova star and the
metallicities that can make PPISN. To first order though, they do not
affect the explosion itself.

It also may be that the stars rotate sufficiently rapidly to
experience CHE in which case giant formation is avoided altogether and
the formula of \citet{Nie90} is not applicable. The treatment of mass
loss for such stars is deferred to \Sect{rotate}.

\begin{deluxetable*}{cccccccc} 
\tablecaption{HELIUM CORE EXPLOSIONS} 
\tablehead{ Mass & M$_{\rm CO}$ & Pulses & Duration & KE-pulse &
   M$_{\rm Fe}$ & M$_{\rm eject}$ & M$_{\rm remnant}$ \\
  (\Msun) &  (\Msun) &    & (sec) & (10$^{51}$ erg)  &
  (\Msun)  & (\Msun) & (\Msun) }
\startdata
30  & 24.65 & stable      &    -     &     -   &  2.34 &    -   & 30.00 \\
32  & 26.30 & stable      &    -     &     -   &  2.38 &    -   & 32.00 \\
34  & 28.01 & 5 weak      &  2.3(3)  &  0.0012 &  2.51 &   0.13 & 33.87 \\
36  & 29.73 & 33 weak     &  1.8(4)  &  0.0037 &  2.53 &   0.18 & 35.82 \\
38  & 31.40 & $>$100 weak &  4.2(4)  &  0.0095 &  2.65 &   0.34 & 37.66 \\
40  & 33.05 & 9 strong    &  7.8(4)  &  0.066  &  2.92 &   0.97 & 39.03 \\
42  & 34.77 & 18          &  2.0(5)  &  0.26   &  2.68 &   2.65 & 39.35 \\
44  & 36.62 & 11          &  7.7(5)  &  0.83   &  3.18 &   5.02 & 38.98 \\
46  & 38.28 & 11          &  1.2(6)  &  0.77   &  2.40 &   5.51 & 40.49 \\
48  & 40.16 &  8          &  3.8(6)  &  0.94   &  2.53 &   6.65 & 41.35 \\
50  & 41.83 &  6          &  1.2(7)  &  0.86   &  2.76 &   6.31 & 43.69 \\
51  & 42.59 &  6          &  1.9(7)  &  1.00   &  2.37 &   7.80 & 43.20 \\
52  & 43.52 &  5          &  1.4(8)  &  0.99   &  2.47 &   7.87 & 44.13 \\
53  & 44.34 &  4          &  7.8(8)  &  0.86   &  2.68 &   4.73 & 46.70 \\
54  & 45.41 &  4          &  4.7(9)  &  0.94   &  2.16 &   6.85 & 47.15 \\
56  & 47.14 &  3          &  3.4(10) &  0.56   &  2.04 &   7.99 & 48.01 \\
58  & 48.71 &  3          &  8.0(10) &  1.1    &  2.00 &  12.14 & 45.86 \\
60  & 50.54 &  3          &  8.5(10) &  0.75   &  1.85 &  12.02 & 47.98 \\
62  & 52.45 &  7          &  2.2(11) &  2.3    &  3.19 &  27.82 & 34.18 \\
64  & 54.14 &  1          &    -     &  4.0    &    -  &   64   &    -  \\
\enddata
\lTab{hecore}
\end{deluxetable*}

\section{Helium Cores}
\lSect{helium}

Many general characteristics of PPISN can be understood from a simple
study of pure helium stars evolved to the supernova stage at constant
mass without rotation. Not only are the properties of PPISN most
sensitive to the helium core mass at death, but often in nature, most
or all the envelope of a hydrogenic star is lost, either to a binary
companion or a wind, so these models should have observable
counterparts in nature. CHE (\Sect{rotate}) will also produce stars
whose late stages of evolution closely resembles that of bare helium
cores.

\subsection{General Characteristics of the Pulsational Pair Instability}
\lSect{taukh}

In the weakest case, for the helium cores less than about 40
\Msun, the PPI manifests as a small amplitude, vibrational instability
brought about by the temperature sensitivity of the nuclear reactions
and the proximity of the structural adiabatic index to 4/3
(\Sect{hecoreexp}). As the core mass increases or the abundance of
nuclear fuel declines, however, the instability becomes more
pronounced. The amplitude of the pulses increases and they become
non-linear. A major readjustment of the core structure occurs during
each pulse that requires a Kelvin-Helmholtz time scale to recover. The
most interesting explosions happen in this non-linear regime.

There, the characteristics of the PPI can be understood from an
examination of the contraction, through its Kelvin-Helmholtz phase, of
a helium star of constant representative mass. The relevant helium
core masses for the PPI are in the range 35 to 65 \Msun \ and final
remnant masses are typically 35 - 45 \Msun. \Fig{taukh} shows the
evolution of a 40 \Msun \ helium core in which nuclear burning, but
not neutrino losses, has been suppressed. The evolution of a
carbon-oxygen (CO) core of the same mass would be very similar. The pair
instability has a strong onset around $3 \times 10^9$ K, and time in
the figure is measured prior to that point. After reaching $3 \times
10^9$ K, the instability develops on a time scale of less than a
minute.

% fig 1 - Kelvin Helmhotz time o40a1 
\begin{figure}
\includegraphics[width=0.49\textwidth]{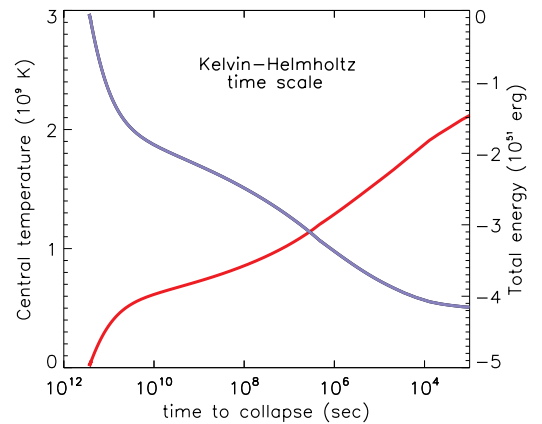}
\caption{Kelvin-Helmholtz evolution of a 40 \Msun \ helium star in
  which nuclear burning has been suppressed. The red curve gives the
  central temperature in billions of K as a function of time. Time is
  measured backwards from that point when the central temperature
  reaches $3 \times 10^9$ K and the core becomes dynamically
  unstable. Also shown is the net binding energy (internal plus
  gravitational binding energy) of the helium star. This is
  a negative number that is zero when the star is unbound. The change
  in slope at $\sim10^{10}$ s ($T_9 = 0.65, \rho = 8000$ g cm$^{-3}$)
  reflects the change from radiation dominated to neutrino dominated
  cooling and an acceleration of the contraction. The Kelvin-Helmholtz
  time starting from very low density is $2.7 \times 10^{11}$ s. This
  is an upper bound to the recurrence time for any single pulse, and
  an approximate upper bound to the total duration of pulsational
  activity. During each pulse the total energy becomes less negative,
  the core expands, cools, and moves to the left to a new point on the
  curve. The time between pulses is the Kelvin-Helmholtz time at this
  new binding energy. It is very short for weak pulses and very long
  for strong ones. \lFig{taukh}}
\end{figure}

During the explosive burning, typically of oxygen, an amount of energy
is released that depends on the mass and composition of the core. More
massive cores require more burning in order to reverse their
implosion. Cores that have already burned some oxygen also bounce
deeper and burn more, provided there is still fuel left to
burn. Explosive burning leads to rapid expansion and cooling. Part of
the energy released powers a shock wave that can eject matter from the
edge of the core, but, qualitatively, the core's evolution is not
altered, so long as its total mass stays roughly constant and nuclear
reactions during the contraction are negligible. After a brief period
of large amplitude oscillation, the core settles back down into a new
state of hydrostatic equilibrium at a less negative net binding energy
and commences a new stage of Kelvin-Helmholtz contraction.

The time until the next pulse depends upon the net binding energy
following the prior pulse. A single pulse appreciably over $4.2 \times
10^{51}$ erg in the 40 \Msun \ model shown in \Fig{taukh} would
completely disrupt the core, producing a regular PISN. The abrupt
generation of $2 \times 10^{51}$ erg, on the other hand, would put the
star back into a state similar to what existed 10$^{10}$ s earlier. Of
course, some energy would be lost to mass ejection, and the core
entropy would change due to the burning and radiative losses
(\Fig{ent50}), but qualitatively the evolution would be similar to the
first contraction. The core explodes, stays bound, but relaxes to a
less tightly bound configuration that experiences another stage of
Kelvin-Helmholtz contraction until it becomes unstable again.  The
process repeats until all fuel is exhausted, the mass is reduced below
a critical value, or the loss of entropy from repeated
Kelvin-Helmholtz episodes removes the instability.  Small explosions
thus recur after a short time, while violent explosions initiate a
longer wait. The interval between pulses is given by the energy of the
prior pulse. Violent pulses also burn more fuel, so there are fewer of
them before the core becomes stable.

% fig 2 - Entropy evolution 50 Msun helium core pulses 
\begin{figure}
\includegraphics[width=0.49\textwidth]{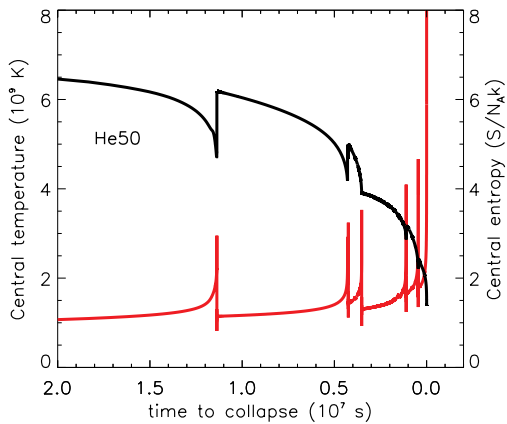}
\caption{Central entropy (black line) and central temperature (red
  line) in a 50 \Msun \ helium core contracting and experiencing the
  PPI. Both quantities are plotted as a function of time until
  iron core collapse in units of 10$^7$ s. The dimensionless entropy
  (S/$N_A k$) rises in response to burning, but decreases due to
  neutrino losses. Spikes in temperature show major core
  pulsations. After the burning moves off-center, the central entropy
  no longer rises significantly during a flash, and the overall
  entropy continues to decline due to neutrino losses. Eventually the
  global entropy becomes small enough that the core becomes sable and
  evolves, without further flashes, to iron core collapse.
  \lFig{ent50}}
\end{figure}

The duration of the pulsations ranges from very short,
essentially the hydrodynamic time scale of the helium core, or 10 
minutes, to the Kelvin Helmholtz time for the marginally bound core,
or several thousand years (\Fig{taukh}). As we shall see, this broad
range of energies and time scales results in a diverse set of observable
phenomena.

Eventually, the core settles into stable silicon burning that produces
an iron core of 2 to 3 \Msun \ (\Tab{hecore}) that collapses to a
proto-neutron star in the usual way. This final evolution is very
unlike ordinary PISN where no silicon or iron core is ever produced in
hydrostatic equilibrium. In lower mass helium cores the PPI is mild
and only afflicts the oxygen burning shell during the last hours and
days of the stars life. For larger cores, though, violent pulses burn
oxygen in roughly the inner 6 solar masses before the star settles
into stable silicon burning. For the highest masses, some silicon has
already burned to iron in the star's center and, during the last
phase, silicon burns in a shell.

% fig 3 - helium core pulsations
\begin{figure*}
\begin{center}
\leavevmode
\includegraphics[width=0.48\textwidth]{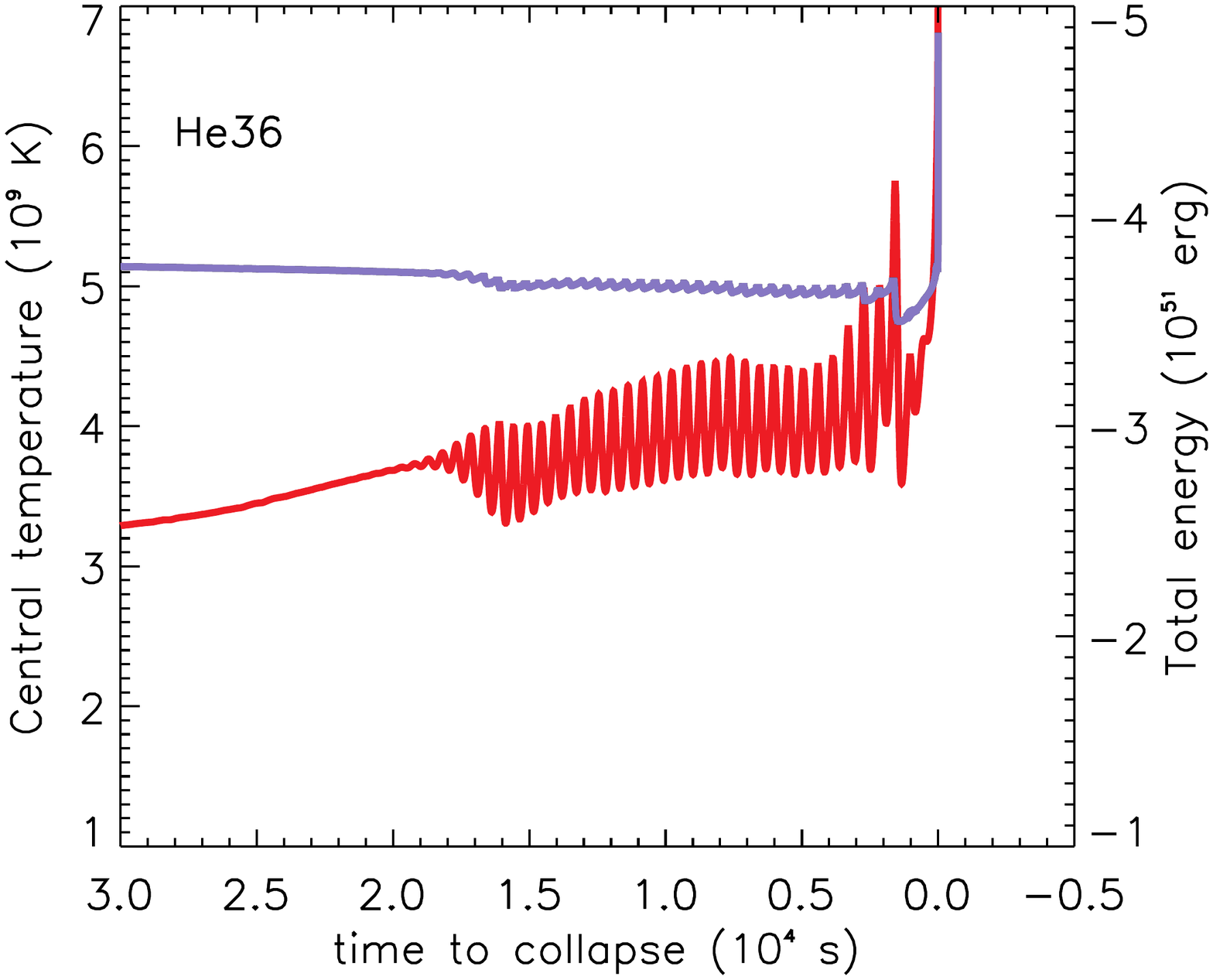}
\includegraphics[width=0.48\textwidth]{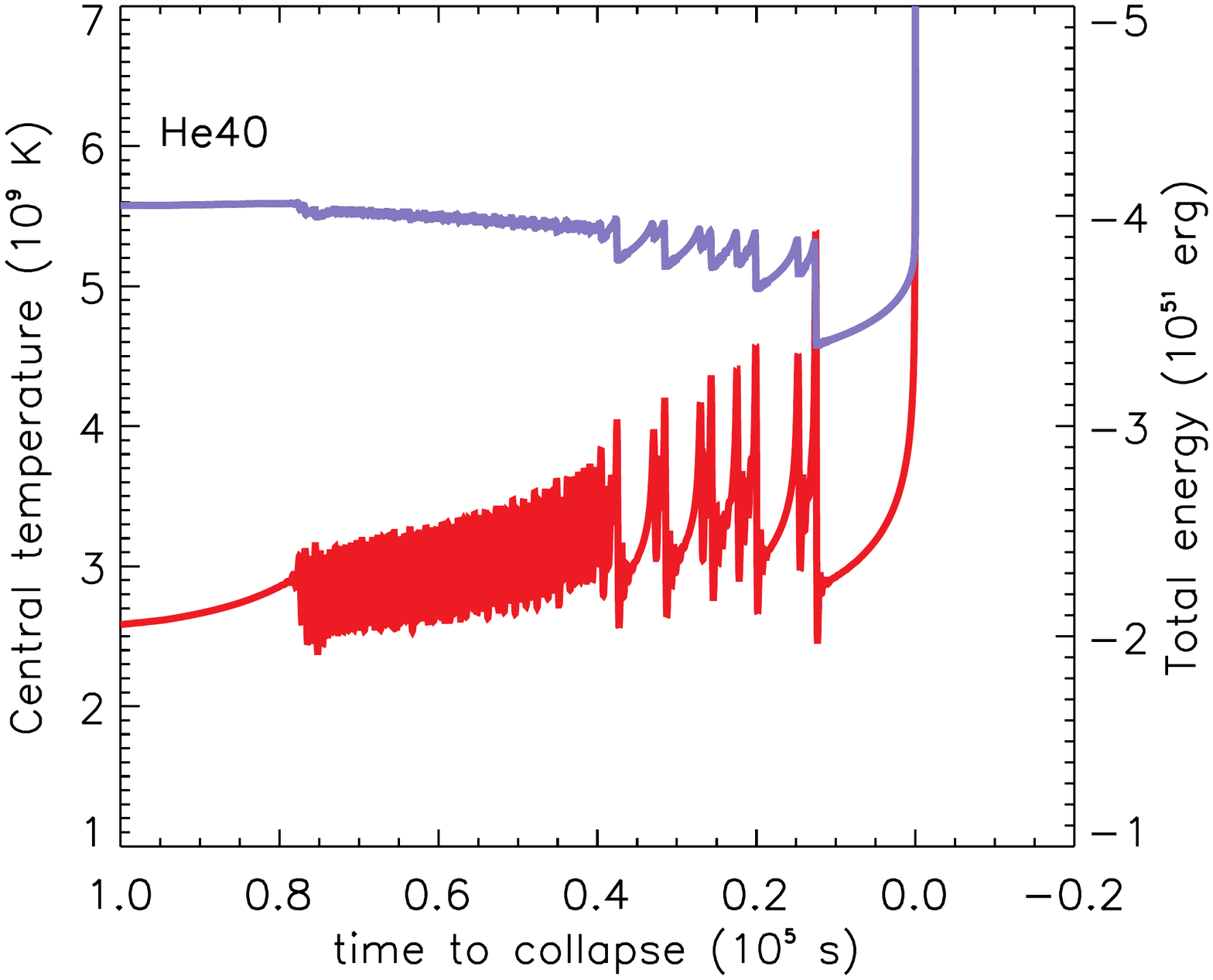}
\vskip 24pt
\includegraphics[width=0.48\textwidth]{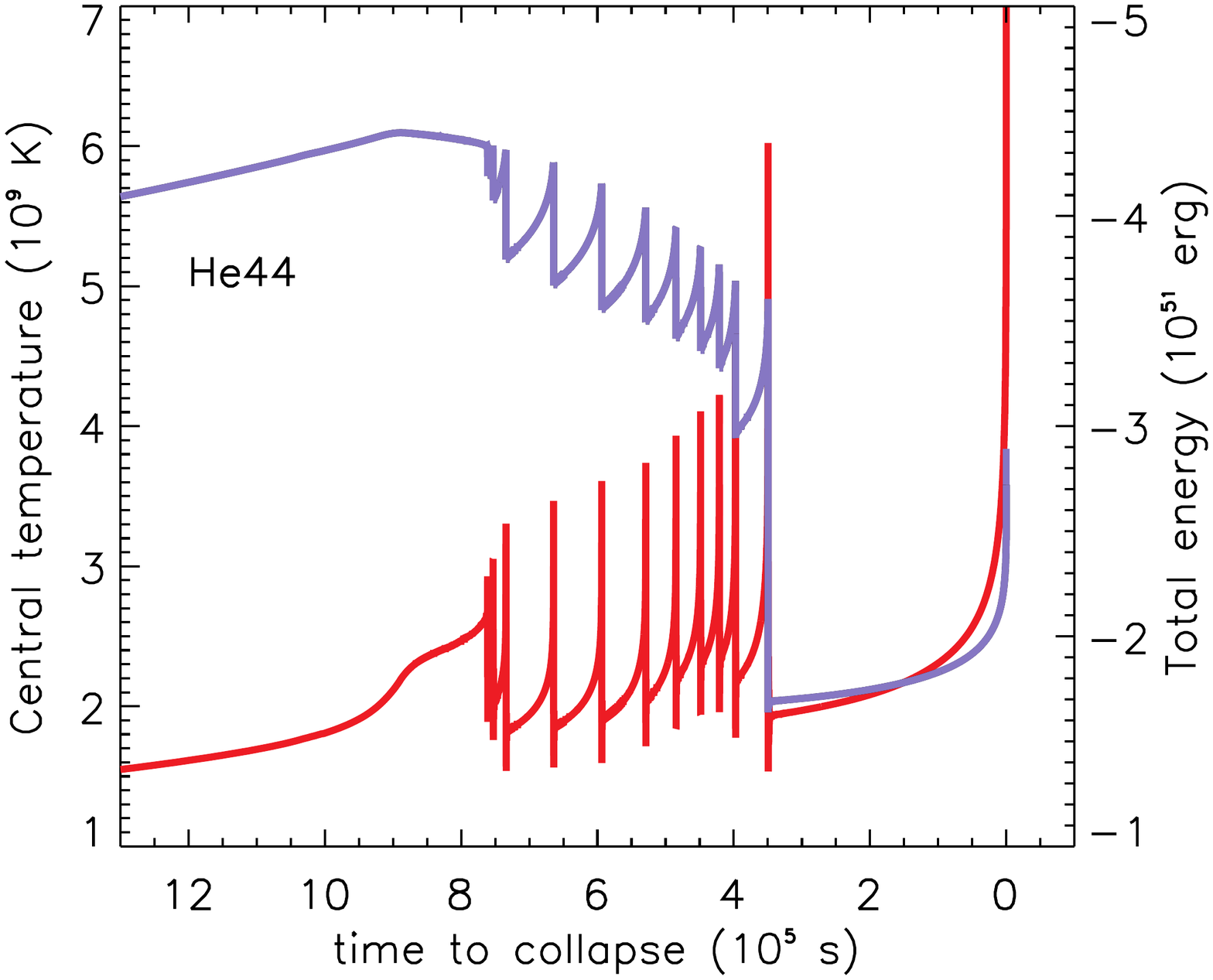}
\includegraphics[width=0.48\textwidth]{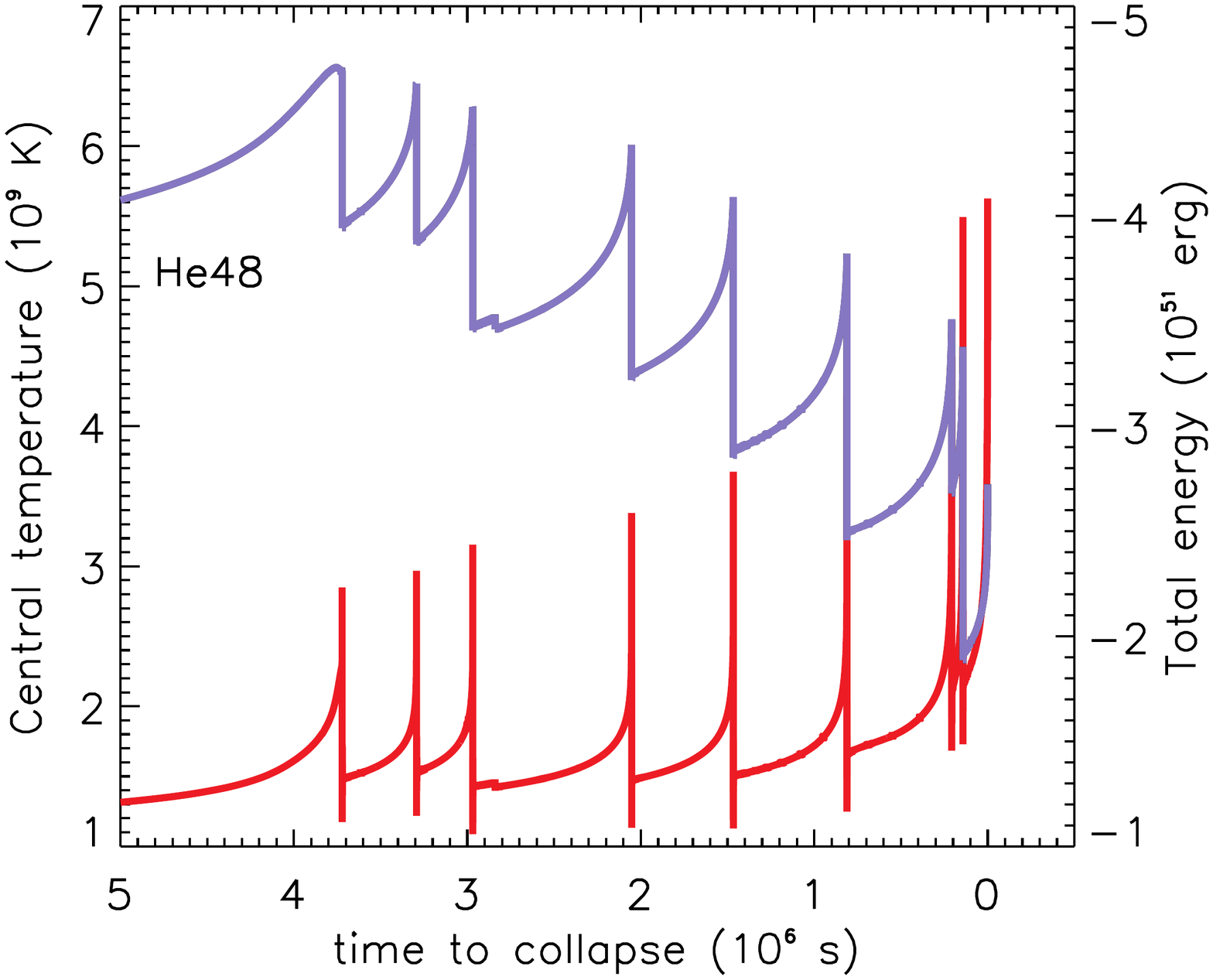}
\vskip 24pt
\includegraphics[width=0.48\textwidth]{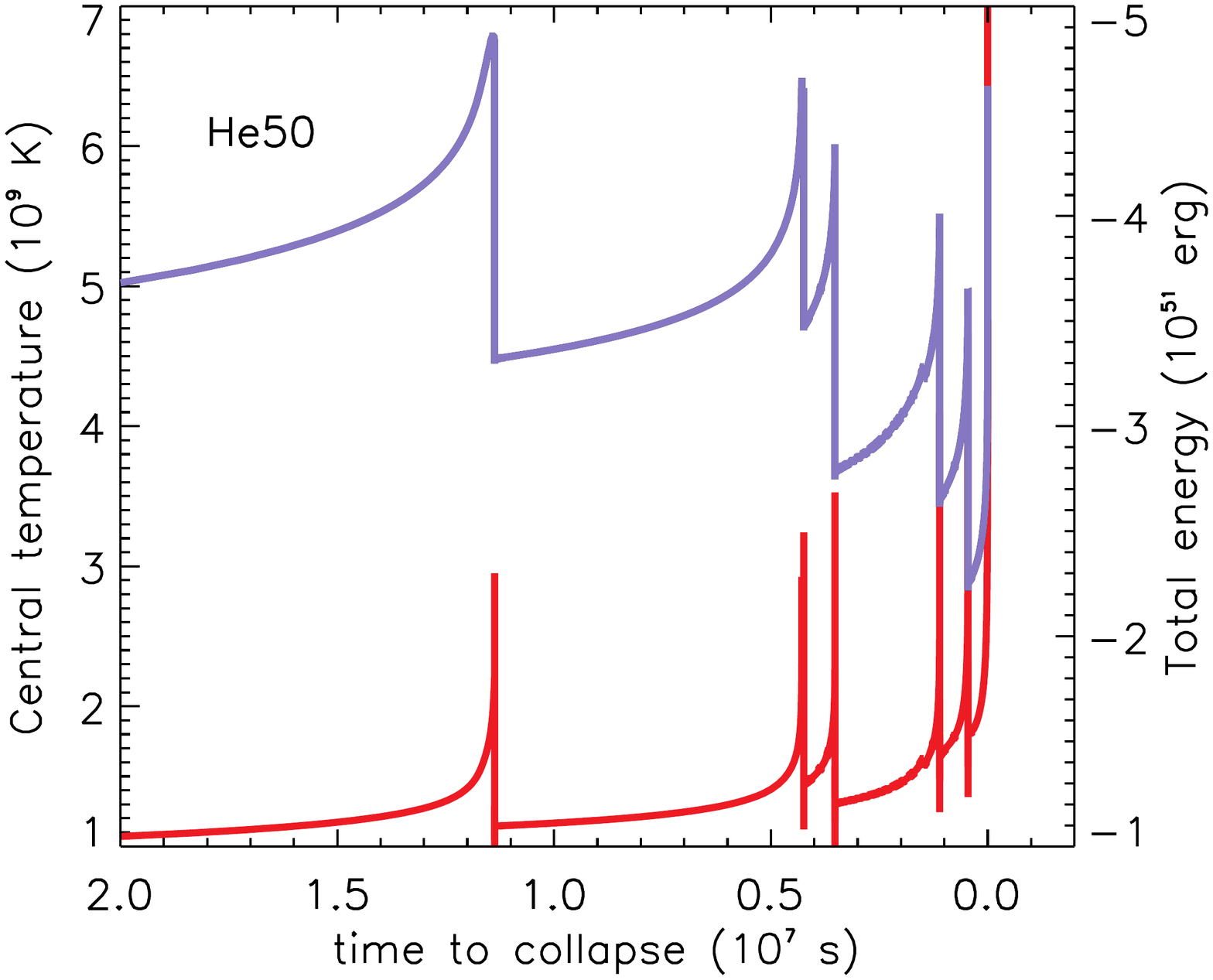}
\includegraphics[width=0.48\textwidth]{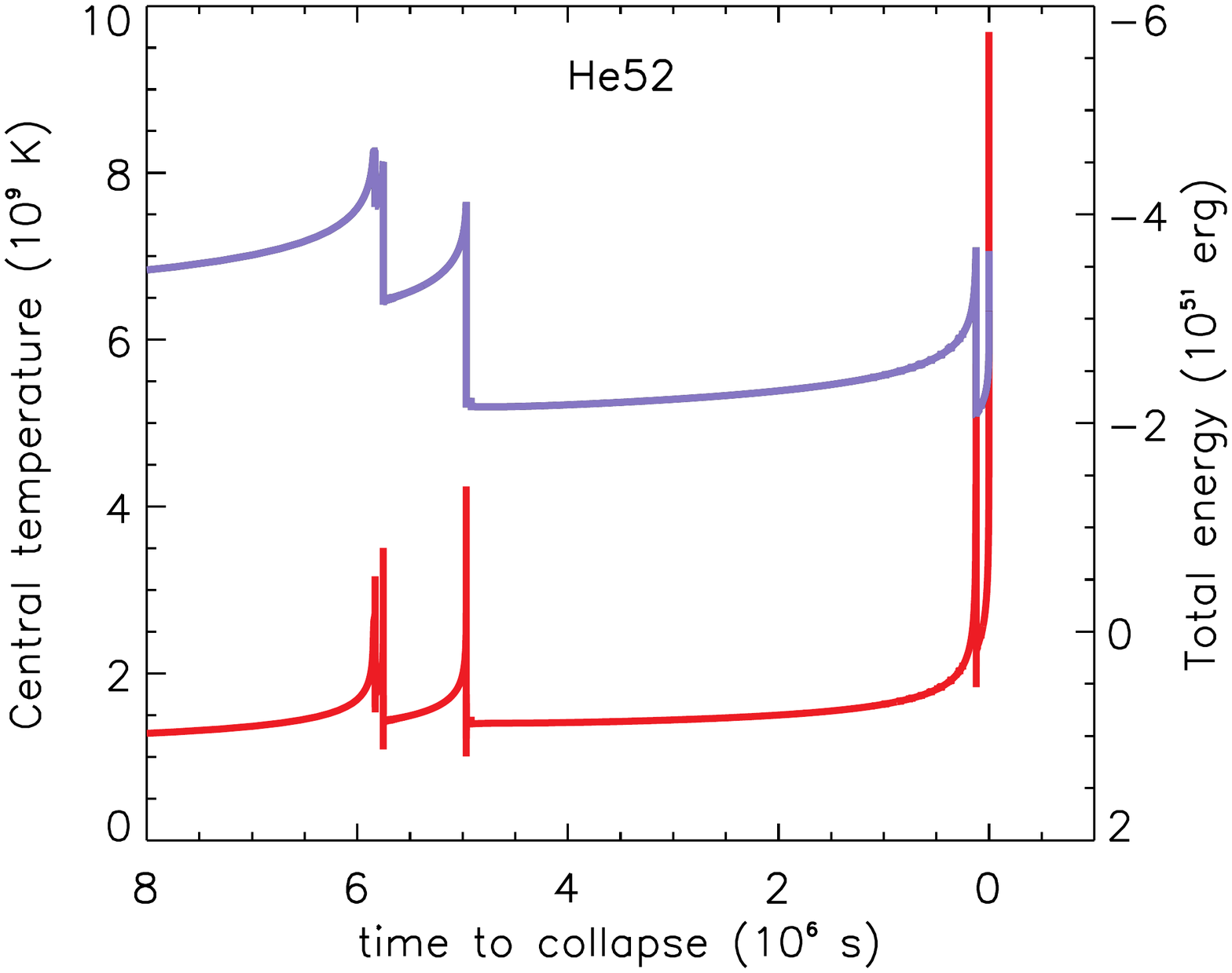}
\caption{Pulse history in bare helium stars of 36, 40 44, 48, 50
  and 52 \Msun. Central temperature in 10$^9$ K is the red line and
  net binding energy in 10$^{51}$ erg is blue. Time, measured prior to
  final core collapse to a black hole, is in units of 10$^{4}$ s for
  the 36 \Msun \ model; 10$^5$ s for the 40 and 44 \Msun \ models;
  10$^6$ s for the 48 \Msun \ and 52 \Msun \ models; and 10$^7$ s for
  the 50 \ model. In the 52 \Msun \ model a strong flash (not shown)
  occurred 4.6 years prior to the final collapse. For that model, the
  panel only shows the activity during the last few months when
  several pulses in rapid succession occurred towards the
  end. \lFig{hepulses}}
\end{center}
\end{figure*}

% fig 4 - helium core pulsations - 56, 60, 62
\begin{figure}
\includegraphics[width=\columnwidth]{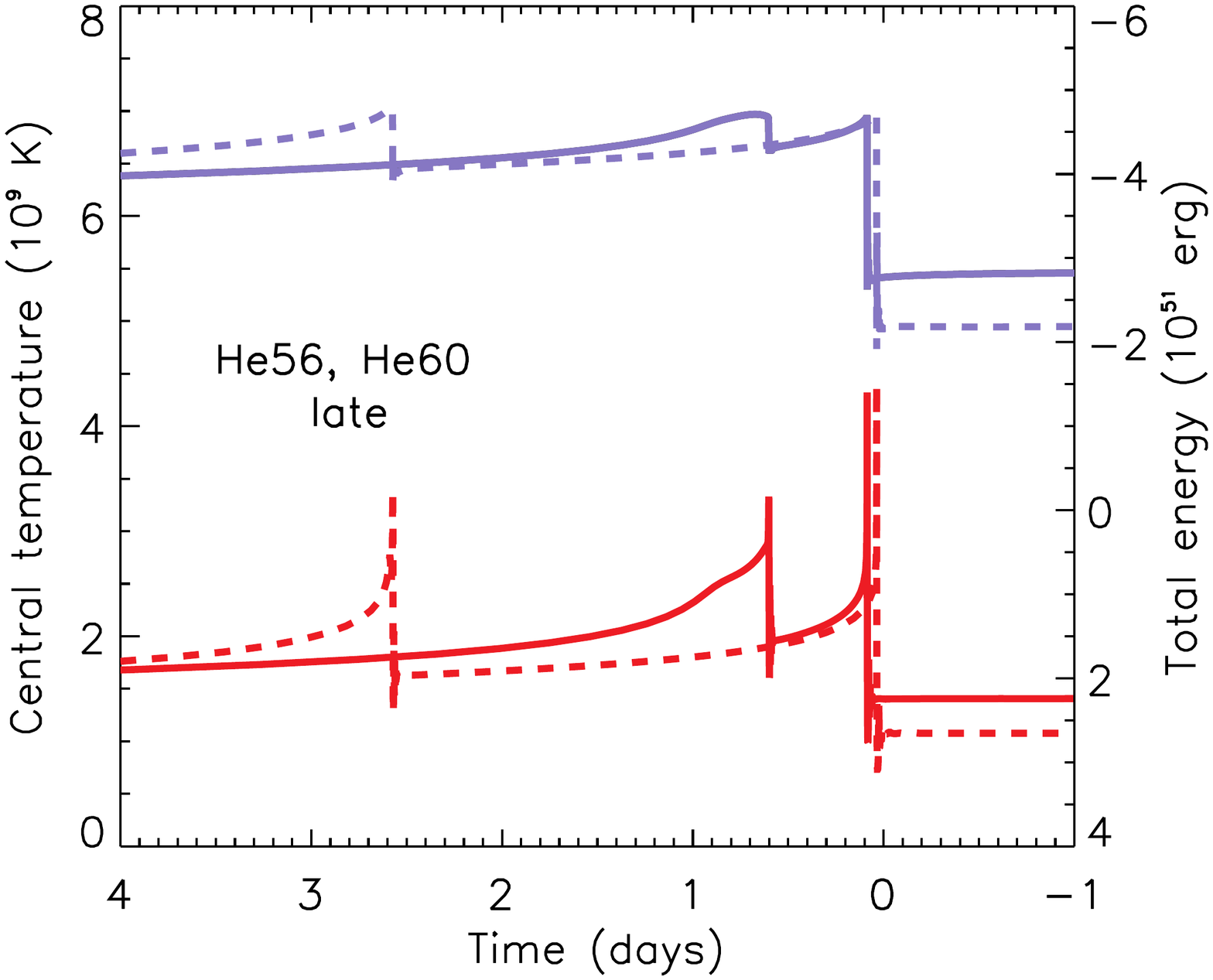}
\includegraphics[width=\columnwidth]{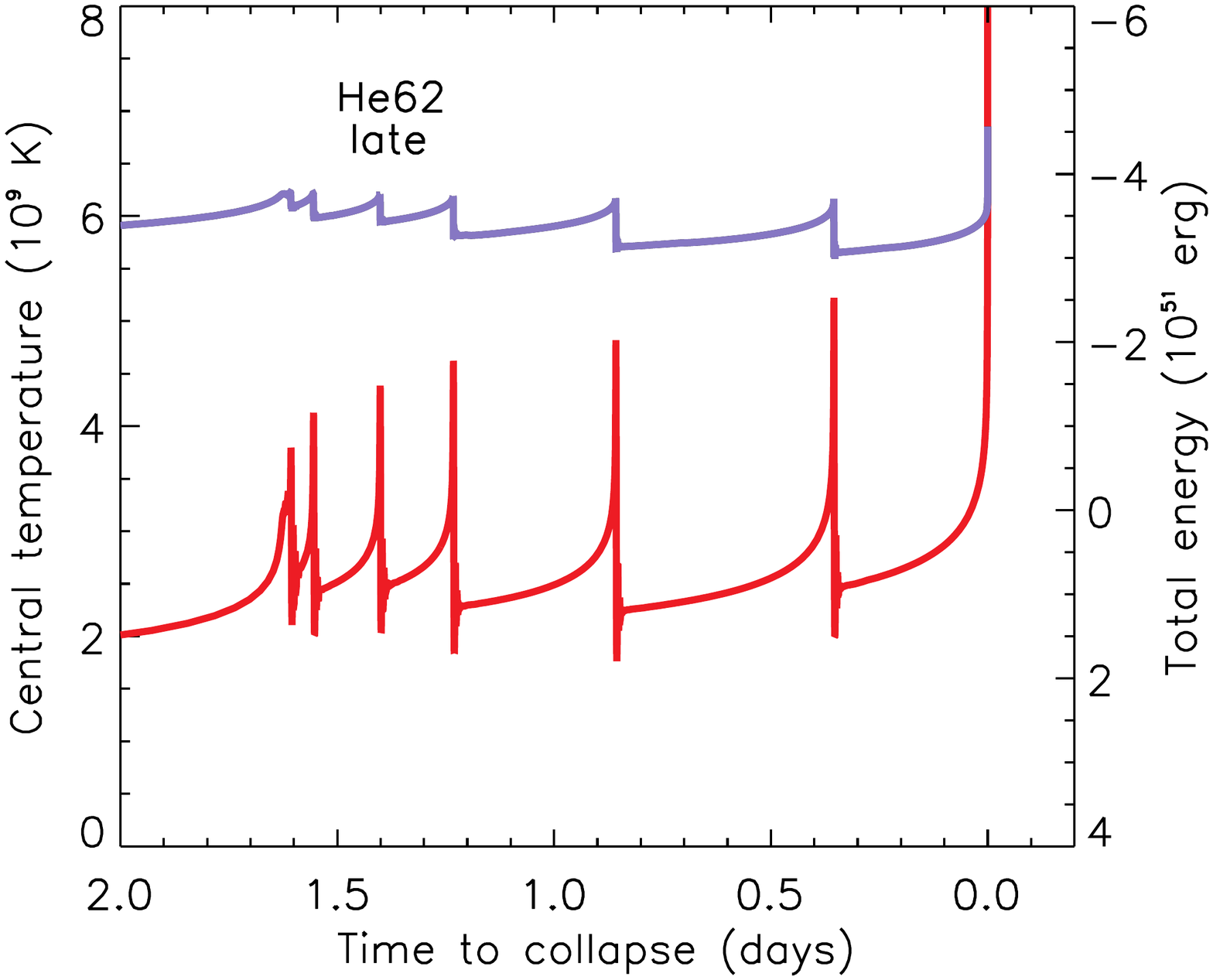}
\caption{(top) History for the second and third pulses of helium
  stars of 56 and 60 \Msun.  Central temperature (red) is in units of
  10$^9$ K; net binding energy (blue) is in units of 10$^{51}$. Time,
  measured with respect to the final pulse before core collapse. For
  the 56 \Msun \ model (solid line), the two pulses commence 1060
  years after the first pulse (not shown) and end 91 days before core
  collapse (not shown). For the 60 \Msun \ model (dashed lines), the
  first pulse was 2680 years earlier and the pulses shown here end 6.0
  years before core collapse. Weak final pulses occurred in both both
  models at the time of core collapse. (bottom) Similar final pulsing
  activity for Model He62. 7000 years have elapsed since the first
  pulse ejected 36 \Msun. Only in the bottom plot is Zero the time of
  iron core collapse. \lFig{hepulses1}}
\end{figure}

\subsection{The Evolution and Explosion of Helium Stars}
\lSect{hecoreexp}

Bare cores consisting initially of pure helium with masses from 30 to
64 \Msun \ (Z= 0) were evolved from the helium burning main sequence
either to iron core collapse, or, in one cases (64 \Msun), to complete
disruption as a PISN (\Tab{hecore}). Mass loss and rotation were
neglected, though the cores could have resulted from rotating
stars. In all cases, most of the helium core burned to carbon and
oxygen before the explosion, so the inclusion of a small amount of
mass loss would have resulted in a WC or WO Wolf-Rayet star
contaminated with a small fraction of helium rather then a star with a
predominantly helium surface, but the explosion dynamics for the same
presupernova core mass would be very similar. For zero metallicity
helium stars, the mass loss rate is expected to be small
\citep{Vin05}, but even for low metallicities the mass loss from a
helium star of such large mass would not be negligible
\citep[e.g.][]{Woo06}. The masses considered here reflect what remains
after all mass loss is finished and the star dies, not necessarily the
helium core mass at the end of hydrogen burning.

Principal results are given in \Tab{hecore} and Figs 3 - 7 \citep[see
  also][]{Woo07,Woo15}.  The duration of the pulsing phase
(\Tab{hecore}) is measured from the onset of the first pulse to
iron core collapse, even though, in the more massive cases, shell
collisions often finished before iron core collapse or continued long
afterwards. The PPI first becomes noticeable near 34 \Msun \ where it
is encountered in the oxygen burning shell during the last few hours
of the star's life. It begins as a series of weak flashes, each
lasting about 500 s, roughly the sound-crossing time for the helium
star. The central density and temperature vary only slightly during
each pulse, but cumulatively, the piling up of shocks from numerous
weak pulses in Model He34 results in the ejection of a about 0.1 \Msun
\ with a kinetic energy $\sim$10$^{48}$ erg. For bare helium cores,
this small ejection produces a very weak transient that would be
difficult to detect. In a RSG however, even this small amount of
energy would be sufficient to eject a significant part of the hydrogen
envelope and produce a faint supernova (\Sect{tmods}). In the absence
of significant rotation, the remainder of the helium core collapses
into a black hole, so this minor ejection would be the only observable
signal of the star's death (other than its disappearance).

For helium cores up to 40 \Msun, essentially the ``linear regime'',
increasing the mass shifts the onset of the PPI to earlier times,
increasing the time the star spends pulsing, the number of pulses that
occur, and their total energy. More mass is ejected, with the energy
eventually reaching $\sim$10$^{49}$ erg at 40 \Msun. Up to this point,
the pulses are only minor perturbations on a monotonically increasing
central temperature with $\Delta T/T \ltaprx$ 30\% (\Fig{hepulses}).
Starting at about 40 \Msun, a qualitative change in behavior occurs,
as dozens of weak pulses give way to a series of less frequent,
explosions. From this point on, the ``pulsations'' cease to to be
perturbations on the core structure and become discrete explosive
events, each generating a dynamic response and significant mass
ejection. Each explosion is followed by relaxation oscillations and an
extended period of Kelvin-Helmholtz contraction to a new unstable
state (\Sect{taukh}).

By 44 \Msun, the interval between pulses is becoming a week or
more, long enough to discern the effects of individual flashes on
the light curve of a Type I supernova (\Sect{helite}). The combined
energy in the pulses also becomes comparable to that of common
supernovae. Above about 50 \Msun, the duration of the pulses exceeds
the time for material coasting at a few thousand km s$^{-1}$ to reach
10$^{15}$ cm, a typical photospheric radius for a supernova.  Shells
thus collide in a region that is not very optically thick and their
differential kinetic energy can be converted into radiation without
much adiabatic degradation. The time when the iron core collapses
begins to lag appreciably after the onset of the light curve, opening
the possibility of recurrent events and surviving stars in supernova
remnants.  By 50 \Msun, the total kinetic energy approaches 10$^{51}$
erg, but this energy is shared among several pulses and the efficiency
for converting kinetic energy to light remains relatively small.

The models from 52 \Msun \ to 62 \Msun \ share similar characteristics
(\Fig{hepulses} and \Fig{hepulses1}). All have a strong first flash
followed, after a long delay, by other pulses.  As the mass increases,
so does the energy of this first pulse. An increasingly large mass is
ejected promptly and the wait time for the next pulse increases. For
52 \Msun, the mass ejected promptly and its energy are 1.1 \Msun \ and
$0.80 \times 10^{50}$ erg; for 54 \Msun, 2.3 \Msun\ and $1.9 \times
10^{50}$ erg; for 58 \Msun, 7.4 \Msun\ and $6.6 \times 10^{50}$ erg;
and for 62 \Msun, 36 \Msun\ and $2.1 \times 10^{51}$ erg. In this last
case, the star is very nearly unbound. Indeed, Model He62 marks the
transition to a full PISN at 64 \Msun. The ejected matter carries no
radioactivity though and, neglecting any interaction with
pre-pulsational mass loss, the light curve this first pulse makes is
faint, hot, and brief.

The object left behind typically has a mass near 51 \Msun \ (except
for the 62 \Msun \ model) with oxygen depleted in its center. The
flash has left it substantially extended and cool inside. For 54 \Msun
\ the central temperature after the first pulse and relaxation
oscillations are over is $6.6 \times 10^8$ K; for 58 \Msun \ it is
$3.69 \times 10^8$ K; and for 62 \Msun \ the central temperature is
only $4.0 \times 10^7$ K, again showing how marginally bound the 62
\Msun \ model is after its first pulse.  For cores above 54 \Msun
\ energy loss from this extended core is dominated by radiation from
the stellar surface, not by neutrinos in the core, and it takes a long
time, up to 7000 years to become unstable again.  During this time, an
observer would see the remnant of a faint supernova, with a brightly
glowing Wolf-Rayet star in its center. If there was appreciable
pre-pulsational mass loss, the object might also be a bright radio or
x-ray source \citep{Che82,Che12,Svi12}. These objects, having
experienced a first outburst with more to come at a much later time
will be referred to as ``dormant supernovae'' (a more suggestive name
might be ``zombie supernovae'', or supernovae Type Z). The time spent
in hibernation is approximately the time required to radiate the
change in binding energy of the core brought about by the first pulse
at either the Eddington luminosity (radiative case) or the global
neutrino loss rate in the expanded state.

In all these models from 52 to 62 \Msun, pulsational activity,
consisting of two or more strong pulses in rapid succession, resumes
as the star approaches its final death. These terminal pulses, which
often come in pairs, are capable of producing bright transients
(\Sect{helite}). In Model He52 pulsations resumed after 4.6 years
(\Fig{hepulses}). In Models He56 and He60, activity resumed after 1060
and 2680 years respectively (\Fig{hepulses1}). There, pulses 2 and 3
ejected a combined 4.6 \Msun \ with $2.5 \times 10^{50}$ erg and 2.0
\Msun \ with $1 \times 10^{50}$ erg, respectively. Model He62 ejected
1.45 \Msun \ with an energy of $2.1 \times 10^{50}$ erg just before
dying (\Fig{hepulses1}).  These delays are very nearly equal to the
total time of pulsational activity (\Tab{hecore}). If the final
iron core collapse produces no outgoing shock, these late time pulses
may be the most readily detected signals of PPISN in this mass range
(though see \Sect{superl}).

The values in \Tab{hecore} suggest an upper limit to the kinetic
energy of purely thermonuclear PPISN of Type I which may be shared by
several pulses of $\sim2 \times 10^{51}$ erg. Most of this energy
comes out in the first pulse of Model He62, however, and unless the
star had very substantial mass loss before becoming pulsationally
unstable. is not available for making radiation.  For events that
might produce optical supernovae, the upper bound is closer to $1
\times 10^{51}$ erg.  Full star models in \Sect{tmods} show a larger
upper bound of about $4 \times 10^{51}$ erg because of the efficient
coupling of the large amplitude bounce to more matter and the stronger
secondary explosions caused by the increased tamping. As will be
discussed, the amount of light radiated is only a fraction of these
kinetic energies.

% fig 5 - helium core light curves
\begin{figure*}
\begin{center}
\leavevmode
\includegraphics[width=0.48\textwidth]{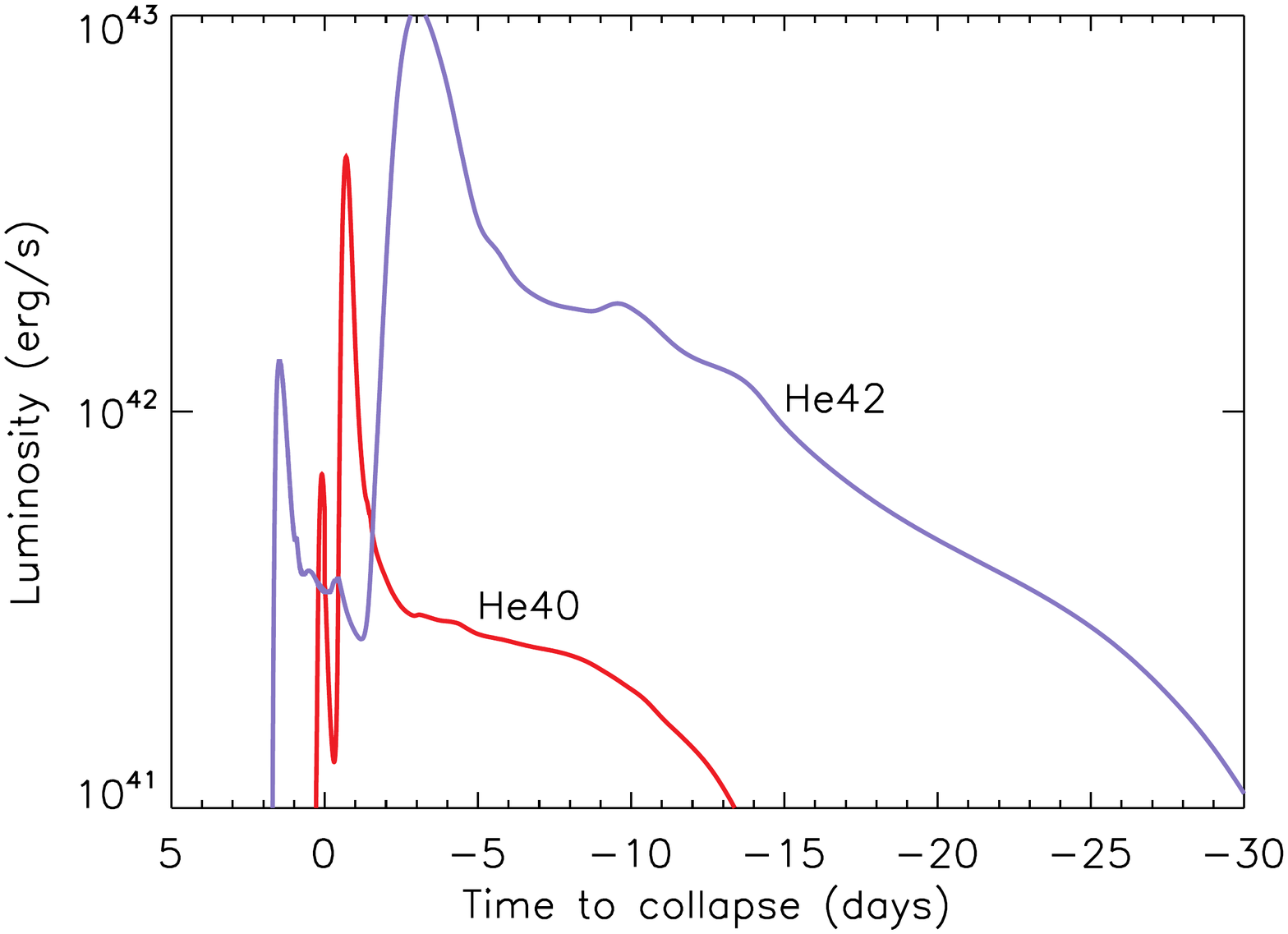}
\includegraphics[width=0.48\textwidth]{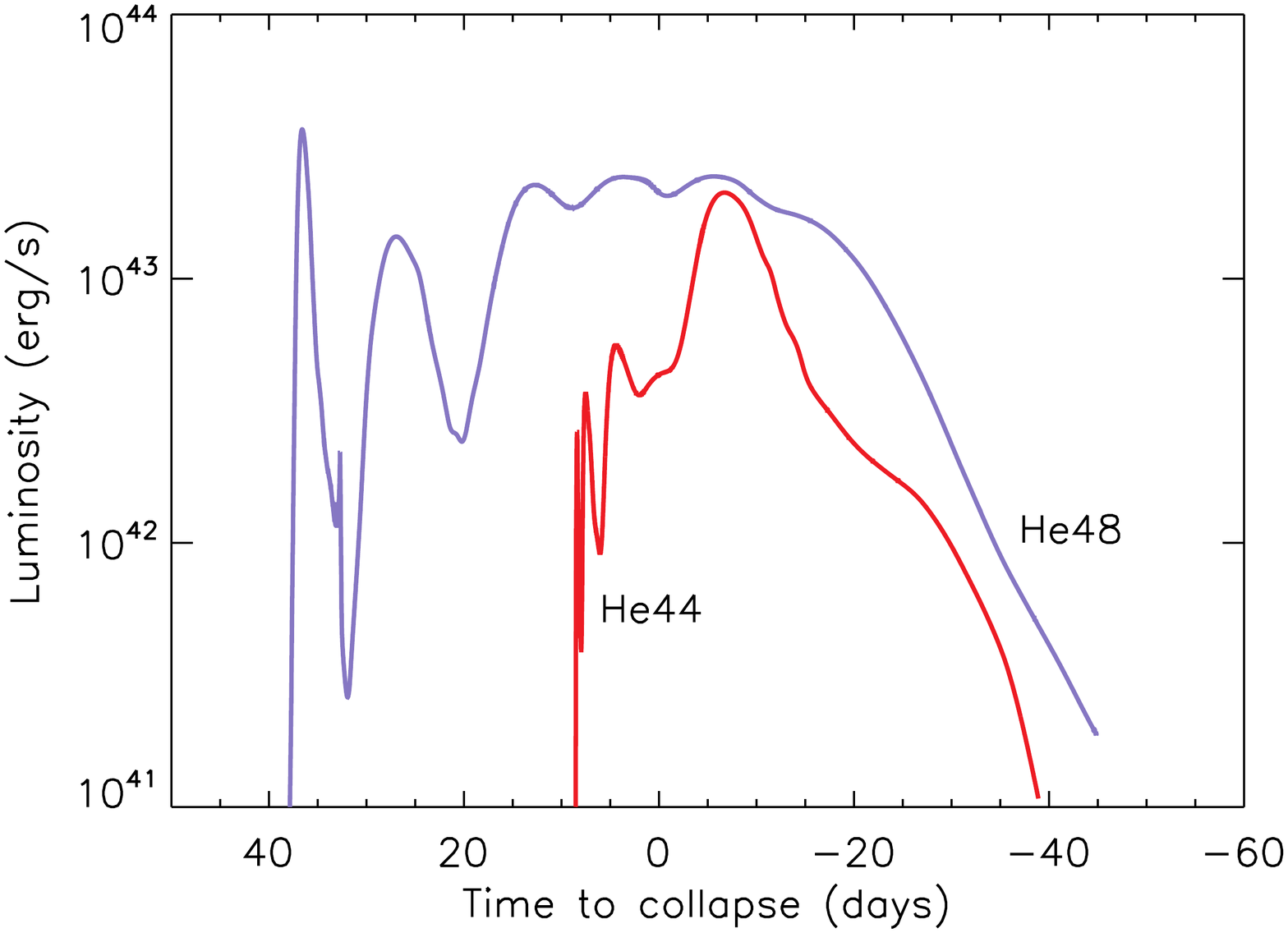}
\vskip 24pt
\includegraphics[width=0.48\textwidth]{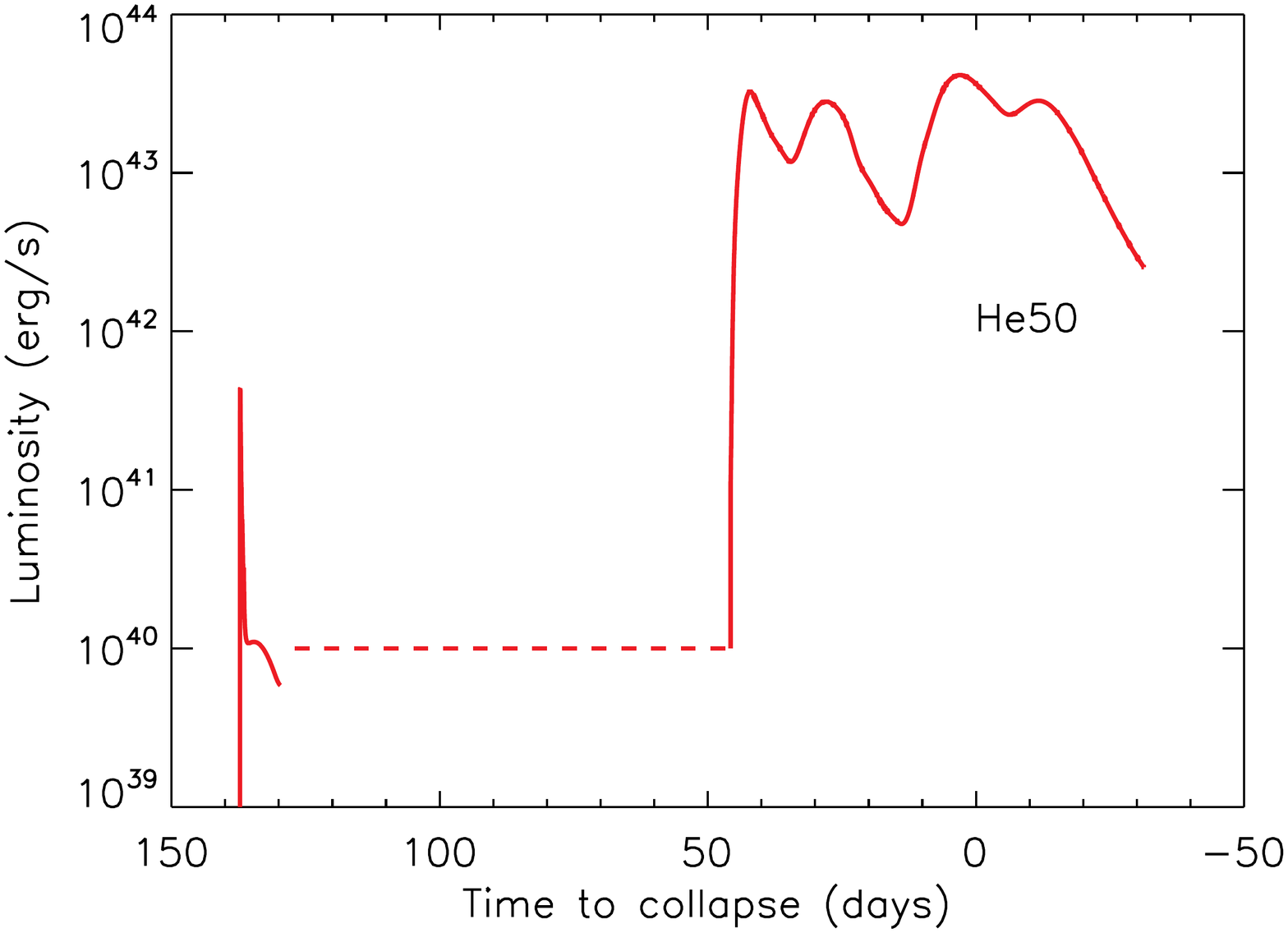}
\includegraphics[width=0.48\textwidth]{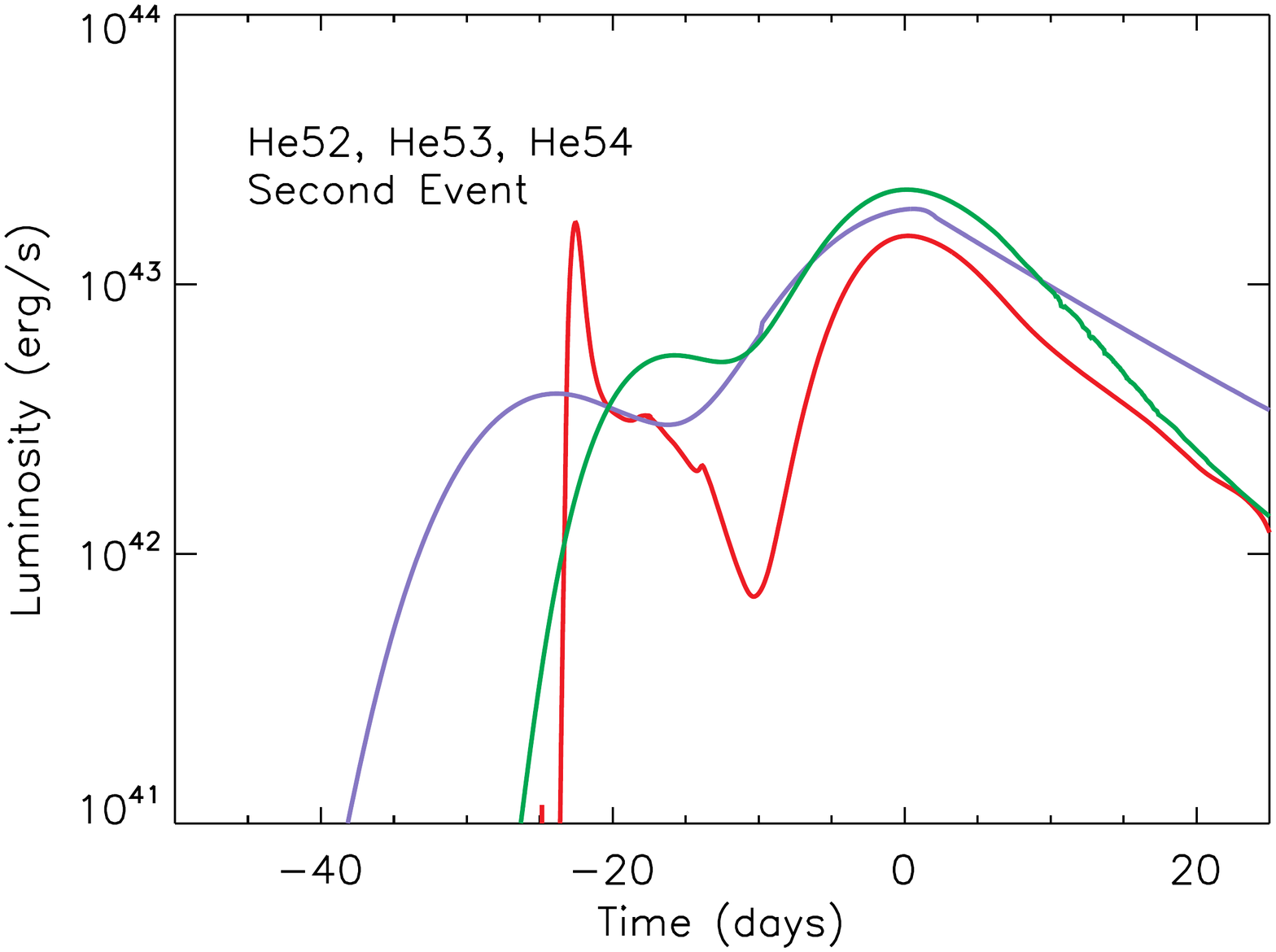}
\caption{Bolometric light curves for the explosion of bare helium
  stars of various masses. See also \Fig{hepulses} and
  \Fig{helight1}. In the top two and bottom left frames, time is given
  in units of days, and measured relative to iron core collapse with
  negative values indicating time elapsed after collapse. The light
  curve for He50 shows two components, an early plateau as the mass
  ejected by the first pulse expands and recombines and a brighter
  complex display produced by the collisions of subsequent pulses with
  the ejecta of the first and each other. The dashed line (not
  calculated) indicates an approximately Eddington luminosity for the
  remaining core during the two pulsationally powered outbursts. In
  the bottom right frame, the second, brighter outbursts are given for
  the 52 \Msun \ (blue), 53 \Msun \ (green) and 54 \Msun \ (red)
  models which began their pulsing activity with a faint outburst 4.6,
  24.5, and 149 years earlier.  In this case, zero time arbitrarily
  corresponds to light curve maximum.  Below 40 \Msun, the optical
  transient produced by the PPI is very faint. \lFig{helight}}
\end{center}
\end{figure*}

All of the helium core models that did not completely disrupt as PISN,
produced massive iron cores (\Tab{hecore}) surrounded by dense shells
of silicon and oxygen that would be very difficult to explode with
neutrinos. Typical net binding energies for the oxygen and silicon
mantles around the iron cores were 3 - 5 $\times 10^{51}$ erg. While
very rapid rotation might still power an explosion (\Sect{superl}), it
seems likely that many of these stars will make black holes.  Thus
stars in this mass would generate a population of 35 - 50 \Msun
\ black holes and nothing lighter or heavier.

\subsection{Helium Core Light Curves}
\lSect{helite}

Light curves for the exploding helium cores were calculated using the
KEPLER code. KEPLER uses flux-limited radiative diffusion and
  operates, while conserving energy and momentum, in both optically
  thick and thin regimes. The bolometric light curves that KEPLER
  calculates have been compared favorably with those from other more
  powerful radiation-transport codes running similar problems
  \citep[e.g.][]{Eas94,Sca05,Woo07,Kas11}. Unfortunately, KEPLER
  treats the radiation as a blackbody having a single temperature, the
  same as the background matter, and thus provides limited information
  on the brightness in various wavebands. It is also unable to
  calculate a realistic opacity in an optically thin region with a
  large Doppler shear. Here, in the ejected material, opacity is
  assumed to be entirely due to electron scattering with a floor
  assigned for recombined material of either 0.001 or 0.01 cm$^2$
  g$^{-1}$. The smaller value is used in recombining ejecta, the
  latter when fall back is important.

Sample results, given in \Fig{helight} and \Fig{helight1}, reveal a
broad range of possibilities. These bolometric curves may require
appreciable correction before comparing with optical light curves. Not
only are the bolometric corrections frequently large, especially at
early times near shock breakout, but the rapid time variations
resulting from colliding shells (e.g., Models He44 and He48) would be
much smoother in a multi-dimensional simulation where the shells would
be substantially broadened by instabilities \citep{Che14}. Any
interaction with pre-pulsational mass loss is ignored and might also
contribute a substantial background luminosity.

Presupernova Model He40, for example, has a radius of $3.5 \times
10^{10}$ cm, so even a moderate breakout luminosity of 10$^{42}$ erg
s$^{-1}$ implies an effective emission temperature of about 10$^6$
K. The matter expands rapidly though, so the ``plateau'' stage in He40
lasts about a week at helium recombination temperatures, $\sim10,000 -
20,000$ K.  Model He40 would thus appear as a fast, faint, blue
transient with relatively slow photospheric speeds 2000 - 3000 km
s$^{-1}$. It might be categorized as a faint Type Ibn or Icn event.

For cores lighter than 48 \Msun, the pulsing activity goes on for a
short time (\Tab{hecore}) and, since the matter has not expanded
greatly beyond $\sim10^{15}$ cm, there is a well-defined photosphere
during the brighter parts of the light curve.  At peak luminosity
after breakout, Model He42 (\Fig{helight}) has a photospheric radius
of $2 \times 10^{14}$ cm and an effective temperature of 23,000 K;
Model He44 has a radius of $7 \times 10^{14}$ cm and a temperature of
15,000 K; and Model He48 has a radius of $\sim10^{15}$ cm and an
effective temperature of 12,000 K. Shell velocities are typically 2000
- 4000 km s$^{-1}$.

Most likely these events would be categorized as Type Ibn and Icn
supernovae \citep{Fol07,Pas08a,Pas08b,Smi12}. Their luminosity,
duration, colors, and velocities are similar, though the post-peak
decline rate is difficult to predict with any accuracy in these 1D
models because of mixing, circumstellar interaction, and bolometric
corrections. It is interesting that SN 2006jc had a faint ``LBV-like''
outburst 2 years before its major display \citep{Pas07,Pas08a}, which
might associate it with a helium core of about 51 \Msun
\ (\Fig{helight}).  Although very rare, other events like SN 2006jc
have been observed, for example SN1999cq \citep{Mat00}; SN2002ao
\citep{Fol07}; SN 2010al \citep{Pas15a}; and ASASS-15ed
\citep{Pas15b}. The models here, by design, all lack hydrogen which
may play an important role in some of these events, especially SN
2011hw \citep{Smi12}, but hydrogen would be present in structurally
similar WNL stars that would have similar light curves upon exploding.

Models heavier than 48 \Msun \ have more complex light curves
with several components. First comes a faint, brief transient
similar to that in the lighter stars, resulting from the ejection of the outer
part of the core by a single pulse. Lacking any radioactivity and
neglecting circumstellar interaction, that explosion is not bright.
Even in the brightest case (\Fig{helight2}), the luminosity produced
as the helium expands beyond 10$^{14}$ cm and recombines does not
exceed $~10^{41}$ erg s$^{-1}$. Because of the low luminosity and
rapidly increasing radius, the transient may evolve rapidly in
color. All are initially blue, but some can become red at late times,
especially the more massive models. Model He54, He58, and He64 were
all near 12,000 K at age one day, but they declined to 11,000 K, 8000
K, and 4500 K respectively at peak.  Velocities were typically a few
thousand km s$^{-1}$, though higher in He62 and in the outermost
layers of the other models. While optically faint, these initial
outbursts carry considerable kinetic energy, up to $2 \times 10^{51}$
erg, and could power bright optical, radio, and x-ray transients if,
as seems likely, the pre-pulsational star had experienced substantial
mass loss.

The second pulsationally-powered display from stars in this mass range
is brighter (\Fig{helight} and \Fig{helight1}). Two or more pulses
occur in rapid succession, shortly before the iron core collapses
(\Fig{hepulses} and \Fig{hepulses1}). The light curve has two stages,
a faint plateau as the first pulse of the delayed series (``pulse 2'')
ejects more core material, and a brighter second peak as subsequent
pulses collide with that ejecta and with themselves. If two pulses
occur sufficiently rapidly, the first may simply inflate the star to a
larger radius, while next shock traverses that still optically thick
``envelope'' and produces a sharp peak due to breakout, as in Models
He54 through He60. Or the shells may collide after becoming almost
optically thin producing broad peaks like in He52 and He53
(\Fig{helight}). The light curve is blue, especially at early
times. Typical temperatures at the bright peak are 10,000 - 12,000 K,
though the photosphere is not always well defined in the heavier
models where the collision happens in a medium thin to electron
scattering. Velocities are 2000 - 4000 km s$^{-1}$ though a small
amount of material moves slower and faster.

This sudden rise to a secondary maximum is similar to what has been
observed in several unusual supernovae. Consider SN 2005bf
\citep{Fol06} as compared with Model He52 (\Fig{helight}). The
luminosity, duration, spectral type (Ic), and ``double-peaked'' shape
are all roughly similar. The model photospheric temperature on the
first bolometric peak was 8,000 K and on the second peak 10,000 K,
within the bounds of the observations at similar times. Model He 52
also had two velocity components, one from its first mass ejection (up
to above 10,000 km s$^{-1}$ at its outer edge at $\sim10^{17}$ cm),
and a much larger mass from the later outburst that made the bright
light curve moving at about 4000 - 7000 km s$^{-1}$.  This is not
necessarily to say that SN 2005bf was a PPISN. It would probably
require a low-metallicity region to make a PPISN, and the metallicity
of the host of SN 2005bf was not specified. The very high velocity
material (over 10,000 km s$^{-1}$) in Model He 52 only existed very
far out in about 0.001 \Msun \ of ejecta, and by design, contained no
hydrogen. But if Model He52 were detected today, observers would
probably call it ``SN 2005bf-like''. Indeed, as will be discussed
further in \Sect{lbv} and \Sect{conclude}, there are also several {\sl
  Type IIn} supernovae that display this sort of pause before
dramatically brightening - among them SN 1961v \citep{Smi11c,Koc11},
SN 2009ip \citep{Fra15}, and SN 2010mc \citep{Ofe13}. The structural
distinction between WRC, WRO, WCN, and LBV progenitors is not great
and this double peaked structure may be a common signature of the PPI
operating in compact progenitors.

During the long dormant phases between the initial outburst and the
bright second display, the luminosity for models between 50 and 60
\Msun, is provided by the central star and is close to the Eddington
value, approximately 10$^{40}$ erg s$^{-1}$. The energy comes from a
combination of Kelvin-Helmholtz contraction and the fall back of
incompletely ejected supernova material. Even though the energy source
is not thermonuclear, such stars would have radii of a few times
10$^{11}$ cm, similar to WR stars, which they might closely
resemble. In all likelihood the pre-pulsationally unstable star had a
strong wind or episodic mass loss, so these dormant supernovae could
be very bright radio and x-ray sources while awaiting their next
outburst.

Even lacking this pre-pulsational mass loss, the final explosions
would eventually catch up with the mass ejected by the very first
pulse. Characteristic radii would be a few thousand km s$^{-1}$ times
the ``Duration'' given in \Tab{hecore}, or 10$^{16}$ - 10$^{19}$ cm.
The collision could give a very bright, long lasting transient with
uncertain properties, especially for the lighter models where the
collision happens earlier when the density is higher. The emission
might be strong in radio and x-rays.

% fig 6 - light curves - 56,58, 60
\begin{figure}
\includegraphics[width=\columnwidth]{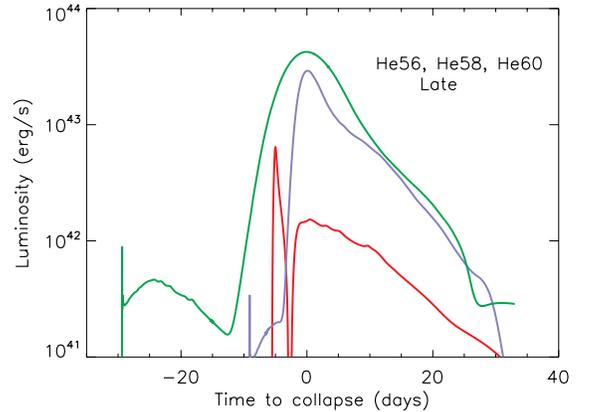}
\caption{Light curves during the second outbursts of Models He56
  (red), He58 (green), and He60 (blue) caused by pulses 2 and 3 (see
  \Fig{hepulses1}). Zero time is arbitrarily set to the post-outburst
  maximum of the curve. The core collapsed 84 days after 0 in the plot
  for He56, 3.4 days later for He58, and 5.9 years later for
  He60. The initial spikes caused by shock break out are very hot and
  would be faint in the optical band. \lFig{helight1}}
\end{figure}

% fig 7 - first light He62
\begin{figure}
\includegraphics[width=\columnwidth]{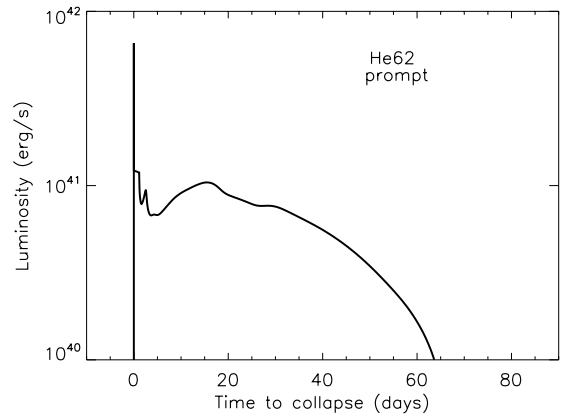}
\caption{First explosion of Model He62. The ejection of the outer
  24 \Msun\ gives rise to a faint display. Lighter models from 50 to
  60 \Msun \ have similar, but fainter and shorter initial light
  curves (see \Fig{helight}). The effective temperature at peak
  luminosity (neglecting break out) is about 5000 K.  \lFig{helight2}}
\end{figure}

None of the bare helium cores studied here produced an exceptionally
brilliant SLSN. Models He48 and He50 in (\Fig{helight}) had a total
light output of 0.84 and $1.2 \times 10^{50}$ erg, respectively, so an
upper limit of about $1 \times 10^{50}$ erg of radiated light seems
reasonable for pure PPISN coming from bare helium or CO cores. Later
it will be shown that full stars with hydrogen envelopes can produce
appreciably brighter light curves with up to $5 \times 10^{50}$ erg of
light.

\section{Full Stars Without Rotation - Red Supergiants}
\lSect{tmods}

Consider now the evolution of PPI unstable helium cores evolving
inside of stars that have not lost their hydrogenic
envelopes. Depending upon their mass loss histories, metallicities,
and rotation rates, such stars will die as red or blue supergiants,
LBVs, or, in the case of rapid rotation and CHE, compact WR
stars. Each case will be considered, but we begin with the most common
result for the mass loss rates and metallicities assumed, PPISN
occurring in RSGs.

The presence of a tenuous envelope, however massive, does not greatly
alter the hydrodynamic behavior of a helium core encountering the
pair-instability for the first time. Once the first explosion is
underway, however, the envelope has major consequences for both the
light curve and the subsequent evolution.  The envelope tamps the
expansion of the core and absorbs momentum, resulting in more of it
falling back. This increases the mass of the bound remnant over what
it would have been without an envelope and hinders its expansion to
low density, thus shortening the interval between pulses. It also
makes the remnant core larger and subsequent pulses more
energetic. Overall, it broadens the mass range for the PPI and shifts
the masses of helium cores in \Tab{hecore} where various phenomena are
expected upwards by a few solar masses. At the upper end, this means
that helium cores that might have been completely unbound if they were
bare, still leave behind bound remnants when embedded in
envelopes. The maximum energy produced by a PPISN is modestly
increased.

With a hydrogen envelope, a greater diversity of observable transients
is also possible. Assuming for now that the final core
collapses to a black hole, PPISN eject no radioactive $^{56}$Ni, so
their displays are entirely a consequence of recombining,
pulse-ejected envelopes and colliding shells.  For the lower energy
pulses in light PPSN, the luminosity on the plateau may be faint since
only part of the envelope is ejected, and even that at low speed. For
higher energy pulses, but still with a duration of less than a few
months, brighter, longer lasting, ``normal'' Type IIp supernovae
result. For the energetic, infrequent pulses that characterize high
mass PPISN, a mixture of Type IIp supernovae and IIn occur. The
structure of the light curves can be complex. Each pulse can make
from one to several light curve peaks as the mass it ejects expands
and cools and runs into shells present from previous
outbursts.

To illustrate and quantify these outcomes, the evolution of stars with
a variety of masses and mass loss rates are considered. The key
quantities are the helium core mass of the presupernova star and the
mass and radius of the hydrogen envelope. Many uncertain factors -
mass loss rates, opacities, rotational mixing, and convective
overshoot mixing - enter into determining these three quantities, but
to first order, two presupernova stars with the same helium core mass,
hydrogen envelope mass, and radius will have similar light curves and
leave similar remnant masses.  The results given are thus generic for
other choices of stellar parameters that give these final masses,
though the main sequence masses and metallicities responsible for these
final states will shift.

\begin{deluxetable*}{cccccccccc} 
\tablecaption{LOW METALLICITY MODELS} 
\tablehead{ Mass & Mass Loss & M$_{\rm preSN}$ & M$_{\rm He}$ & M$_{\rm CO}$ & M$_{\rm Si}$ & M$_{\rm Fe}$ &  Duration & M$_{\rm final}$ & KE$_{\rm eject}$ \\
(\Msun) &  & (\Msun) & (\Msun)  & (\Msun)  & (\Msun) & (\Msun) & (10$^7$ sec) & (\Msun) & (10$^{50}$ erg)}
\startdata
T70   & 1   & 47.31 & 29.42 & 25.62 & 7.58 & 2.54 & 0.00066& 47   &  -    \\
T70A  & 1/2 & 51.85 & 30.10 & 26.41 & 7.87 & 2.58 & 0.00065& 52   &  -    \\
T70B  & 1/4 & 59.62 & 30.50 & 26.84 & 8.28 & 2.57 & 0.00072& 60   &  -    \\
T70C  & 1/8 & 64.66 & 30.72 & 27.14 & 8.22 & 2.54 & 0.00068& 65   & 0.0005\\
T70D  & 0.  & 70    & 31.57 & 28.00 & 8.41 & 2.57 & 0.0012 & 52   & 0.015 \\
T75   & 1   & 48.46 & 32.47 & 28.36 & 7.41 & 2.54 & 0.00075& 41   & 0.0028\\
T75A  & 1/2 & 54.24 & 31.90 & 27.97 & 8.64 & 2.52 & 0.0014 & 42   & 0.024 \\ 
T75B  & 1/4 & 62.97 & 33.07 & 29.15 & 8.71 & 2.64 & 0.0015 & 51   & 0.021 \\
T75C  & 1/8 & 68.61 & 33.41 & 29.67 & 8.91 & 2.61 & 0.0016 & 51   & 0.029 \\
T75D  & 0.  & 75    & 33.82 & 30.20 & 8.71 & 2.67 & 0.0019 & 50   & 0.11  \\
T80   & 1   & 50.79 & 34.70 & 30.81 & 7.90 & 2.65 & 0.0019 & 39.6 & 0.19\\
T80A  & 1/2 & 55.32 & 34.59 & 30.74 & 8.38 & 2.62 & 0.0061 & 39.2 & 0.39\\
T80B  & 1/4 & 66.04 & 35.30 & 31.37 & 8.44 & 3.00 & 0.0098 & 34.7 & 0.92\\
T80C  & 1/8 & 72.76 & 36.24 & 32.28 & 8.03 & 3.29 & 0.014  & 34.8 & 1.3 \\
T80D  &  0  &  80   & 36.40 & 32.56 & 7.93 & 3.09 & 0.015  & 34.9 & 1.5 \\
T90   &  1  & 55.32 & 38.77 & 34.58 & 7.16 & 2.73 & 0.039  & 37.3 & 2.6 \\
T90A  & 1/2 & 60.62 & 39.69 & 35.37 & 9.54 & 2.57 & 0.11   & 35.9 & 4.1 \\
T90B  & 1/4 & 72.16 & 40.41 & 36.16 & 9.54 & 2.84 & 0.18   & 36.4 & 5.2 \\
T90C  & 1/8 & 80.61 & 40.21 & 36.00 & 6.22 & 2.87 & 0.20   & 37.4 & 4.9 \\
T90D  &  0  & 90    & 40.92 & 36.78 & 8.35 & 2.86 & 0.19   & 37.1 & 4.9 \\
T100  &  1  & 57.58 & 44.85 & 39.65 & 4.56 & 2.48 & 1.0    & 38.9 & 7.0 \\
T100A & 1/2 & 62.20 & 44.46 & 39.74 & 5.24 & 2.73 & 0.74   & 39.3 & 7.7 \\
T100B & 1/4 & 78.58 & 45.11 & 40.61 & 4.64 & 2.44 & 0.92   & 39.9 & 7.6 \\
T100C & 1/8 & 88.11 & 45.71 & 41.23 & 4.67 & 2.53 & 1.7    & 40.4 & 6.9 \\
T100D &  0  & 100   & 45.13 & 40.70 & 6.44 & 2.87 & 0.45   & 40.4 & 6.6 \\
T105  &  1  & 59.54 & 47.52 & 42.00 & 4.78 & 2.79 & 7.34   & 43.6 & 7.8 \\
T105A & 1/2 & 66.88 & 46.04 & 41.45 & 4.78 & 2.62 & 1.22   & 40.8 & 8.0 \\
T105B & 1/4 & 81.18 & 47.34 & 42.55 & 5.75 & 2.92 & 2.20   & 42.5 & 7.8 \\
T105C & 1/8 & 91.94 & 48.33 & 43.56 & 4.70 & 2.73 & 4.38   & 44.2 & 7.0 \\
T105D &  0  & 105   & 49.45 & 44.67 & 4.87 & 1.97 & 10.7   & 44.8 & 7.8 \\
T110  &  1  & 63.31 & 49.89 & 44.39 & 4.92 & 1.98 & 17     & 45.1 & 8.6 \\
T110A & 1/2 & 68.41 & 49.68 & 44.58 & 4.88 & 1.95 & 39     & 44.5 & 7.6 \\
T110B & 1/4 & 84.13 & 49.50 & 44.67 & 4.70 & 2.18 & 9.5    & 44.7 & 7.4 \\
T110C & 1/8 & 95.98 & 48.91 & 44.19 & 4.53 & 2.59 & 5.8    & 44.8 & 7.1 \\
T110D &  0  & 110   & 50.49 & 45.44 & 4.75 & 2.08 & 30     & 45.0 & 7.7 \\
T115  &  1  & 63.23 & 53.09 & 47.11 & 5.51 & 1.85 & 2600   & 49.3 & 11.5\\
T115A & 1/2 & 71.40 & 50.47 & 45.40 & 4.78 & 2.38 & 13     & 45.7 & 7.9 \\
T115B & 1/4 & 86.39 & 50.72 & 45.80 & 4.69 & 2.16 & 120    & 45.1 & 7.8 \\
T115C & 1/8 & 99.74 & 51.35 & 46.50 & 4.55 & 2.07 & 670    & 45.6 & 8.3 \\
T115D &  0  & 115   & 51.96 & 46.71 & 5.88 & 3.01 & 200    & 47.5 & 8.6 \\
T120  &  1  & 66.99 & 55.01 & 50.10 & 5.75 & 2.61 & 4000   & 47.7 & 16  \\
T120A & 1/2 & 79.55 & 55.08 & 49.16 & 4.60 & 2.60 & 460    & 50.6 & 15  \\
T120B & 1/4 & 90.11 & 53.41 & 48.21 & 4.65 & 2.52 & 250    & 48.2 & 8.0 \\
T120C & 1/8 & 103.3 & 54.94 & 49.79 & 4.31 & 2.03 & 350    & 51.8 & 11  \\
T120D &  0  & 120   & 56.11 & 50.52 & 4.75 & 2.18 & 1200   & 51.8 & 14  \\
T121A & 1/2 & 73.09 & 54.67 & 49.14 & 4.74 & 2.03 & 460    & 50.9 & 11  \\
T122A & 1/2 & 73.94 & 56.06 & 49.76 & 6.05 & 2.24 & 12000  & 44.9 & 31  \\
T123A & 1/2 & 74.38 & 55.79 & 50.38 & 5.36 & 1.74 & 3900   & 50.2 & 17  \\
T124A & 1/2 & 74.39 & 56.85 & 50.58 & 6.24 & 2.30 & 12000  & 46.9 & 35  \\
T125  &  1  & 69.21 & 57.49 & 51.75 & 5.49 & 1.78 & 6500   & 50.3 & 13  \\
T125A & 1/2 & 81.38 & 57.12 & 51.20 & 5.79 & 1.90 & 8600   & 51.8 & 16  \\
T125B & 1/4 & 92.24 & 57.08 & 51.53 & 5.44 & 1.70 & 4900   & 50.9 & 15  \\
T125C & 1/8 & 107.1 & 57.58 & 52.08 & 5.69 & 2.43 & 11000  & 49.0 & 14  \\
T125D &  0  & 125   & 56.20 & 51.75 & 4.89 & 2.58 & 7400   & 47.8 & 11  \\
T130  &  1  & 71.00 & 60.50 & 54.62 & 6.75 & 2.41 & 15000  & 50.8 & 23  \\
T130A & 1/2 & 79.69 & 60.20 & 54.28 & 6.03 & 1.81 & 10000  & 51.3 & 33  \\
T130B & 1/4 & 94.26 & 58.28 & 53.48 & 8.16 & 3.75 & 13000  & 48.4 & 27  \\
T130C & 1/8 & 110.6 & 61.91 & 56.10 & 8.99 & 3.95 & 16000  & 49.0 & 31  \\
T130D &  0  & 130   & 59.96 & 54.28 & 2.04 & 2.04 & 25000  & 38.8 & 41  \\
T135  &  1  & 71.37 & 64.04 & 56.60 & 5.43 & 3.83 & 140    & 18.9 & 42  \\ 
T135A & 1/2 & 85.71 & 65.42 & 56.36 & 5.56 & 3.27 & 19000  & 43.3 & 38  \\
T135B & 1/4 & 97.54 & 61.15 & 55.30 & 5.39 & 3.05 & 18000  & 42.9 & 35  \\
T135C & 1/8 & 107.2 & 60.14 & 54.71 & 2.41 & 2.07 &  4500  & 23.2 & 31  \\
% very thick O shell mostly Si driven by Ni decay?
T135D &  0  & 135   & 63.91 & 57.54 & 4.37 & 2.84 &  4300  & 35.0 & 39  \\
T140  &  1  & 75.29 & 65.63 & 58.32 & 5.48 &  -   &  -     &  0   & 44  \\
T140A & 1/2 & 89.64 & 65.90 & 59.55 & 5.54 & 1.95 & 200    & 4.5  & 41  \\
T140B & 1/4 & 99.08 & 65.01 & 59.06 & 4.25 & 2.65 & 110    & 29.2 & 38  \\
T140C & 1/8 & 108.6 & 63.87 & 57.96 & 6.04 &  -   &  -     &  0   & 48  \\
T140D &  0  & 140   & 65.24 & 59.19 & 5.20 & 2.63 & 21000  & 37.4 & 33  \\
T150  &  1  & 76.38 & 71.63 & 64.73 & 6.83 &  -   &  -     &  0   & 120 \\
T150A & 1/2 & 95.98 & 70.89 & 64.20 & 5.99 &  -   &  -     &  0   & 70  \\
T150B & 1/4 & 106.4 & 69.05 & 62.76 & 6.11 &  -   &  -     &  0   & 60  \\
T150C & 1/8 & 113.4 & 70.17 & 63.94 & 5.93 &  -   &  -     &  0   & 71  \\
T150D &  0  & 150   & 70.18 & 64.86 & 6.41 &  -   &  -     &  0   & 98  \\
\enddata
\lTab{tmodels}
\end{deluxetable*}

\Tab{tmodels} shows the results for a grid of 10\% solar metallicity
stars in the main sequence mass range 70 to 150 \Msun. For the assumed
stellar physics, these give helium core masses in the range 30 to 70
\Msun \ and thus span the range where PPISN are expected. Since mass
loss on the main sequence is relatively small, the helium core mass
for these models is mostly determined by the main sequence mass, but
the mass of the hydrogen envelope (if any) that surrounds the
presupernova core depends on the mass loss rate. Here the standard
mass loss rates (\Sect{mdot}) have been multiplied by factors of 1,
1/2, 1/4, 1/8 and 0. The modified cases, with multipliers 1/2, 1/4,
etc. are referred to as the ``A'', ``B'', ``C'' and ``D'' series of a
given mass. The models are thus named by their metallicity (``T'' for
tenth solar), their main sequence mass, and their mass loss
rate. Model T100D was a 100 \Msun \ star on the main sequence with no
rotation, a metallicity one-tenth that of the sun, and no mass
loss. Smaller multipliers also correspond to the results expected for
stars of lower metallicity since, for a given structure, the mass loss
is just proportional to Z to some power. Very low metallicity stars
might also be blue rather than RSGs and have still lower mass loss rates.

The helium core masses and CO core masses rise roughly monotonically
with main sequence mass. Small variations are expected due to the
complex interplay of mass loss, convection, convective overshoot
mixing and semiconvection. Larger variations are seen for the silicon
and iron core masses due to the interaction of burning of multiple
convective shells of carbon, neon and oxygen \citep{Suk16}. Major
differences exist for some pulse durations and explosion energies,
e.g., Models T115 and T115A, because the tamping effect of the
hydrogen envelope influences the expansion of the exploding core and
the interval to the next pulse. The helium and CO cores are also
typically a bit larger for the models with full mass loss since so
much of the envelope is lost that the convective dredge up of helium
near the end of the star's life is reduced.  Generally though, 
explosion energies and durations increase with mass.

% fig 8 - Pulse duration as function of mass1 
\begin{figure}
\includegraphics[width=0.49\textwidth]{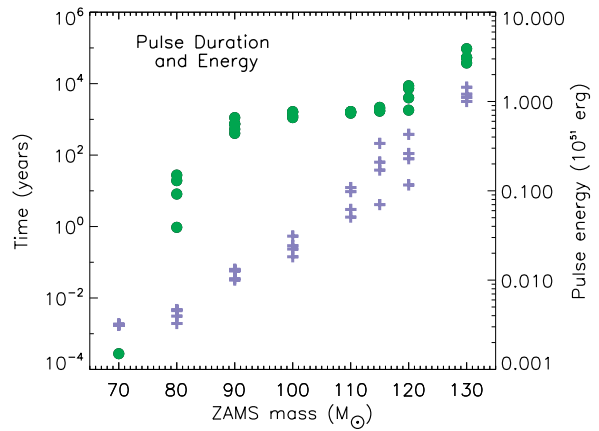}
\caption{Pulse duration in years (blue crosses) and total kinetic
  energy in all the ejected shells in units of 10$^{51}$ erg (green filled
  circles) as a function of the main sequence mass in \Msun \ for full
  stars of 10\% solar metallicity evolved until iron core collapse.
  The duration of pulsing activity and total energy are highly
  correlated with weak explosions also having short duration. From 90
  to 120 \Msun \ the explosion energy is nearly constant even though
  the time scale varies significantly. Fewer pulses each carrying more
  energy happen for the heavier stars. \lFig{tevsm}}
\end{figure}

\Tab{tmodels} and \Fig{tevsm} give the presupernova mass, prior to any
pulsing activity, and the masses of helium, CO, silicon, and iron cores
where they existed in hydrostatic equilibrium. The hydrogen envelope
mass is the presupernova mass minus the helium core mass. Also given
is the total duration of the pulses, again measured from the first
pulse until core collapse, the final mass of the bound remnant after
all pulses ceased, and the total kinetic energy of all matter ejected
by the pulses.

\subsubsection{70 - 80 \Msun \ - Faint Type IIp Supernovae}
\lSect{t70}

Regardless of envelope mass, the helium core mass for non-rotating 70
\Msun \ stars is in the range 29 - 32 \Msun
\ (\Tab{tmodels}). Similarly, for 75 \Msun \ stars, the helium core
mass is 32 - 34 \Msun. Helium cores of this mass are marginally stable
(\Tab{hecore}), but a more relevant quantity is the CO core mass,
which is larger for Models T70 and T75 than for the equivalent helium
cores evolved at constant mass. \Fig{t70cen} shows that the central
temperature history for Model T70A is actually intermediate between
those those of Models He32 and He33 even though the actual helium core
mass of T70A is 30.1 \Msun. This implies an offset in helium core mass
of about 2 \Msun. The boundary pressure of the hydrogen shell is
small. One must go less than 0.1 \Msun \ into the core before the
pressure rises by a factor of 2. The offset reflects more the growth
of the helium core by hydrogen shell burning in the full stars while
convective central helium burning is in progress. As a result, the the
helium convection zone grows and, in the end, a larger CO core mass is
produced.  The CO core in T70A is 26.41 \Msun, more like the CO cores
In He32 and He33 (26.3 \Msun \ and 27.2 \Msun, respectively) and
considerably larger than the CO core in He30 (24.65 \Msun). Similar
behavior was noted by \citet{Woo07}.

% fig 9 - 70  msun T-central
\begin{figure}
\includegraphics[width=0.49\textwidth]{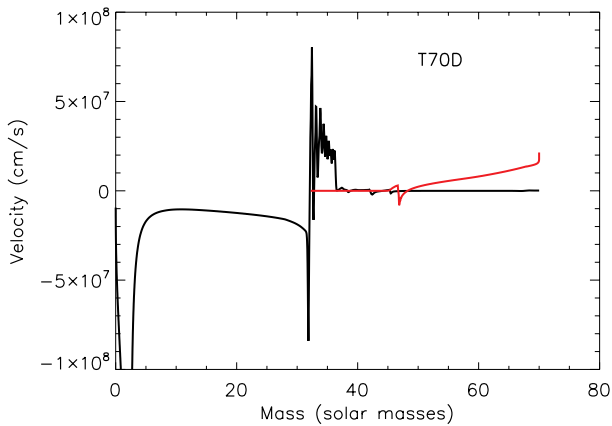}
\caption{Central conditions during the last 3000 seconds before core
  collapse for Model T70A, which has a helium core mass of 30.1 \Msun,
  and the 32 and 33 \Msun \ helium cores studied in
  \Sect{helium}. He33 is the red line. Blue is He32 and green is
  T70A. \lFig{t70cen}}.
\end{figure}

\Fig{t70un} shows the velocity structure in Model T70D at the time
when the iron core collapses (note the high negative speed in the
inner 2 \Msun). Numerous low energy pulses have already steepened into
shocks in the density gradient at the edge of the helium core and are
accumulating at the base of the hydrogen shell. After the helium core
collapses, presumably to a black hole, these pulses continue out
though the envelope, eventually merging into a single shock wave. The
momentum of the small amount of matter that initially moves with high
speed must be shared with the large mass of the envelope though, so
the speed slows. Peak velocities are only $\sim100$ km s$^{-1}$ (red
line, \Fig{t70un}).

% fig 10 - 70  msun presn and postsn velocity
\begin{figure}
\includegraphics[width=0.49\textwidth]{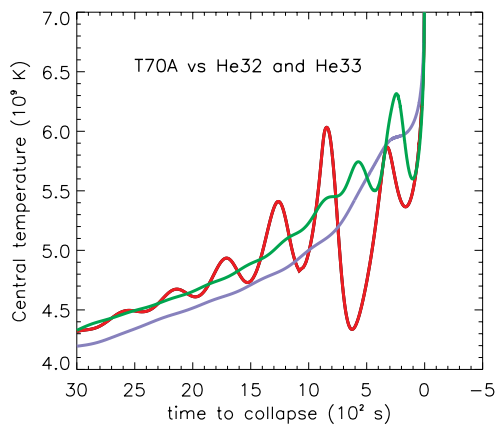}
\caption{Velocity in Model T70D at the time the iron core collapses
  to a protoneutron star. Multiple pulses have resulted in the
  accumulation of energy and momentum in the hydrogen envelope just
  outside the helium core edge at 30.1 \Msun. These pulses merge into
  a weak shock that propagates through the envelope and ejects about
  10 \Msun \ of material with kinetic energy $1.5 \times 10^{48}$
  erg. \lFig{t70un}}.
\end{figure}

In Models T70A and T70B, the pulses were so weak that the shock died
in the envelope without ejecting any discernible matter. The implicit
hydrodynamics in KEPLER damps very weak shocks numerically, so the
possibility of some small, low velocity ejection is not ruled out.  In
the other two 70 \Msun models though, and in all of the 75 \Msun
\ mass models, part of the hydrogen envelope was ejected, about 1
\Msun in Model T70C and about 18 \Msun in Model T70D. These ejections
had very little kinetic energy (\Tab{tmodels}), e.g., 5 $\times
10^{46}$ erg in Model T70C and $1.5 \times 10^{48}$ erg in Model
T70D. These energies were far less than the binding of the entire
envelope, about $3 \times 10^{49}$ erg, so most of the envelope may
fall into the black hole. The mass ejection did power some faint,
light curves, however (\Fig{t70lite}). Typical temperatures for Models
T70C and T70D were 3000 - 4000 K on the ``plateau'' with photospheric
speeds of only 50 - 150 km s$^{-1}$. For the 75 \Msun \ models
(\Fig{t75lite}), the temperatures were more like typical Type IIp
supernovae $\sim6000$ K. The light curves were also a bit brighter,
though still fainter than normal Type IIp supernovae. The expansion
speeds were still very slow, 100 - 200 km s$^{-1}$.  Similar low
energy light curves have also been studied by \citet{Lov16},
especially their bright, brief, blue shock break out phases.

% fig 11 - 70  msun light curves
\begin{figure}
\includegraphics[width=0.49\textwidth]{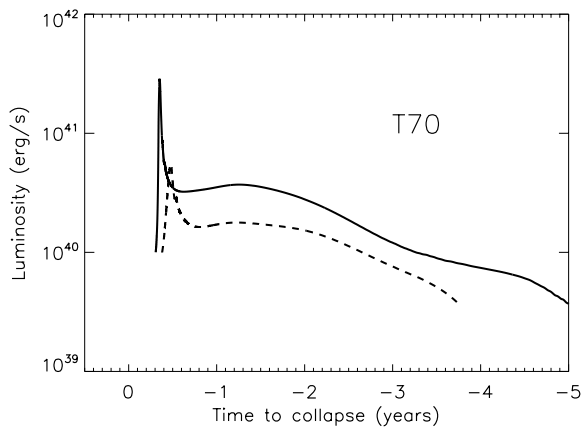}
\caption{Light curves for two of the 70 \Msun \ models (T70C and
  T70D). Time is measured in years relative to the time when the iron
  core collapses. Negative time is post-collapse. These very low
  energy explosions eject only a fraction of their hydrogen envelopes
  and have very faint light curves that, well after shock break out,
  are red and have low velocities, $\sim50 - 150$ km s$^{-1}$.
  \lFig{t70lite}}.
\end{figure}

% fig 12 - 75  msun light curves
\begin{figure}
\includegraphics[width=0.49\textwidth]{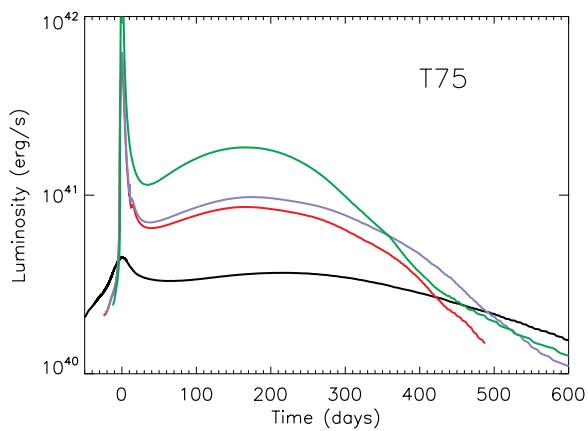}
\caption{Light curves for the four 75 \Msun \ models.  The curves are
  for models with different mass loss rates: T75A (black), T75B (red);
  T75C (blue), and T75D (green). See \Tab{tmodels}. Slightly more
  energetic than the T70 models (\Fig{t70lite}), these low energy
  explosions still eject only a fraction of their hydrogen envelopes
  and have faint light curves that, after shock break out that have
  very low velocities, $\sim100 - 200$ km s$^{-1}$, and last a year or
  more.  The colors on the plateaus are typical of Type IIp
  supernovae, $\sim6000$ K.  \lFig{t75lite}}.
\end{figure}

The PPI is considerably stronger in an 80 \Msun \ model. The total
energy in pulses, $\sim10^{50}$ erg, is still only about 10\% that of
an ordinary Type IIp supernova and the duration of the pulses, roughly
a day, is sort compared with the duration of the light curve. The
result is a single, sub-energetic Type IIp supernova
(\Fig{t80lite}). Typical expansion speeds have risen to 200 - 800 km
s$^{-1}$ (T80A) and 300 - 1000 km s$^{-1}$ (T80B, T80C, T80D). These
would probably be Type IIn supernovae.

Together, these models in the 70 - 80 \Msun \ range should be roughly
half as frequent in nature as the 100 - 130 \Msun \ stars to be
discussed later that might make SLSN. They are obviously more
difficult to detect, but their very low expansion speeds, faint
emission, and long duration are distinctive. Some might be even
classified as ``supernova impostors'' \citep{Smi11c}. The brighter
ones might be Type IIn supernovae, especially if they had appreciable
mass loss before starting to pulse.

% fig 13 - 80  msun light curves
\begin{figure}
\includegraphics[width=0.49\textwidth]{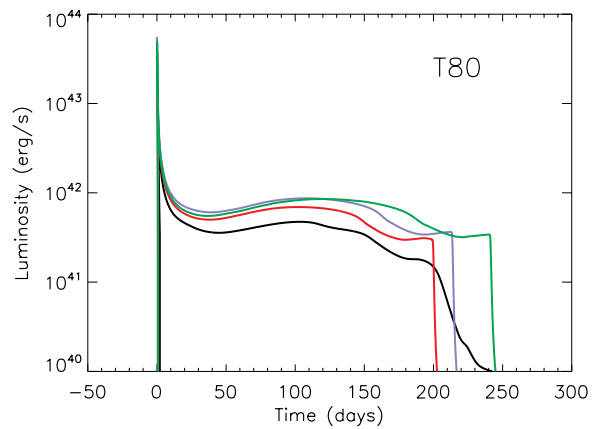}
\caption{Light curves from the 80 \Msun \ low metallicity models.  The
  curves are for models with different mass loss rates: T80A (black),
  T80B (red); T80C (blue), and T80D (green). See \Tab{tmodels}. All
  light curves are single events similar in appearance to faint SN IIp
  with explosion energies $\sim10^{50}$ erg or less
  \citep{Lov16}. Models with larger envelope masses have
  systematically longer plateaus.  \lFig{t80lite}}.
\end{figure}

\subsubsection{80 - 90 \Msun \ - Ordinary Type II Supernovae}
\lSect{t90}

By 90 \Msun, the total energy of the pulses has become an appreciable
fraction of 10$^{51}$ erg and that energy is still being deposited
over a time short compared with the $\sim10^7$ required for the
envelope to expand and recombine (\Fig{t90pulse}). Shells collide
while the star is still very optically thick. The result is a single
ordinary-looking Type IIp supernova, with several exceptions: 1) the
duration for the PPISN may be longer than for typical Type IIp
supernovae depending upon how much of the envelope has been lost; 2)
the photospheric speed is slower, typically varying from 1000 to 2500
km s$^{-1}$ on the plateau; 3) if the progenitor is a RSG, the initial
radius is unusually large, and so too is the initial luminosity; 4)
the mass loss rate may have been unusually high just before the
explosion; 5) the metallicity is low; and 6) no radioactivity is
ejected. The light curve thus has no $^{56}$Co-powered ``tail''. At
the end of the plateau, the light curve plummets. Circumstellar
interaction may add an appreciable late-time component though that
could mimic a tail (\Sect{conclude}). Careful study of the decay time
scale and a spectroscopic search for narrow lines might be necessary
to distinguish this from radioactive decay.

% fig 14 -  90 msun pulses
\begin{figure*}
\begin{center}
\leavevmode
\includegraphics[width=0.49\textwidth]{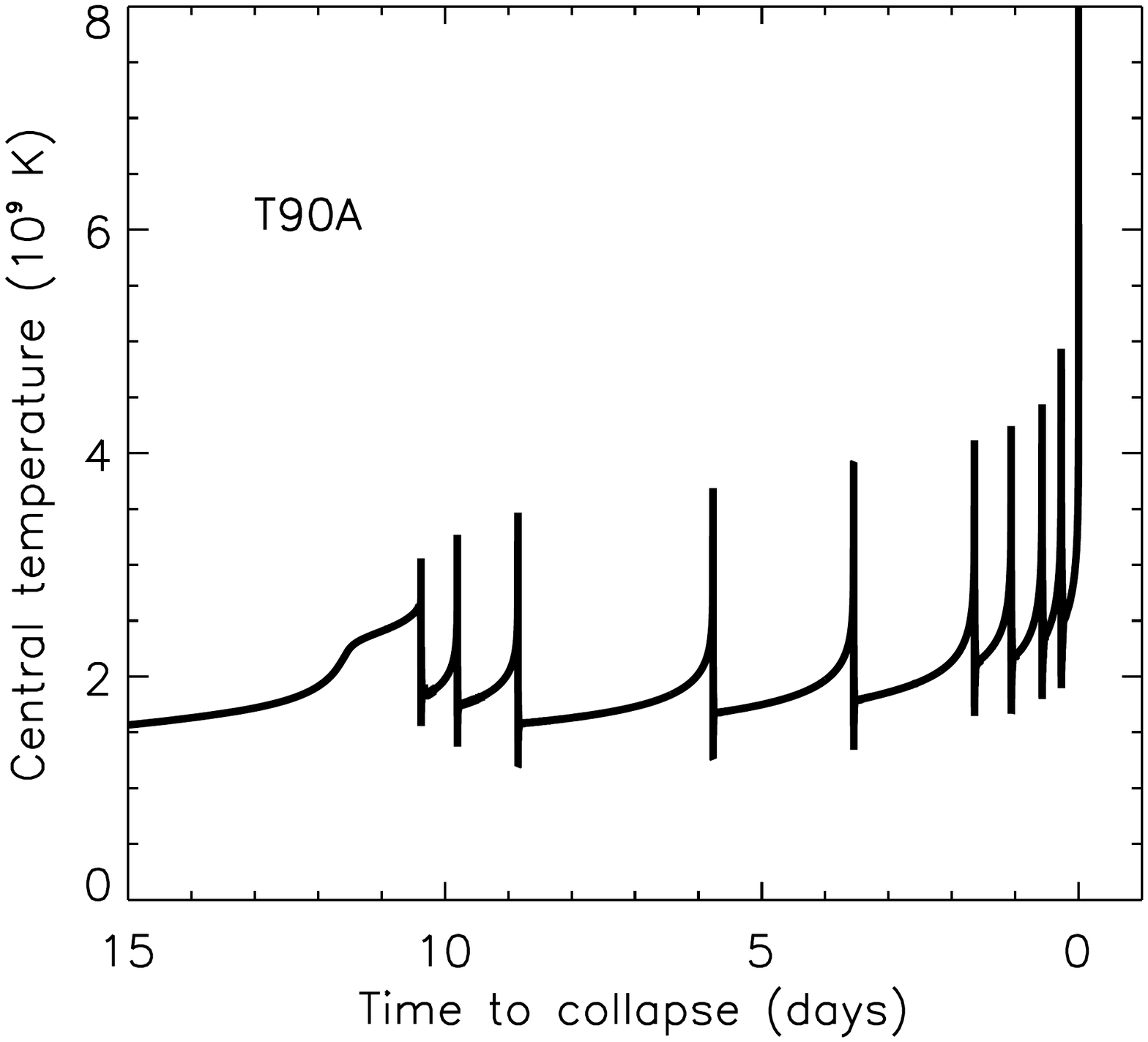}
\includegraphics[width=0.49\textwidth]{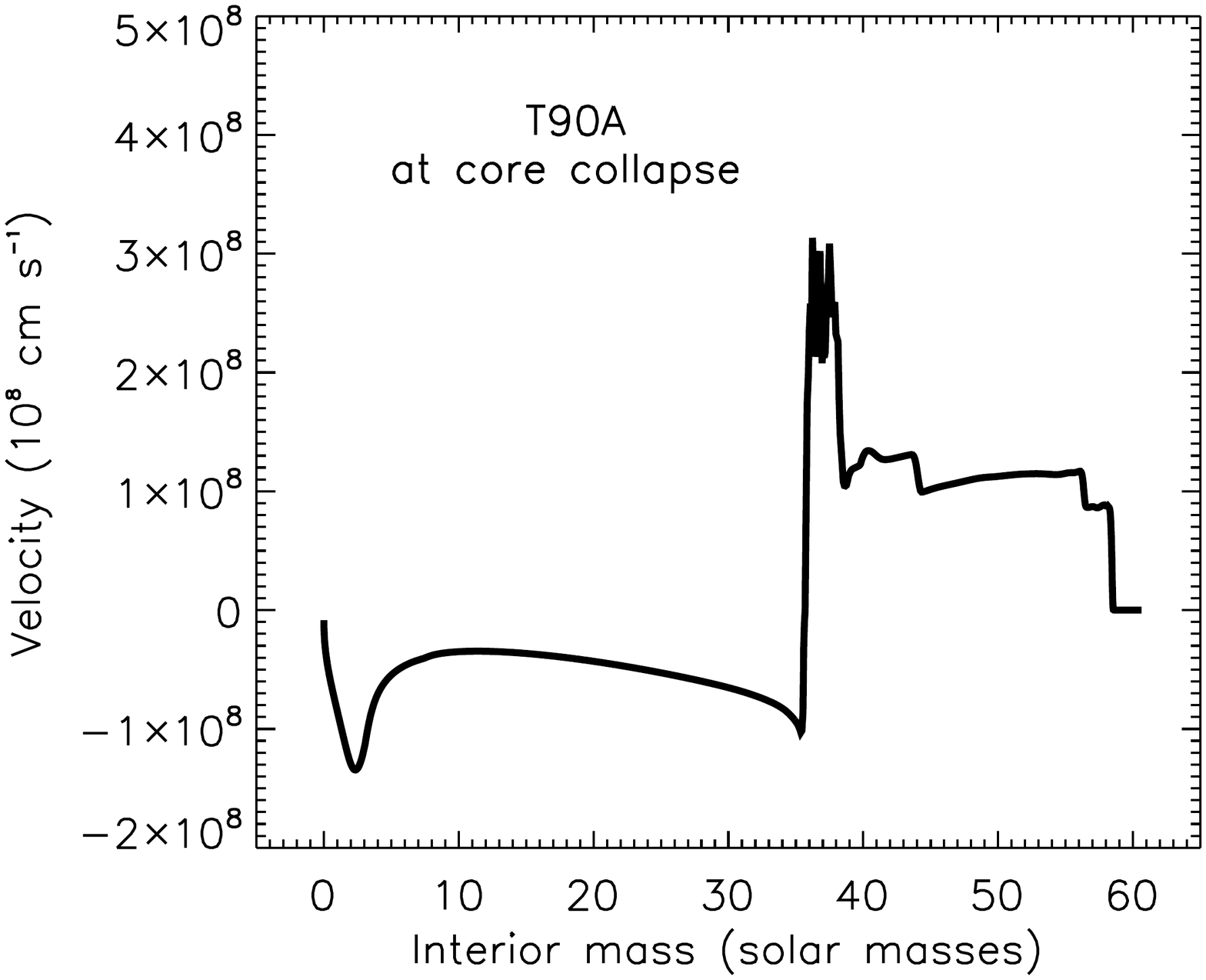}
\caption{Left: The pulsing activity of the low metallicity 90 \Msun
  \ Model T90A is reflected in its central temperature. Time is in
  units of 10$^5$ s measured backwards from the time the iron core
  collapses. The
  total duration of the pulsing activity is $1.1 \times 10^6$ s
  (\Tab{tmodels}) which is less than the shock crossing time for the
  envelope, hence shock waves pile up there and eventually merge into
  a single explosion with a smooth plateau (\Fig{t90lite}). The right
  frame shows the velocity structure in units of 1000's of km s$^{-1}$
  at time zero in the left frame (core collapse). \lFig{t90pulse}}.
\end{center}
\end{figure*}

% fig 15 -  90 msun light curves
\begin{figure}
\includegraphics[width=0.49\textwidth]{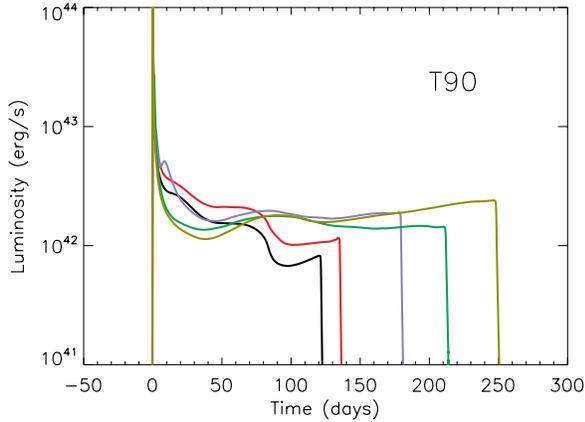}
\caption{Light curves for the 90 \Msun \ low metallicity models.
  The curves are for models with different mass loss rates: T90
  (black), T90A (red); T90B (blue), T90C(green), and T90D (gold). See
  \Tab{tmodels}. The luminosity on the plateau is similar to common
  SN IIp, although a bit faint. Models with larger envelope masses
  have systematically longer plateaus. There are no radioactive tails
  unless one results from circumstellar interaction.}  \lFig{t90lite}.
\end{figure}

\subsubsection{90 - 105 \Msun \ - Long Irregular Type IIp Supernovae}
\lSect{t100}

In this interesting mass range, the duration of the pulses roughly
equals or slightly exceeds the length of the plateau phase of the
supernova ($\sim100 - 200$ d). Depending upon the mass of the hydrogen
envelope, repeated pulses can lengthen, brighten, and add noticeable
structure to the light curves of some events and provide late time
activity in others, but there is still just one
supernova. \Fig{t100lite} shows the light curves for four
representative cases (\Tab{tmodels}). Some models, like T100C, are
very luminous for a long time. Roughly a quarter of the $6.9 \times
10^{50}$ erg of kinetic energy in the pulses is converted into light
here and the supernova might be categorized as ``superluminous''.

For perhaps the last time, the supernova has, throughout its duration, a
well-defined photosphere. Typical effective temperatures, well after
shock breakout, are around 6000 K.

% fig 16 -  100 msun light curves
\begin{figure}
\includegraphics[width=0.49\textwidth]{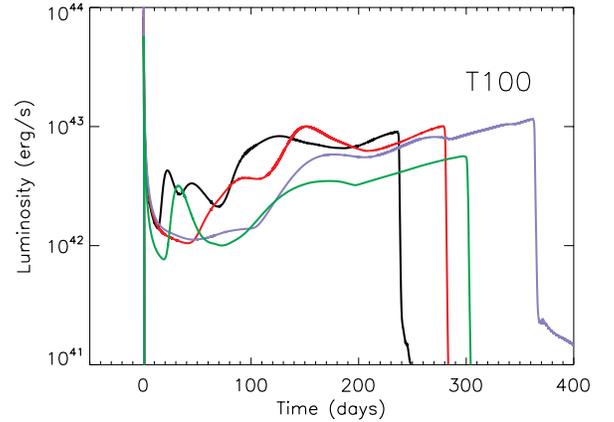}
\caption{Light curves from the 100 \Msun \ low metallicity models. The
  curves are for models with different mass loss rates: T100A (black),
  T100B (red); T100C (blue), and T100D (green). See
  \Tab{tmodels}. Structure from individual pulses is now starting to
  become apparent and the light curve is powered by a combination of
  recombination released shock energy and colliding shells. Generally
  models with larger envelope masses have longer plateaus, but Model
  T100C is exceptionally long due to the longer duration of the
  pulsing phase. The integrated light in these four curves is 1.2,
  1.4, 1.7 and 0.8 $\times 10^{50}$ erg for Models
  T100ABCD. \lFig{t100lite}}.
\end{figure}

\subsubsection{105  - 120 \Msun \ - Multiple Supernova and Long Luminous 
Events} 
\lSect{t110}

A further increase in mass results in energetic pulses that continue
longer than the duration of any single supernova. The first pulse
ejects what is left of the hydrogen envelope. Typically this matter is
helium rich and its ejection results in a light curve (\Fig{t110lite})
that resembles an ordinary Type IIp supernovae.  If the envelope is
massive and the pulse energy not unusually large, most of the ejecta
moves with at a relatively slow speed, around 1000 km s$^{-1}$. This
matter will later provide the ``anvil'' against which later faster
moving ejecta will strike. Most of the emission from these later mass
ejections is from the forward shock of the last shell ejected, though
the reverse shock can contribute to the luminosity since the mass of
the second ejection is usually less than the first. Both forward and
reverse shocks cause the pile up of matter in dense thin shell that
subsequent pulses can encounter.

For example, consider Models T110B and T110C (\Fig{t110lite}).  For
T110B, several energetic pulses in rapid succession impart a kinetic
energy to the envelope of $5.1 \times 10^{50}$ erg. The presupernova
mass was 84.1 \Msun \ and this first explosion
% t110b1 25000 odep 20000
ejects 35.3 \Msun \ at an average speed of $\sim1300$ km s$^{-1}$,
reducing the star's mass to 48.8 \Msun, essentially the bare helium
core.

There follows, in Model T110B, a quiescent period of about 2 years
during which no additional explosion occurs. The ejecta from the first
pulse expand and thin, eventually becoming transparent. At that point,
unless the formation of dust intervenes, one might see directly to the
helium core which would resemble a Wolf-Rayet star, but with several
complications. First the WR star is not shining by nuclear reactions,
but by gravitational contraction. The luminosity may be almost the
same - near Eddington, and the star may even have a wind, but its
radius, at least initially, is larger than a WR star of the same
mass. Moreover, substantial matter from the first ejection falls back
and accretes. This can contribute to the luminosity, but also
partially obscures the star. Generally though, one finds luminosities
near 10$^{40}$ erg s$^{-1}$ and radii of a few to 10 $\times 10^{11}$
cm. The underlying spectrum is thus very hot, $\sim10^5$ K and the
radiation may ionize some of the surrounding material. A more physical
treatment of the radiation transport problem is needed than is
feasible here.

% fig 17 - SLSN
\begin{figure*}
\begin{center}
\leavevmode
\includegraphics[width=0.48\textwidth]{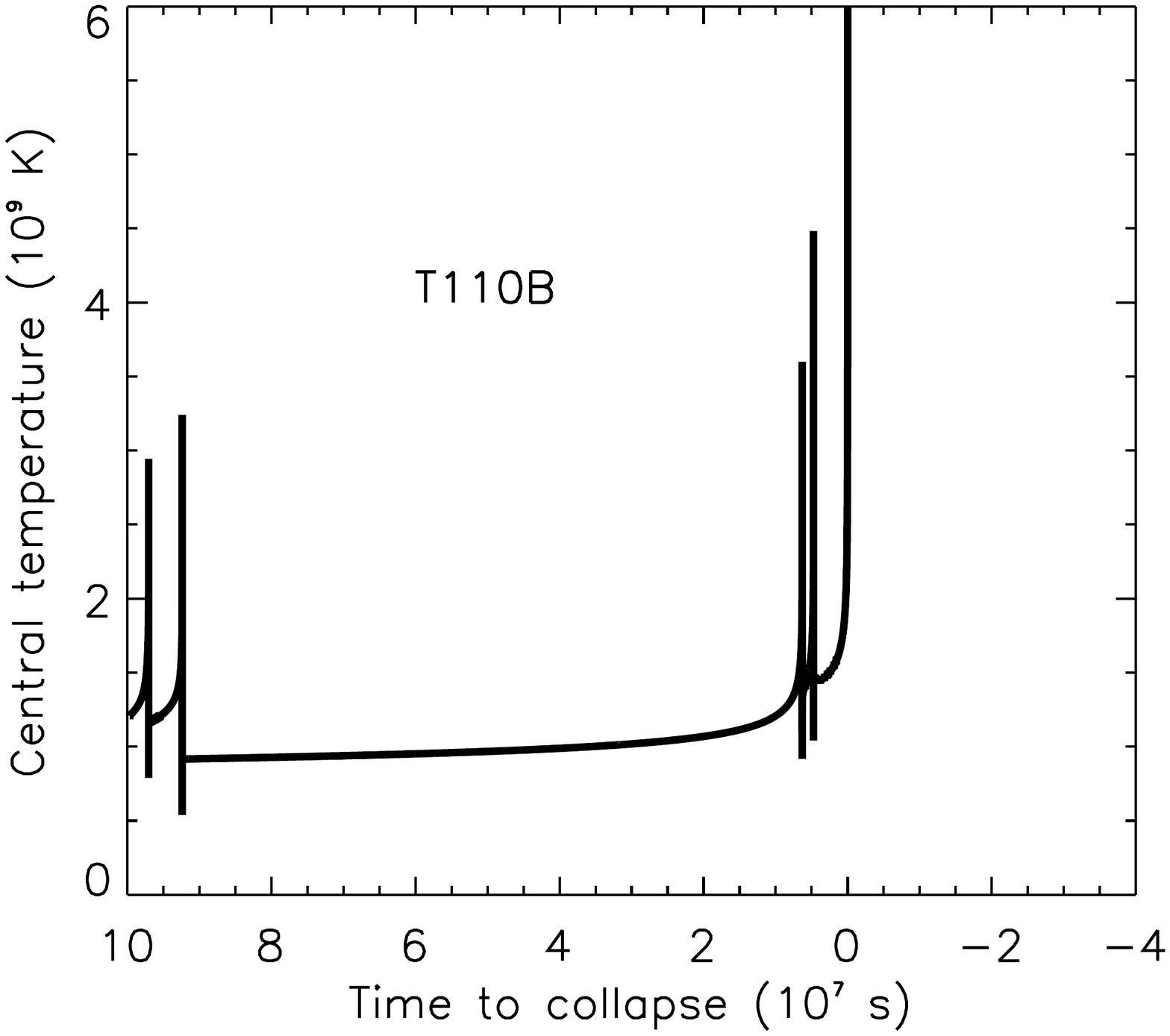}
\includegraphics[width=0.48\textwidth]{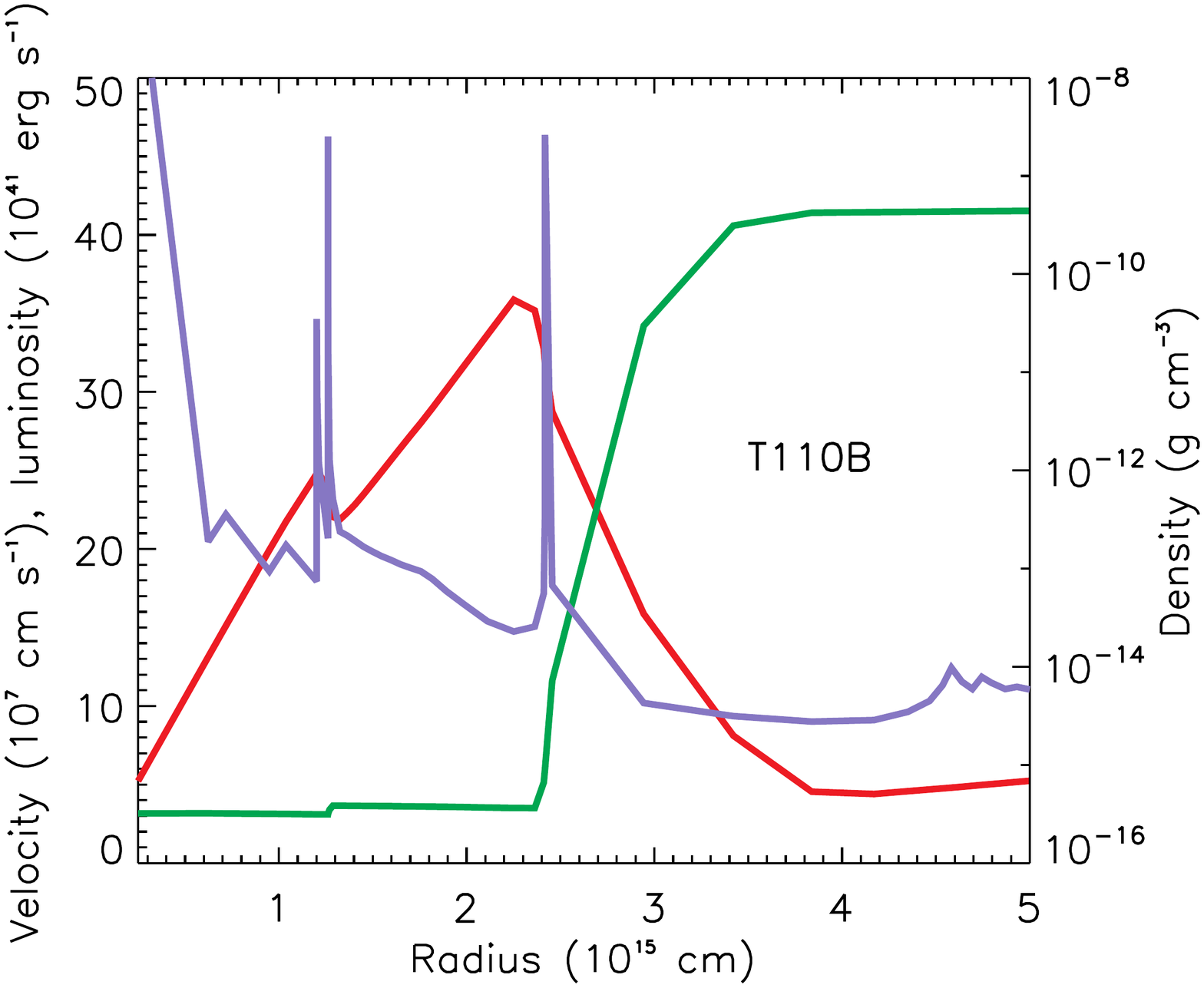}
\vskip 24pt
\includegraphics[width=0.48\textwidth]{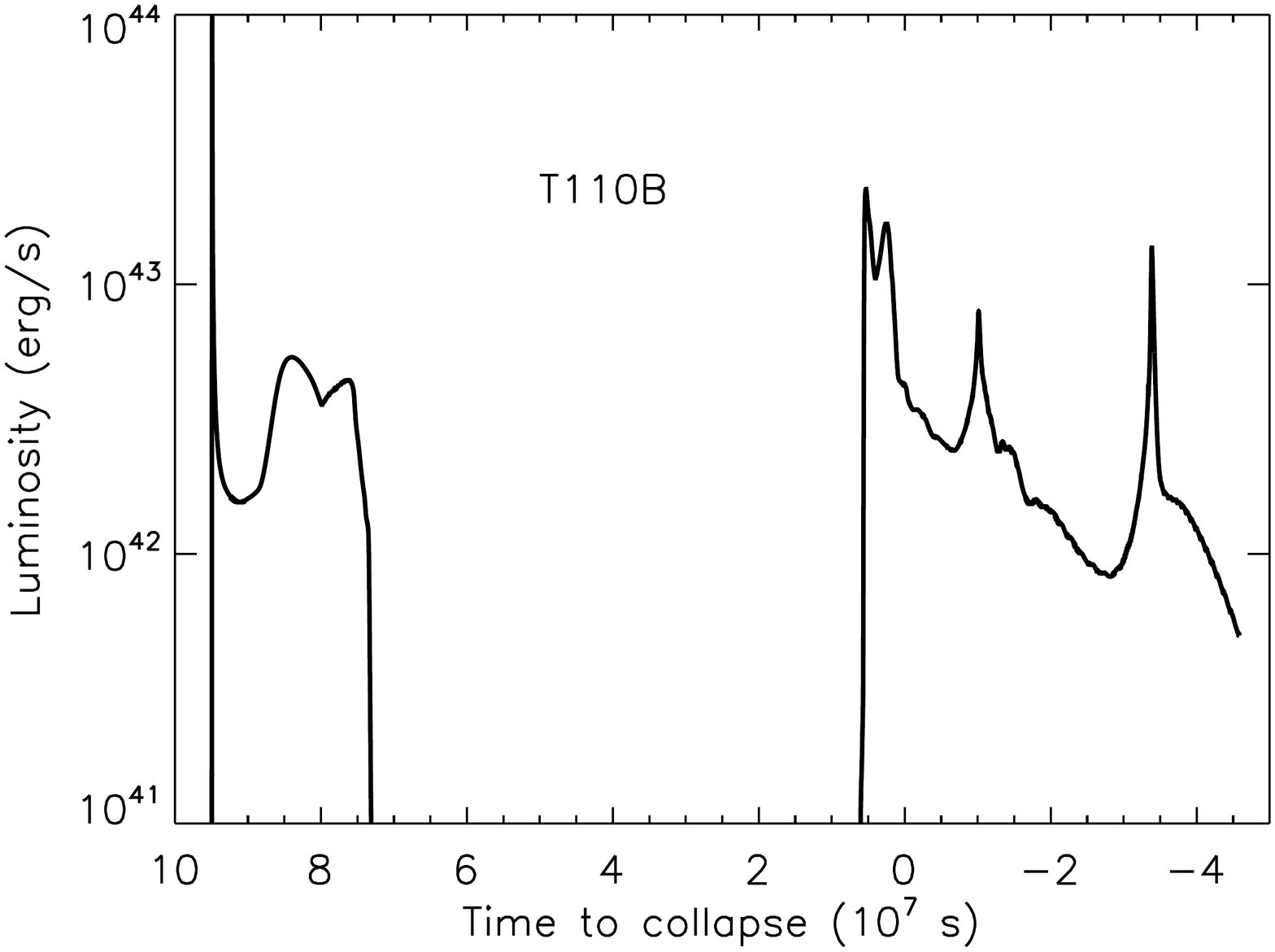}
\includegraphics[width=0.48\textwidth]{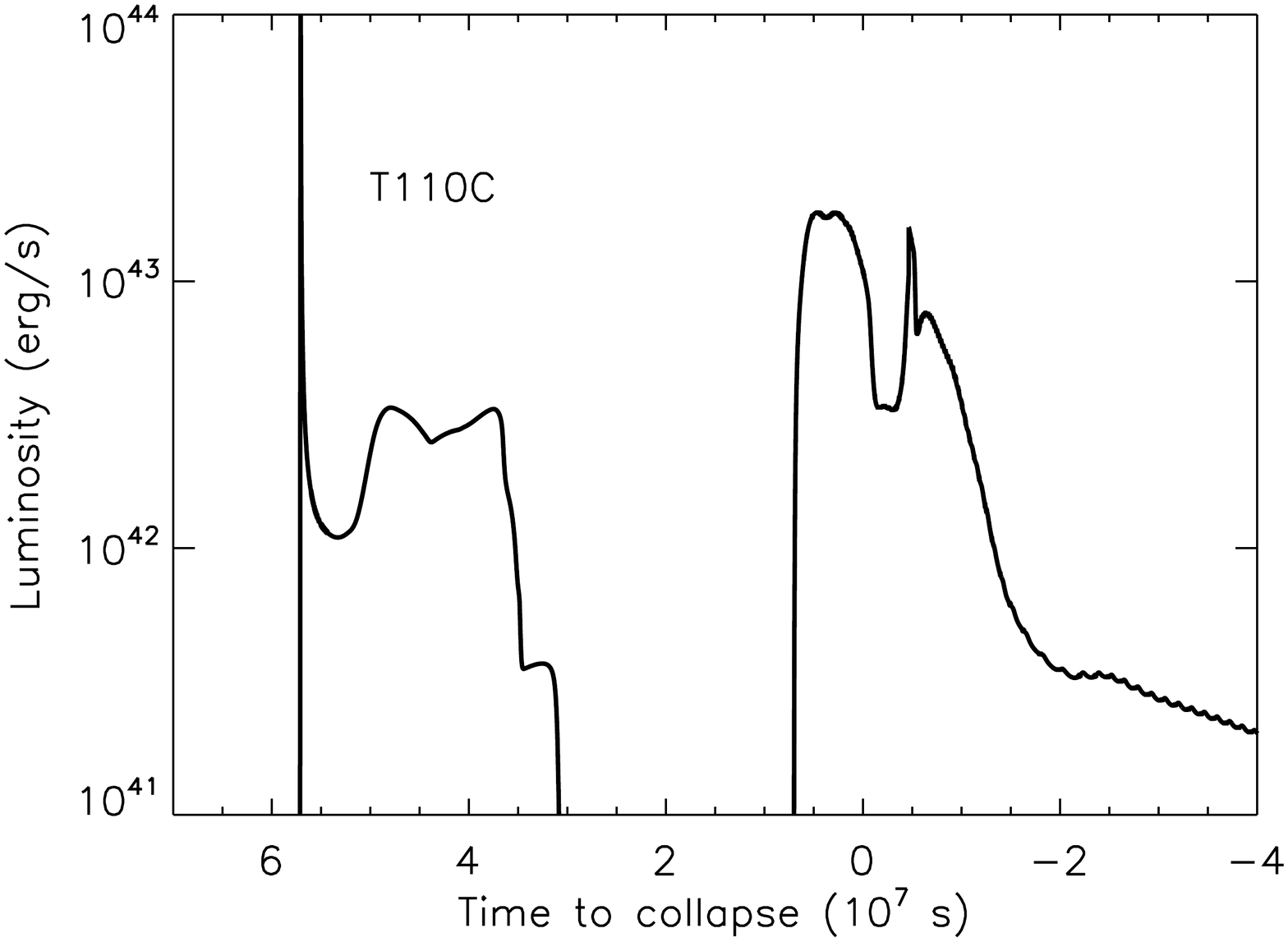}
\caption{Pulses and light curves for Models T110B and T110C. (Upper
  left:) The first two pulses in Model T110B $9.5 \times 10^7$ s
  before iron core collapse eject the hydrogen envelope and then
  impact it again producing a structured Type IIp-like light curve
  with a secondary maximum (lower left). There then
  ensues a 2.7 year delay as the core contracts in a Kelvin Helmholtz
  phase. This is followed by several strong pulses that eject an
  additional 4.1 \Msun, mostly of helium, with an energy of $2.3
  \times 10^{50}$ erg and speed 2000 - 3500 km s$^{-1}$. A short time
  later the iron core collapses, though the light curve continues to
  be powered by collisions for along time afterwards. (Lower left:)
  The light curve reflects the interaction among the ejected
  shells. The emission was rising towards a third sharp peak when the
  code became unstable due to the thin shells. (Upper right:) The
  density (blue; g cm$^{-3}$), velocity (red; 10$^7$ cm s$^{-1}$), and
  luminosity (green; 10$^{41}$ erg s$^{-1}$) structures in the ejecta
  are shown at the time the iron core collapses. The luminosity is
  chiefly originating from the collision at $2.5 \times 10^{15}$
  cm. Note the unphysical pile up of most of the ejected matter in
  very thin, dense shells. (Lower right:) The light curve for Model
  T110C is similar, but smoother because of the larger mass
  ejected. \lFig{t110lite}}.
\end{center}
\end{figure*}

2.7 years later, the contracting core of Model T110B encounters the
PPI again, launching a second set of pulses (\Fig{t110lite}). This
time a smaller amount of mass, 4.1 \Msun, is ejected, but with a
comparable energy, $2.3 \times 10^{50}$ erg, and higher speed,
2000 - 3500 km s$^{-1}$. This fast ejecta overtakes the previously
ejected envelope and slams into it, giving rise to a second bright,
collisionally-powered display. The matter it first impacts is the
inner part of the former envelope moving at about 500 - 700 km
s$^{-1}$ and located at about 5 $\times 10^{15}$ cm, an ideal radius
for converting kinetic energy into optical light and radiating it
without much adiabatic degradation. Depending upon the uncertain
opacity assumed for the matter external to the shock, the emitting
region may be optically thin or nearly so, thus the complete
thermalization of the emitted light is questionable. Even more
problematic is the tendency of the second mass ejection to pile up all
of the matter it encounters into a very thin, high density shell
moving at nearly constant speed. As time passes, a large fraction of
the total ejecta is contained in such shells. This is unphysical and
resolving their progress poses numerical difficulty for the 1D
code. Future radiation transport calculation need to be done in
2D or with some artificial means to keep the shells from becoming
unphysically thin.

Given these difficulties, the effective temperature cannot be
accurately calculated for these and heavier models, though the
bolometric light curve, which is essentially just $L = 2 \pi R_{\rm
  shock}^2 \rho v_{\rm shock}^3$, can. Here $\rho$ is the density
ahead of the shock, a residual of the earlier mass ejection.
\Fig{t110lite} shows several episodes of high luminosity, including
unphysically sharp spikes as the thin shells from the last two pulses
first run into each other and then into the first ejection. Additional
structure is imprinted by a dense shell associated with the formation
of a reverse shock during the first mass ejection. Altogether $1.6
\times 10^{50}$ erg of light is emitted in the lengthy second outburst
shown in \Fig{t110lite}. This is about half of the total kinetic
energy of the second set of pulses.

Model T110C is qualitatively similar, suggesting that the answer is
not very sensitive to the choice of mass loss rate so long as the
envelope is not removed. Initial pulses eject 47.2 \Msun, including
the entire hydrogen envelope, with an energy of $3.6 \times 10^{50}$
erg, leaving a core of 48.8 \Msun. 1.7 years later another 4 \Msun is
ejected with energy $3.5 \times 10^{50}$ erg. This runs into envelope
material at about 10$^{15}$ cm moving at about 400 - 1000 km s$^{-1}$
giving rise to the bright display in \Fig{t110lite}.  $1.8 \times
10^{50}$ erg is radiated in the second display.

Other heavier models in \Tab{tmodels} with total pulsational durations
less than about 10 years show similar behavior and light curves
to the 110 \Msun \ models.  For example, Model T115A, resembles
110B and 110C, but has three well-spaced, strong pulses at 13.2, 7.8,
and 0.67 $\times 10^7$ s before core collapse. Once again the narrow
spikes in the light curve, except for the first break-out transient,
would be broader but contain about the same total energy in a 2D
simulation. For Model T115B though (not shown), the interval between
pulses has become so long (39 years) that the the radius where the
shock interaction takes place was well beyond 10$^{16}$ cm and the
efficiency for optical emission, uncertain. 

% fig 18 -  120 msun light curve
\begin{figure}
\includegraphics[width=0.48\textwidth]{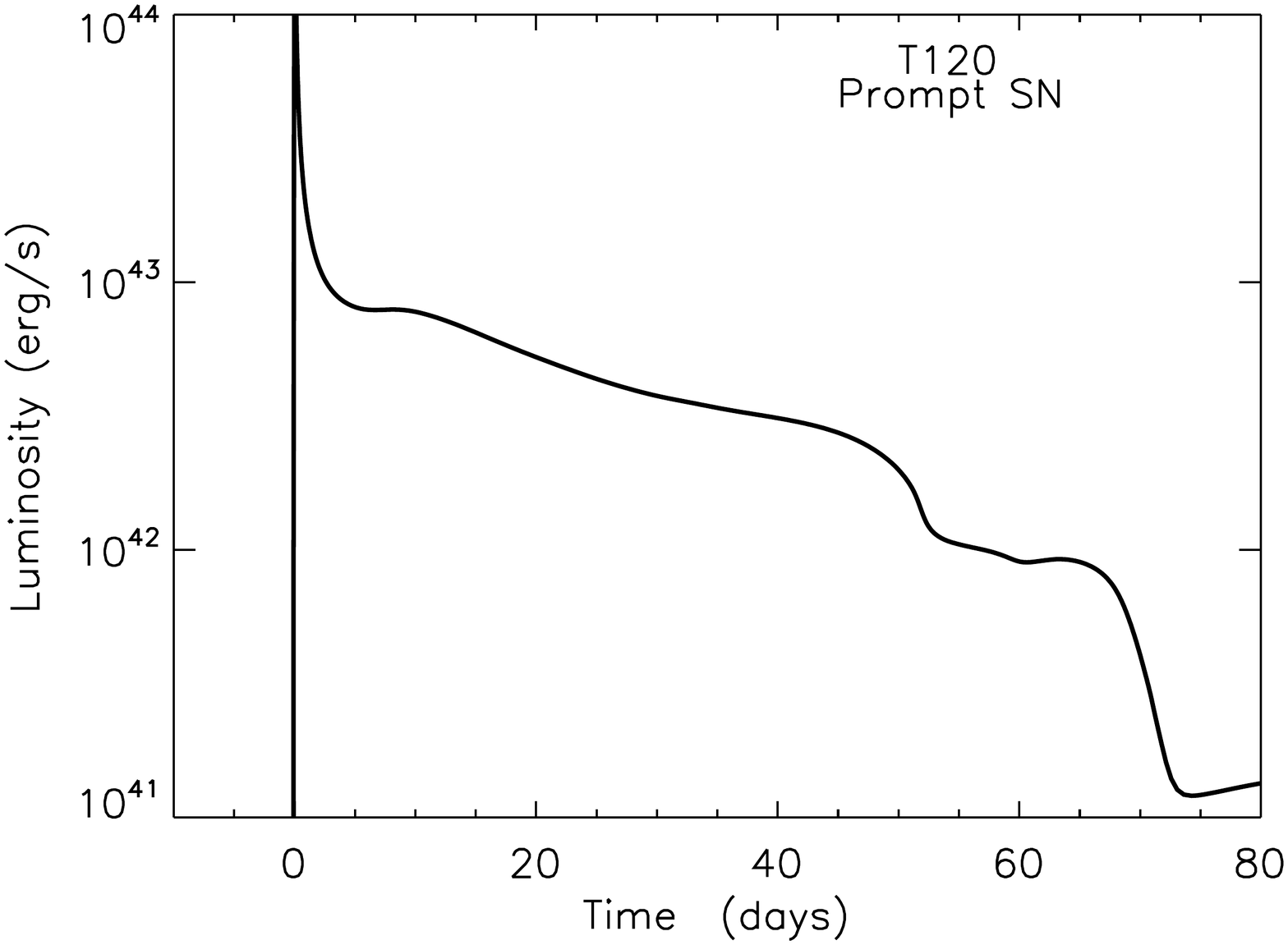}
\includegraphics[width=0.48\textwidth]{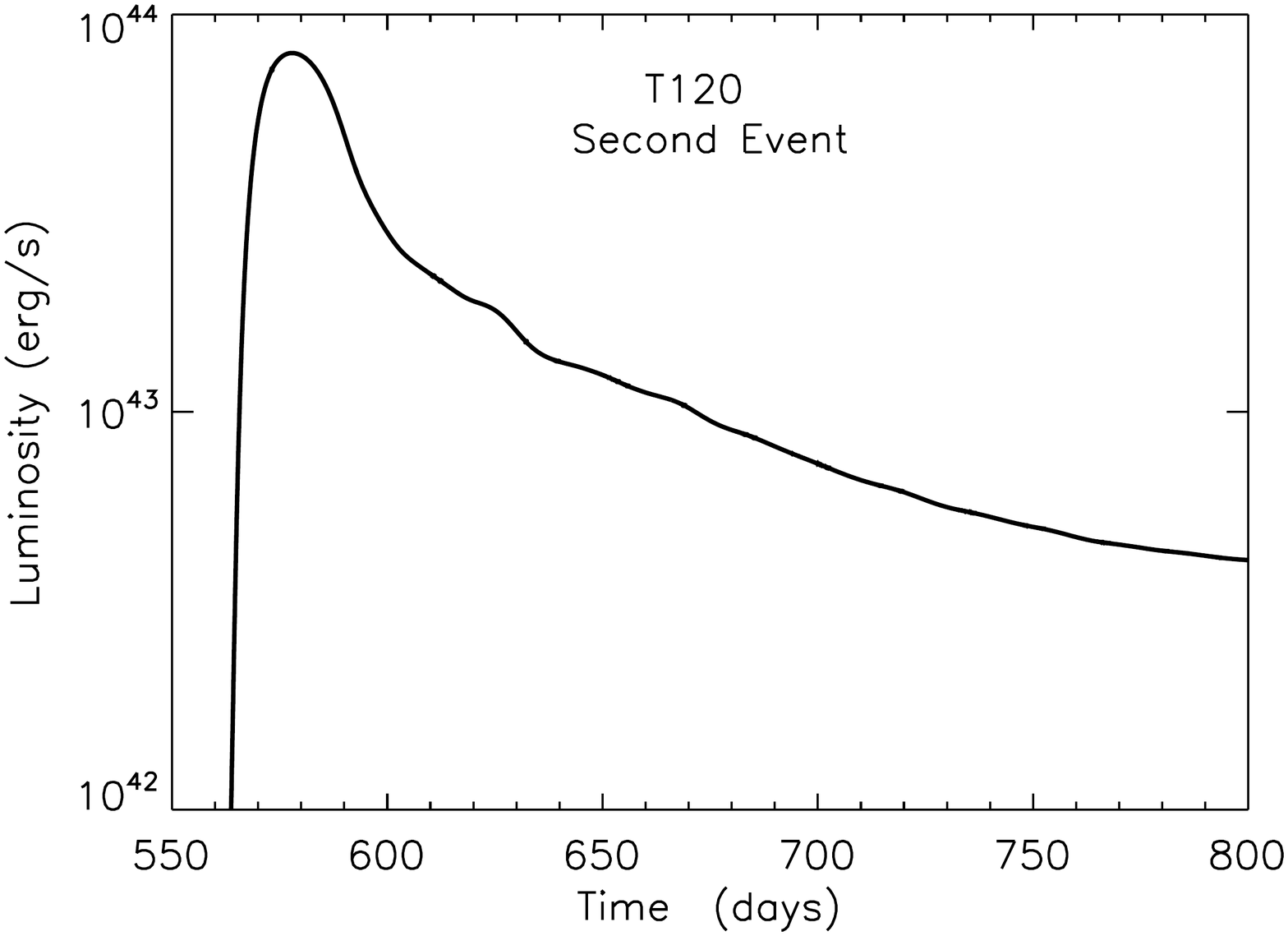}
\caption{The first pulse in Model T120 produces a Type IIp supernova
  (upper frame) which declines rapidly with time once the denser inner
  envelope of the presupernova star is encountered. If the
  presupernova star had been a BSG this initial display would have been
  much fainter (\Sect{lbv}). Eighteen months later, a second very
  bright supernova is produced as the ejecta of additional pulses run
  into the ejected envelope and reheat it. $3.3 \times 10^{50}$ erg is
  radiated during this second outburst. 1200 years after that the iron
  core collapses.  \lFig{t120lite}}.
\end{figure}

Besides producing repeating supernovae, energetic pulses that happen
over a time span of years are of interest for producing SLSN. With
typical speeds of 1000 km s$^{-1}$, the first ejection moves to a few
times 10$^{15}$ cm in a year. That is an optimal radius for converting
kinetic energy into a supernova-like display. Unfortunately, while
some of the light curves are indeed much brighter than common
supernovae, the total light emitted was only a fraction of
10$^{51}$ erg, and did not rise to the level of the brightest
``superluminous'' supernovae. This is a dilemma to which we shall
return in \Sect{superl} and \Sect{conclude}.

PPISN in the upper end of this mass range could also give rise to
supernovae of mixed typology. Since the first pulse ejects the
hydrogen envelope, with a delay of years to centuries until the next
outburst, the supernova might initially present as Ib and later turn
into a IIn. An example would be Model 115B. Here several initial
pulses eject the hydrogen envelope of about 36 \Msun \ with an energy
of $3.8 \times 10^{50}$ erg and typical speeds 1000 km s$^{-1}$. A
model with more mass loss would have ejected less mass with greater
speed. 39 years later, two more pulses with a combined energy of $2.2
\times 10^{50}$ erg eject an additional 5.2 \Msun \ of helium
core. The collision of these two shells produces a supernova,
preseumably of Type Ib, since the photosphere and shock were in the
helium layers. The merged shell then encountered the slowest moving
hydrogen at $3 \times 10^{16}$ cm about 3 years later. The peak
luminosity of the Ib supernova was $4 \times 10^{42}$ erg
s$^{-1}$. These results resemble SN 2014C which, though born a Ib
switched to Type IIn about a year later \citep{Mil15}, although the
authors say that this particular supernova was too light to have been
a PPISN.

\subsubsection{120 - 140 \Msun \ - Bright Circumstellar Interaction 
and Delayed Supernovae}
\lSect{t120}

\paragraph{Interaction with the Ejected Envelope}

For the heaviest PPISN, the interval between the ejection of the
envelope and the later pulses is so long that the envelope expands out
of the radial range where optically bright supernovae are expected
from the interaction.  Any subsequent shocks will interact with a
medium that is optically thin, at least to electron scattering. The
duration of the collision will be years or even centuries, so the
luminosity required to emit the differential kinetic energy is
lower. The collision may produce x-rays and radio emission as well as
optical emission. An example of a Type IIn supernova in this state,
although not necessarily a PPISN itself, is SN 1996er \citep{Meu13},
which is currently interacting with several solar masses of ejecta at
about 10$^{17}$ cm.

The qualitative nature of these events can be inferred from the time
scales for pulsational duration given in \Tab{tmodels}. Assume the
first pulse ejects the envelope at 1000 - 2000 km s$^{-1}$ and later
pulsations eject shells of helium and heavy elements at 4000 km
s$^{-1}$. The collision with the envelope then happens at
$\sim10^{18}$ cm when the time scale is centuries and at $\sim10^{19}$
cm when it is millennia. To dissipate a kinetic energy of 10$^{51}$
erg, a characteristic shock luminosity would be 10$^{40}$ to 10$^{41}$
erg s$^{-1}$. The emission might resemble what is presently
transpiring in SN 1987A where the material ejected in the equatorial
ring about $\sim 10,000$ y ago is being impacted by high velocity
matter \citep[e.g..][]{Sug05}. The display from a PPISN would differ
in that it would be approximately spherically symmetric and the
velocities slower. The supernova remnant would also contain the
central star, still glowing with a luminosity comparable to the
circumstellar interaction.

Consider for example, Model T123A. The first pulse in this model
ejected the star's envelope producing a typical Type IIp supernova
that lasted (L $> 10^{42}$ erg s$^{-1}$) for roughly 120 days. The
mass of the ejected materal was 17.8 \Msun \ and its kinetic energy,
$1.1 \times 10^{51}$ erg. Velocities ranged from 1000 to 3000 km
s$^{-1}$ in most of the ejecta. A bound remnant of 56.6 \Msun \ was
left, the outer few tenths \Msun \ of which still contained
appreciable hydrogen. The central temperature of the remaining star,
after a few brief oscillations, was $5.4 \times 10^8$ K. Over the next
1010 years the core contracted, eventually encountering the PPI a
second time. By then, most of the former envelope had coasted to
between 1 and 10 $\times 10^{18}$ cm, with about 10 \Msun \ inside $5
\times 10^{18}$ cm. This second pulse ejected 6.4 \Msun, consisting
chiefly of helium, with an energy of $6\times 10^{50}$ erg that ran
into the ejected envelope. The collision produced a low luminosity,
$\sim10^{40} - 10^{41}$ erg s$^{-1}$ event that continued for
centuries until most of the kinetic energy of the second pulse was
radiated away.  220 years after the second pulse, the iron core
collapsed, presumably to a black hole. There were no later
pulses. This behavior was typical for stars in the 120 - 130 \Msun
\ range, especially for those models with total explosion energies
(\Tab{tmodels}) below $3 \times 10^{51}$ erg.

%The early evolution of Model T130A was similar. The first pulse
%ejected the envelope, giving a Type IIp supernova \Fig{t130lite}. The
%mass ejected in this initial episode was 19.1 \Msun \ and the kinetic
%energy, $1.57 \times 10^{51}$ erg. Over the next 2800 years, the
%supernova debris expanded and the WR star contracted, as the central
%density and temperature rose to the point where a second flash
%occurred, ejecting 9.3 \Msun \ of helium, carbon and oxygen with an
%energy of $1.69 \times 10^{51}$ erg and velocity around 4000 km
%s$^{-1}$. By this time though most of the envelope ejected in the
%first supernova was outside 10$^{19}$ cm. 330 years later the iron
%core collapsed without further pulsing activity.

%  130A  first pulse t minus 1.00e11 s. 2nd pulse 1.044e10

% see especially model
% medusa:/Volumes/DATA1/q/qdat/woosley/stars/ppsn16/t130/t130a3#49000 
% this is an example of a very deep final bounce that made a lot
% of 56Ni and a bright light curve from CSM collision. Probably rare.

\paragraph{Late Pulsations and Bright Supernovae}

In addition to enduring circumstellar interaction, some of the models
in this mass range also produced a second bright supernova after a
very long delay.  Their evolution was similar to the heaviest helium
cores discussed in \Sect{hecoreexp} that also produced delayed
supernovae, but here the initial explosion was brighter and there
would be hydrogen in the spectrum. A strong initial pulse ejects the
envelope and, after a long delay, two or more pulses, shortly before
the star dies, collide with one another powering a second supernova.
Model T130D is a particularly energetic example.  The large envelope
mass in this model (\Tab{tmodels}) is due to the complete neglect of
mass loss, but other combinations of main sequence mass and mass loss
give a similar core structure and late-time light curve. Three of the
four 135 \Msun \ models were similar, as well as some of the more
energetc cases between 120 and 130 \Msun \ (T122A and T124A for
example).  Model T130D had three strong pulses, the first of which
ejected the 70.0 \Msun \ envelope, giving the usual Type IIp supernova
(\Fig{t130lite}). This first pulse was quite strong, $1.5 \times
10^{51}$ erg, and almost unbound the star. Following relaxation to
hydrostatic equilibrium after the pulse, the central temperature was
only $2.8 \times 10^8$ K.  3300 years later, the core experienced a
second instability and ejected 7.7 \Msun \ of helium, C, and O with an
additional $1.1 \times 10^{51}$ erg of kinetic energy.  Eight months
after that, a third and final very deep bounce ejected 13.5 \Msun
\ with $1.5 \times 10^{51}$ erg. The peak central temperature reached
during this last pulse was very hot, $5.95 \times 10^9$ K, sufficient
to produce 1.8 \Msun \ of $^{56}$Ni (that was not ejected).  This 13.5
\Msun \ collided with the shell ejected by the second pulse at a
radius of $\sim10^{15}$ cm producing a very luminous supernova,
T130D-b in \Fig{t130lite}. Because the collision was between two
shells mostly devoid of hydrogen, this would probably have been a Type
Ibn or Icn supernova, though perhaps with some hydrogen lines from
outer edge of the second mass ejection. The total energy in light was
$4.5 \times 10^{50}$ erg, or about a third of the energy in the last
pulse. {\sl This was the most energy in light found for any PPISN in
  the present study that did not invoke magnetar formation.}

% see model 50000

The post-explosion structure of Model T130D was affected by its large
$^{56}$Ni production.  After a brief stage of adjustment to
hydrostatic equilibrium following the last pulse, the iron core mass,
including the 1.8 \Msun \ of $^{56}$Ni, was 3.54 \Msun, and the
silicon plus iron core was 5.82 \Msun. The time since the onset of
pulsing activity was $1.0 \times 10^{11}$ s. Over the next $1.5 \times
10^{11}$ s, the core experienced a lengthy Kelvin Helmholtz
contraction. During the first few months, the decay of $^{56}$Ni to
$^{56}$Fe powered extensive convection. This resulted in the full
mixing of material from 2.04 \Msun \ out to 29.4 \Msun, i.e., almost
the entire remaining star. As a result the compositional distinction
between ``iron core'', ``silicon core'', and ``oxygen core'' became
blurred. Several other massive models also produced a lot of $^{56}$Ni
and experienced extensive mixing powered by radioactive decay, Model
T120A for example.

\begin{deluxetable}{ccccc} 
\tablecaption{SOLAR METALLICITY MODELS} 
\tablehead{ Mass & Mass Loss & M$_{\rm preSN}$ & M$_{\rm He}$ & M$_{\rm CO}$ \\
(\Msun) &  & (\Msun) & (\Msun) & (\Msun)  }
\startdata
S80B   & 1/4 & 40.47 & 34.71 & 30.25  \\
S80C   & 1/8 & 55.95 & 36.48 & 32.00  \\
S90B   & 1/4 & 48.01 & 40.07 & 35.29  \\ 
S90C   & 1/8 & 65.30 & 41.06 & 36.00  \\
S100B  & 1/4 & 48.92 & 44.90 & 38.86  \\
%primary nitrogen S100C, extra dredge up max he core 45.95
S100C  & 1/8 & 72.59 & 41.00 & 41.00  \\
%S110B  & 1/4 &  64.23 &  49.20 & 44  \\
S110C  & 1/8 & 63.70 & 50.85 & 45.16  \\
%S115B  & 1/4 & 70.34 & 51.45 &       \\
%S115C  & 1/8 & 67.08 & 53.15 & 47.11 \\
%S120B  & 1/4 & 69.52 & 56.13 &       \\
S120C  & 1/8 & 71.31 & 53.67 & 47.27   \\
\enddata
\lTab{smodels}
\end{deluxetable}

% fig 19 -  T130D light curves
\begin{figure}
\includegraphics[width=0.49\textwidth]{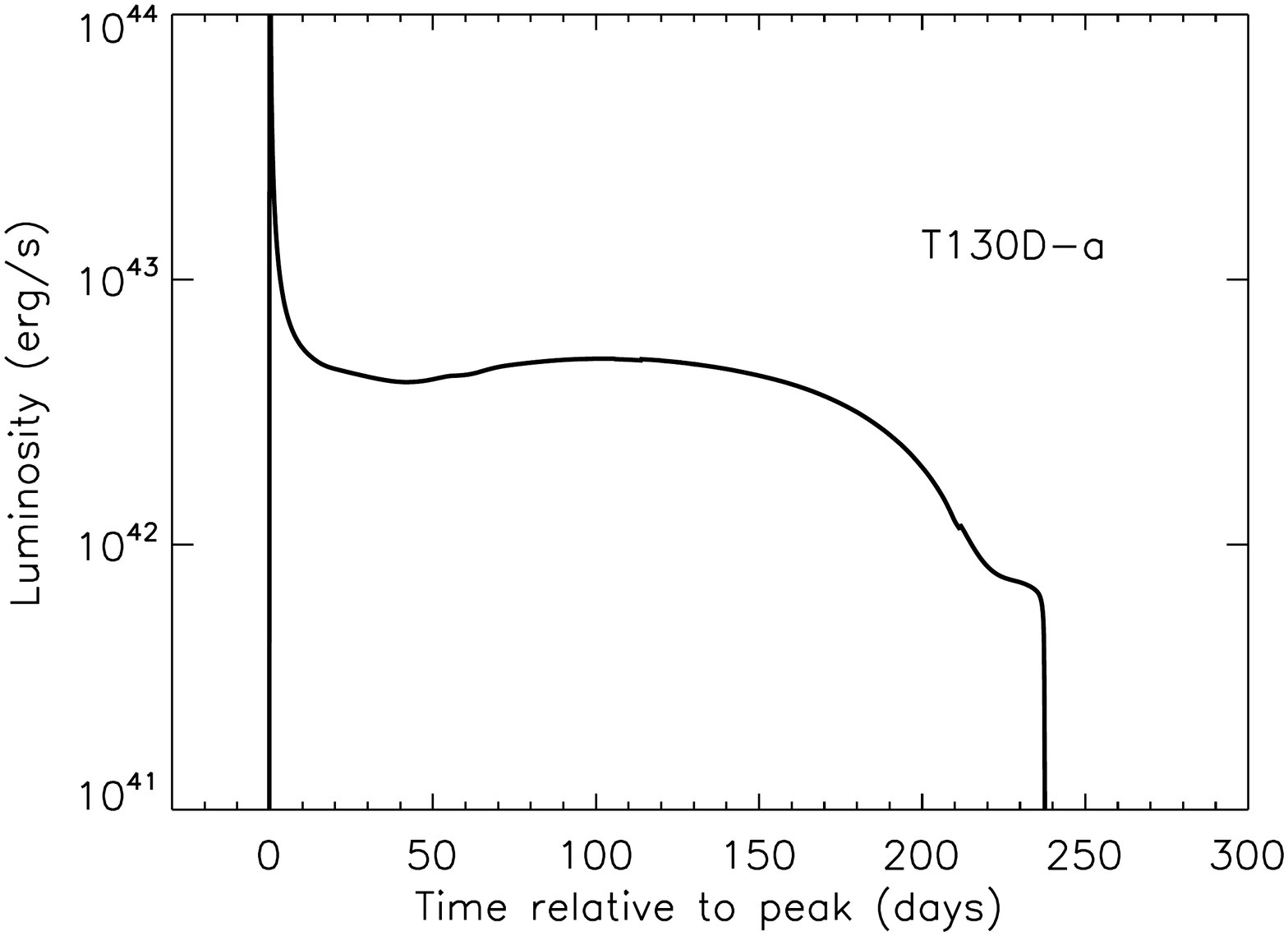}
\includegraphics[width=0.49\textwidth]{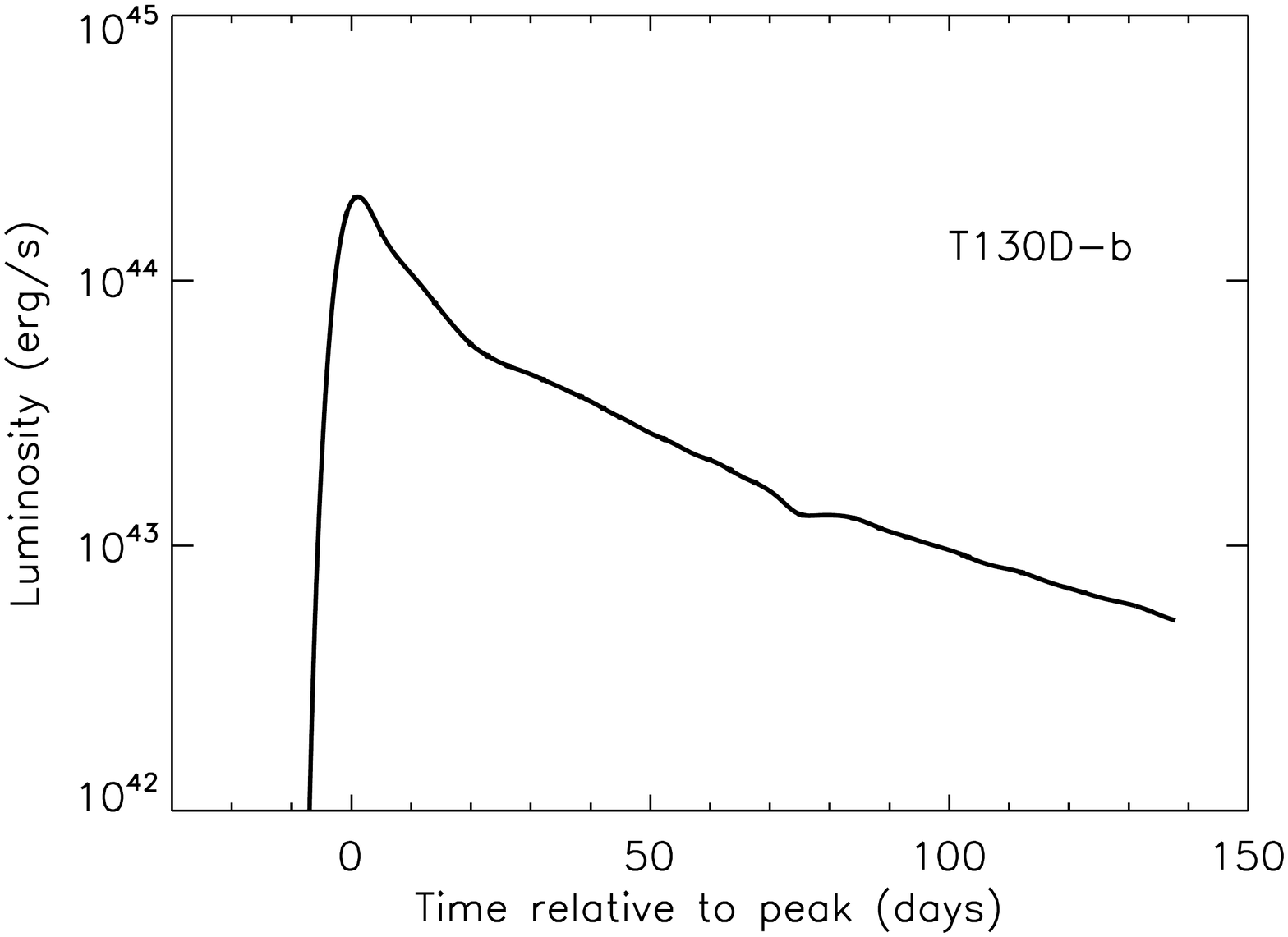}
\caption{Light curves from Model T130D. The top frame (Model T130D-a)
  gives the bolometric light curve resulting from the first pulse and
  envelope ejection. 3260 years later, pulses 2 and 3 in rapid
  succession collided with one another producing the bright light
  curve shown in the lower panel, Model T130D-b. \lFig{t130lite}}.
\end{figure}

\section{Stars with Solar Metallicity}
\lSect{solar}

A smaller grid of solar metallicity models was also calculated
(\Tab{smodels}).  In order to leave helium cores sufficiently massive
that the PPI is encountered, a substantial reduction in the mass loss
rate was required (\Sect{mdot}), roughly a factor of 4 to 8. Given the
low mass loss and a non-trivial hydrogen envelope at the end, the
outcome for non-rotating solar metallicity models is similar to those
with lower metallicity (\Tab{tmodels}).  For example, Models T80 leave
a helium core mass of 35 $\pm 1$ \Msun, the same as Models S80C and
S80D. The CO core masses are also similar. Since the final evolution
for non-rotating stars depends chiefly the helium core mass, the
outcome will be the same.

There are interesting differences in the envelope structure,
however. The effect of the ``iron bump'' on the opacity is more
pronounced in the solar metallicity models. This gives them larger
radii, and in some cases, makes the models difficult to
converge. Large density inversions develop when the mass of the
convective envelope is small and the local luminosity close to
Eddington \citep[e.g.,][]{San15}. These difficulties inhibited the
study of masses above 120 \Msun \ or mass loss rates so large that
most or all of the hydrogen envelope was lost. Models S110B and S120B
(not given in the table) had mass loss rates so high, even with 25\%
of the standard value, that they would have lost their envelopes and
become Wolf-Rayet stars. Some of these cores might become PPISN, but
most would have continued to lose so much mass that they would have
died short of the 30 \Msun \ required for the PPI.

\section{Luminous Blue Progenitors}
\lSect{lbv}

The ``T'' series (\Tab{tmodels}) and ``S'' series (\Tab{smodels})
models ended their lives as RSGs with photospheric radii $\sim
10^{14}$ cm. Other recent theoretical studies \citep[e.g.,][]{Che15}
also show stars with 10\% solar metallicity and masses up to 150 \Msun
\ ending their lives as RSG's. These results might be regarded as
inconsistent, however, with observations showing that stars above
about 35 \Msun, the ``Humphreys-Davidson limit'' \citep{Hum79}, do not
spend a significant part of their lifetime as RSGs, even in the SMC
\citep{Mas03,Lev07} where the metallicity is about one-seventh solar.
Given the possible tension between theory and observations, it is
worth exploring the consequences of both BSG and RSG progenitors for
PPISN.

% fig 20 - LBV
\begin{figure*}
\begin{center}
\leavevmode
\includegraphics[width=0.48\textwidth]{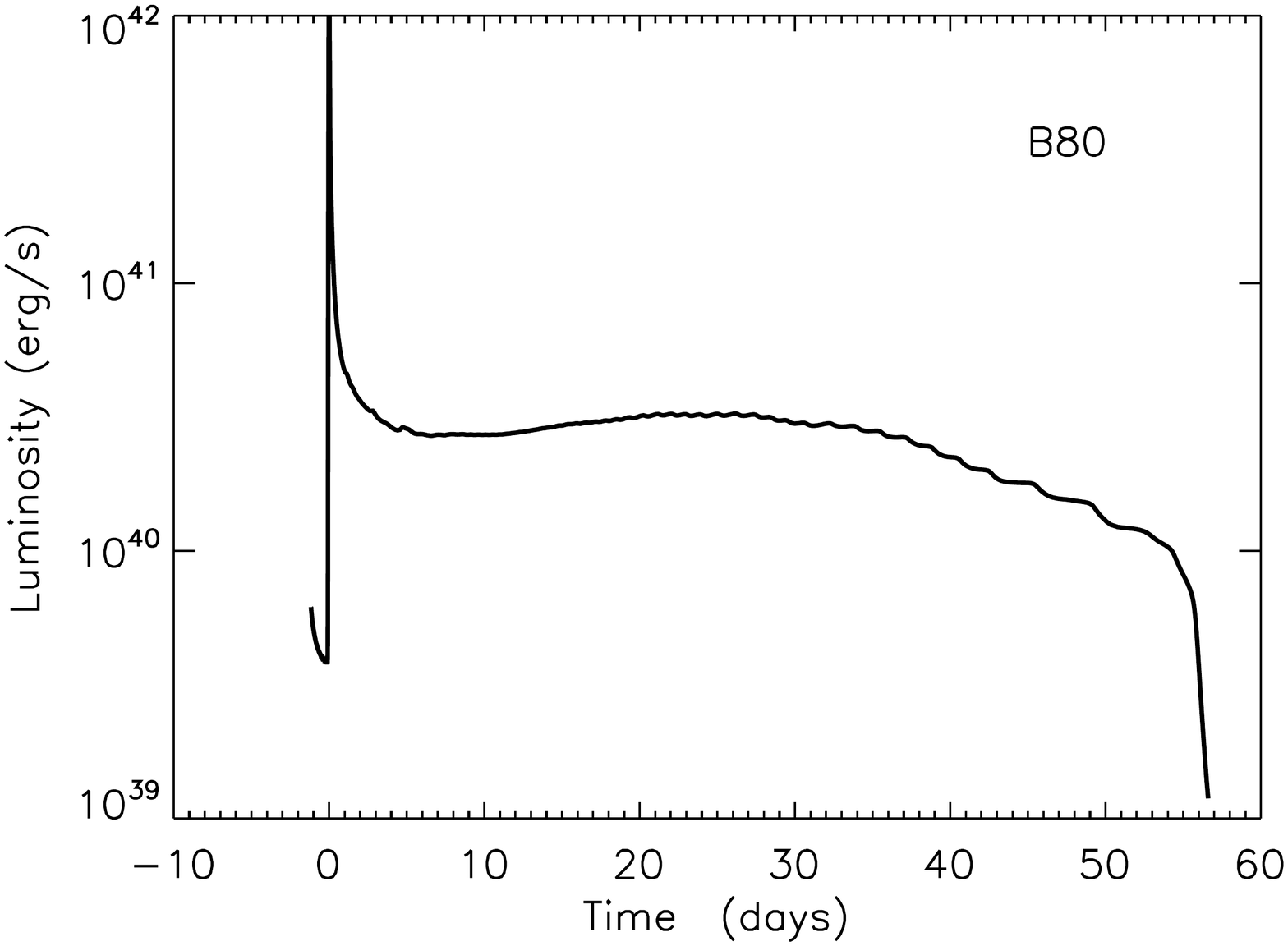}
\includegraphics[width=0.48\textwidth]{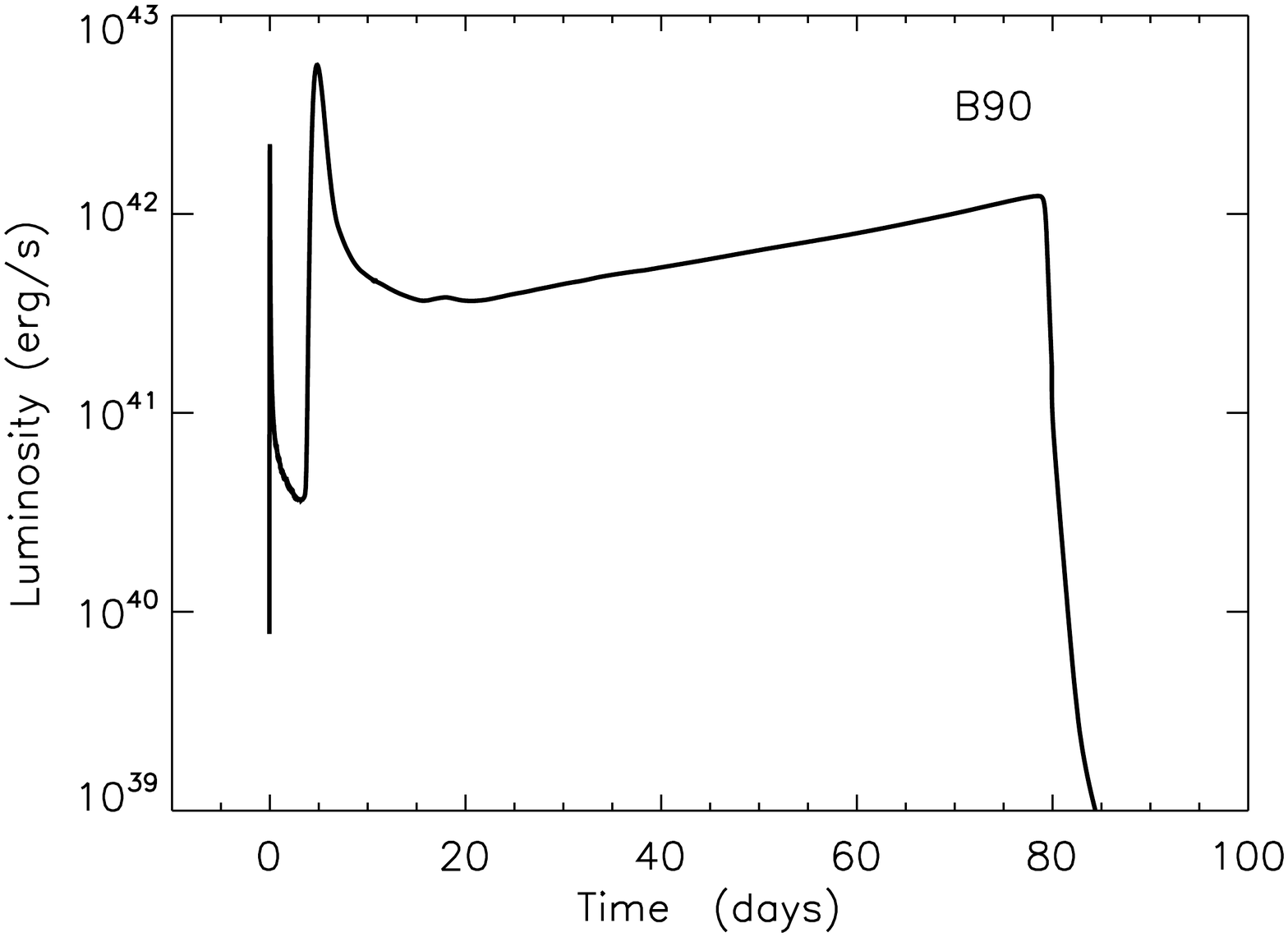}
\vskip 24pt
\includegraphics[width=0.48\textwidth]{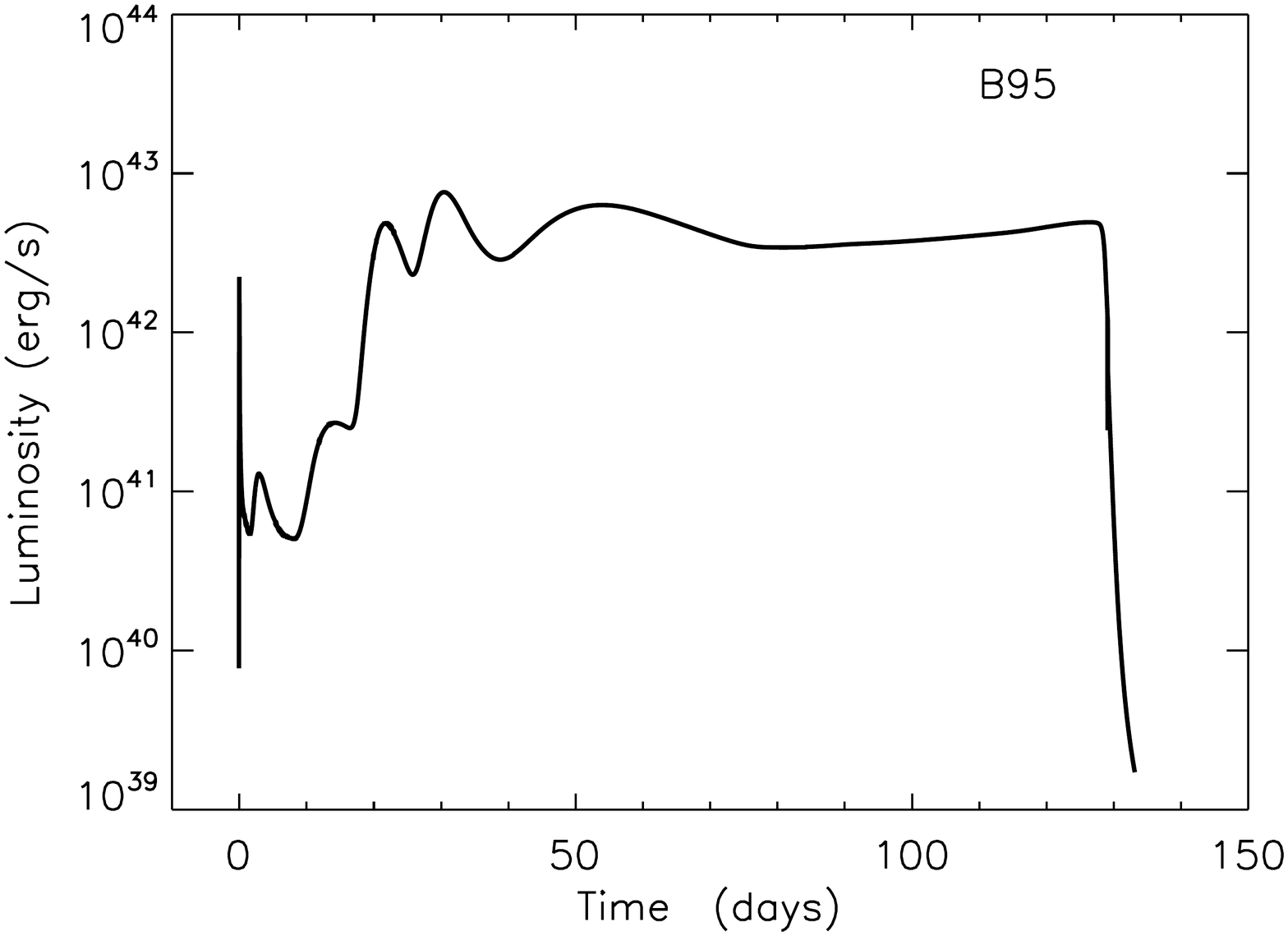}
\includegraphics[width=0.48\textwidth]{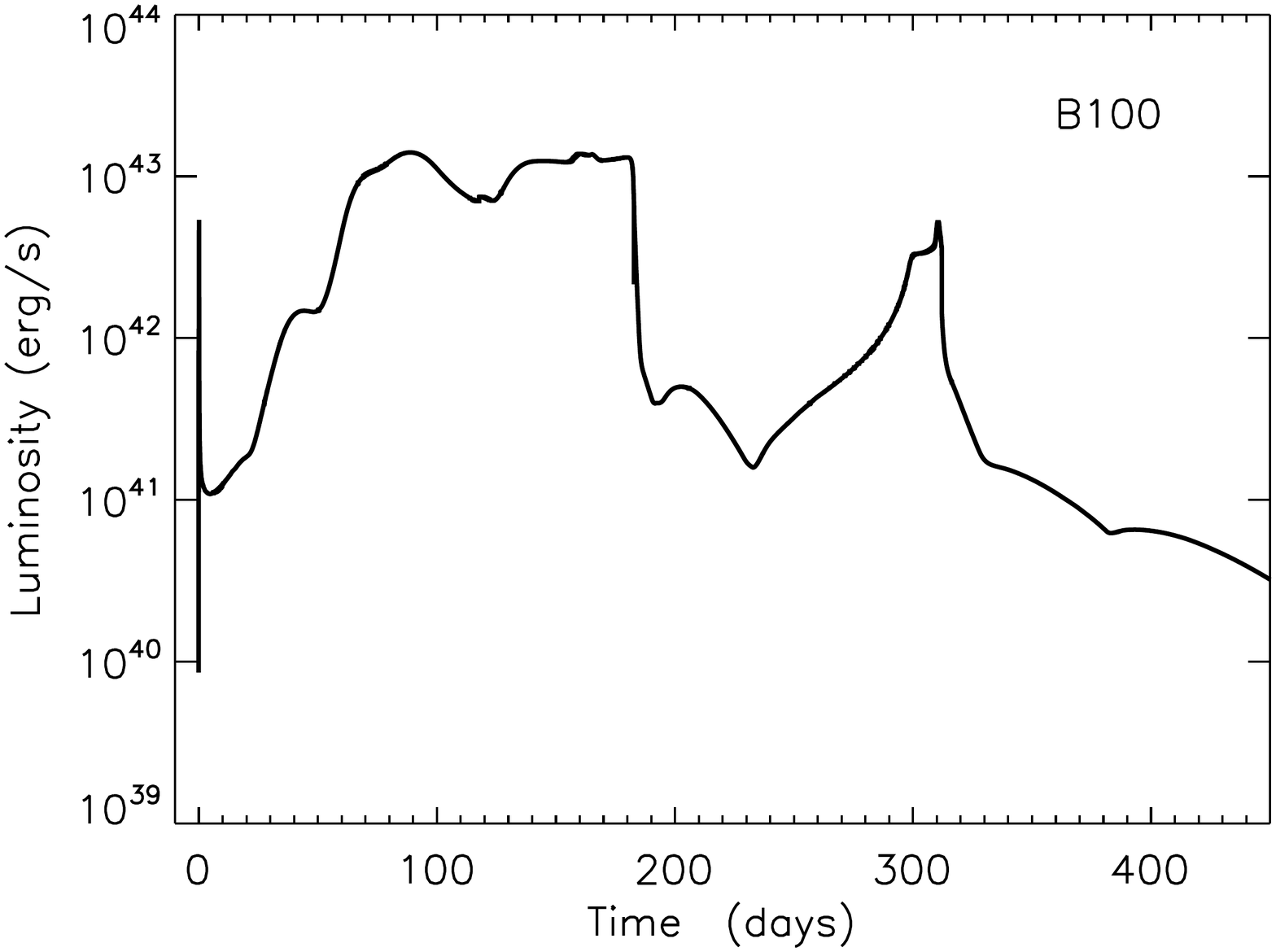}
\vskip 24pt
\includegraphics[width=0.48\textwidth]{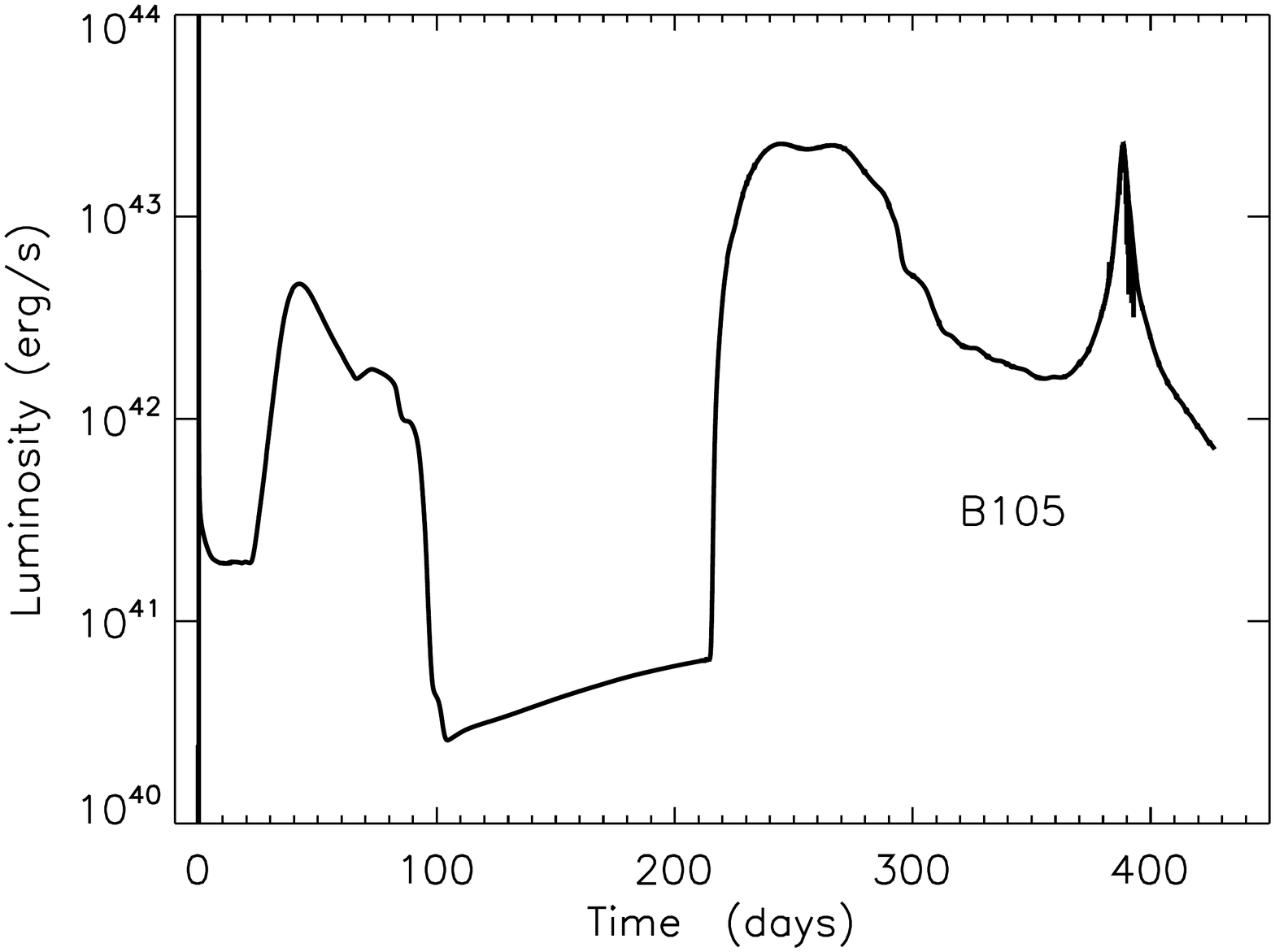}
\includegraphics[width=0.48\textwidth]{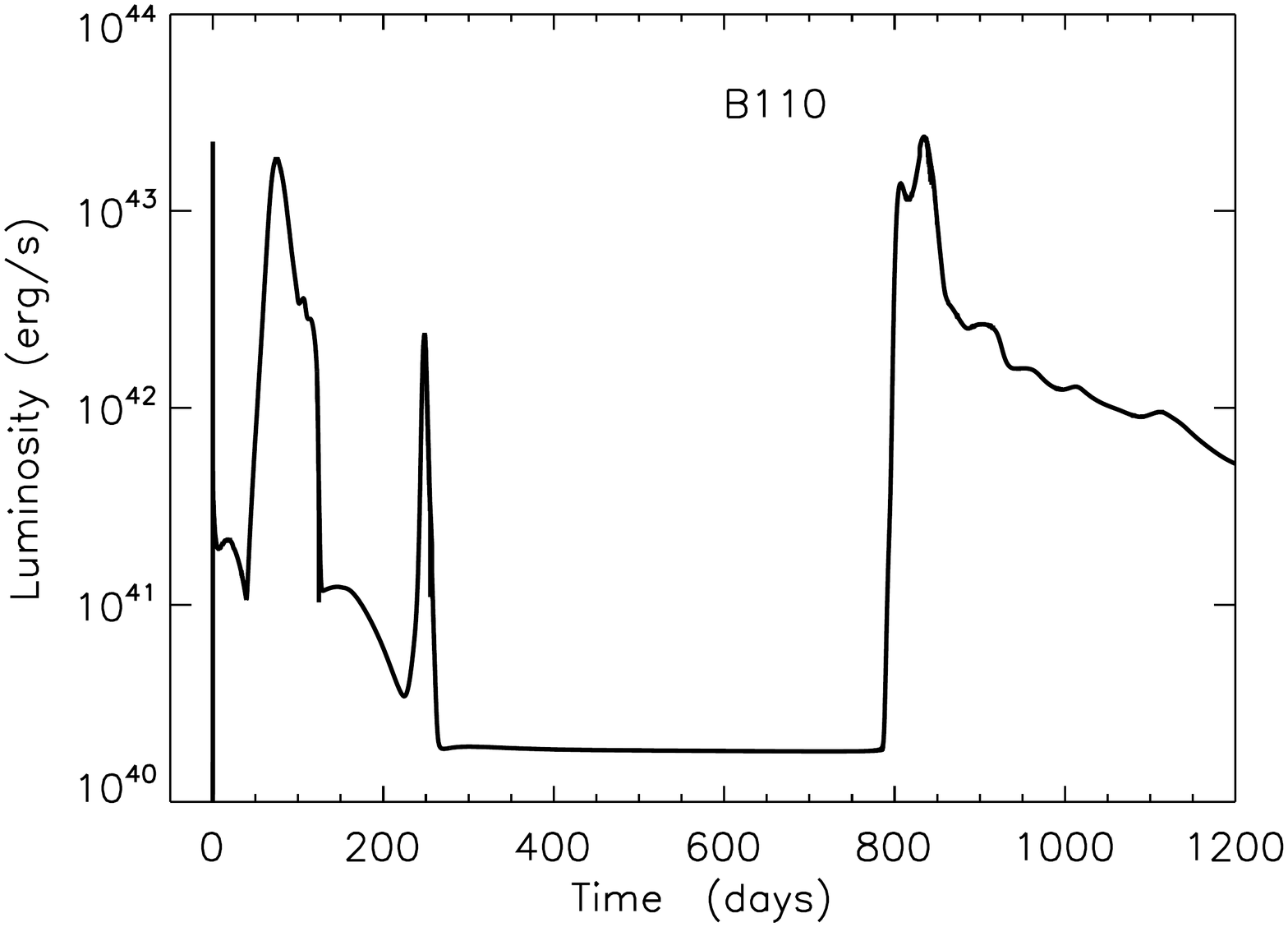}
\caption{Light curves for blue progenitor stars with presupernova
  radii 2 to $5 \times 10^{12}$ cm. Zero time here corresponds to
  shock break out following the first pulse. On this scale, iron core
  collapse occurs at a time given by ``Duration'' in
  \Tab{bmodels}. Any temporal structure after envelope recombination
  would be smoother in nature than in this 1D simulation.  Any
  interaction with a presupernova wind is omitted. Luminosities in the
  low luminosity dormant phases of Models B105 and B110 are partly due
  to fallback and are poorly determined. During these inactive
  periods, a constant of 10$^{40}$ erg s$^{-1}$ might be more
  appropriate. \lFig{lbvlitea}}.
\end{center}
\end{figure*}

% fig 21 - LBV various Menv
\begin{figure}
\includegraphics[width=0.48\textwidth]{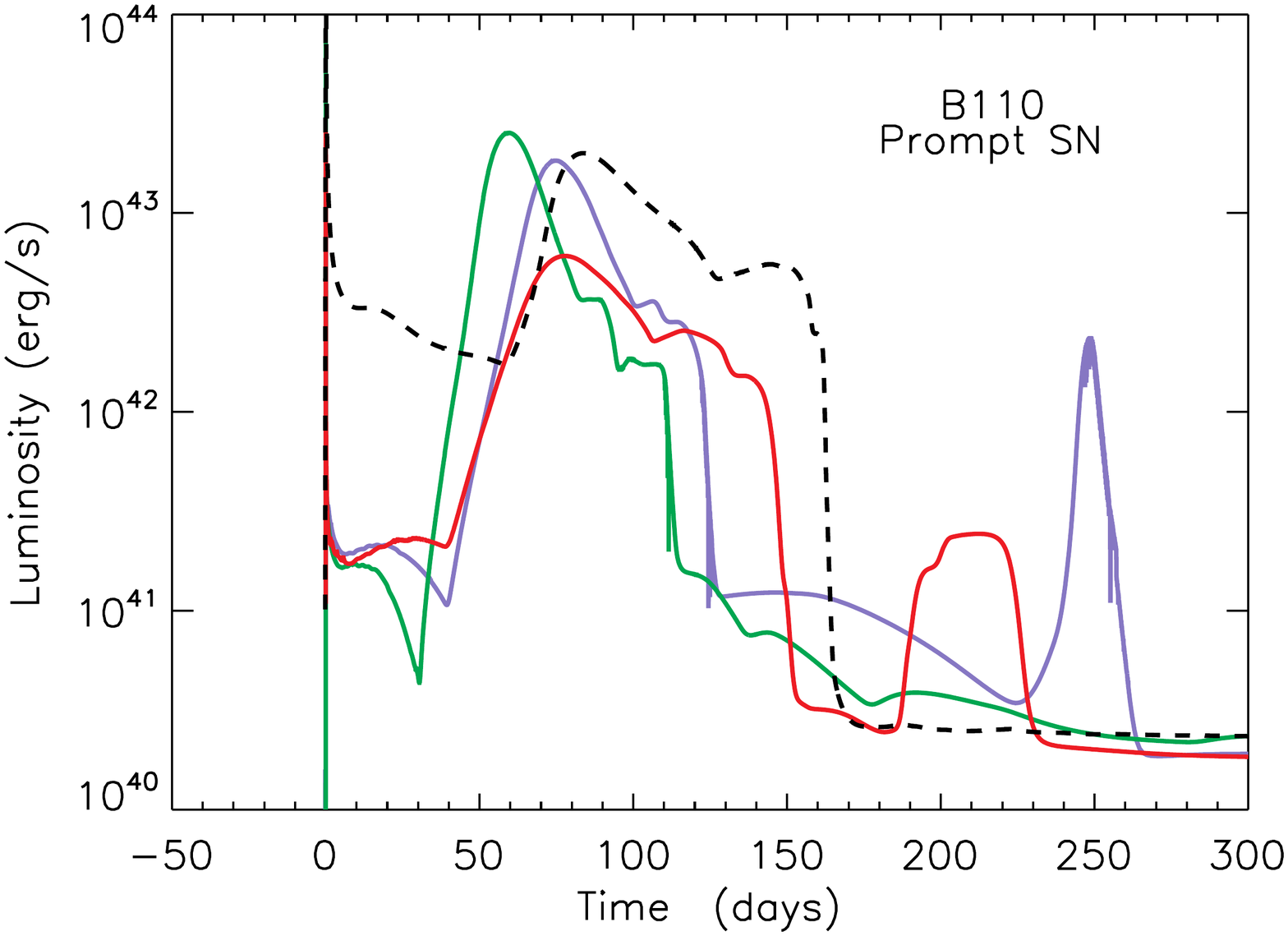}
\includegraphics[width=0.48\textwidth]{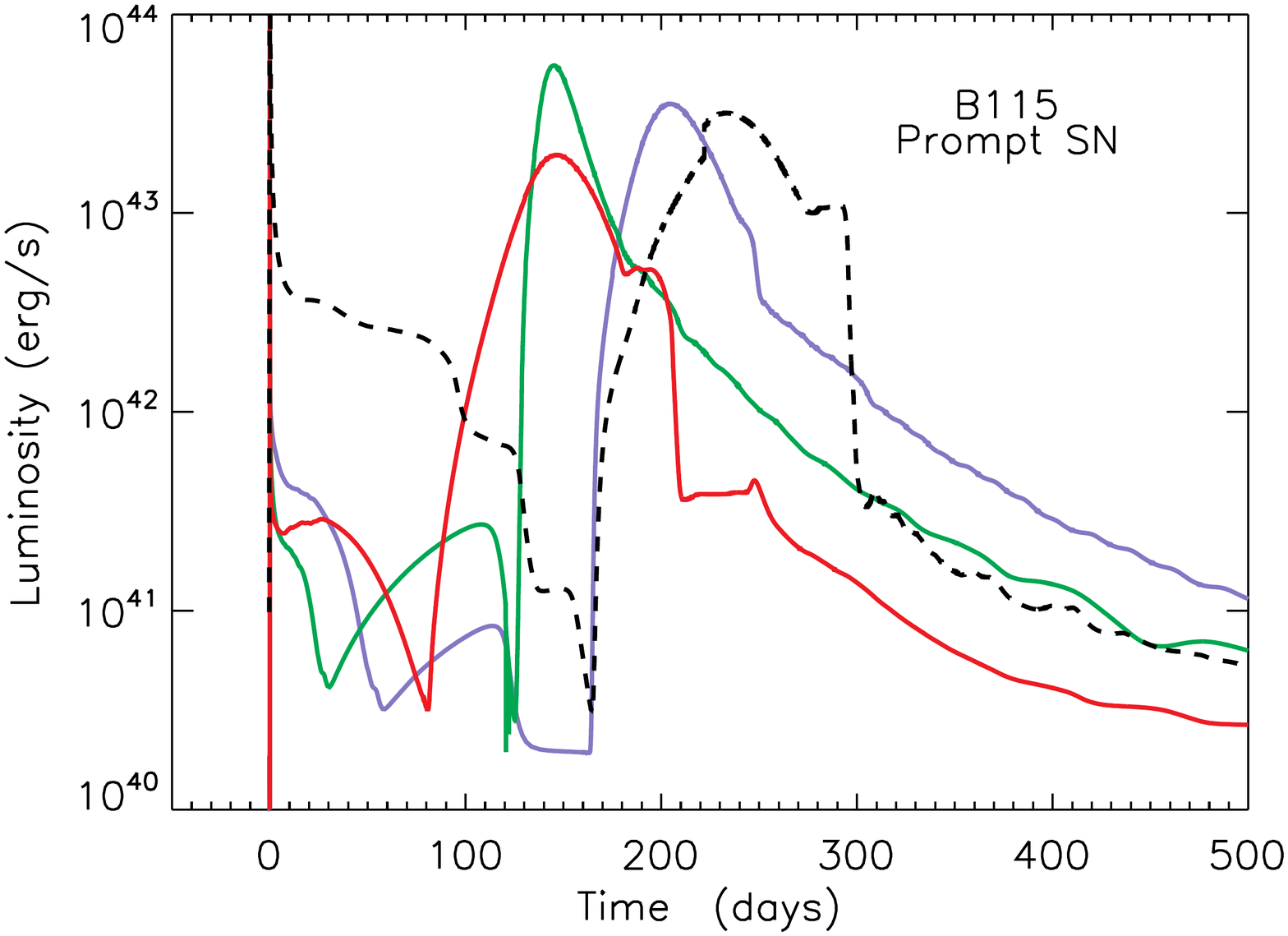}
\caption{Bolometric light curves for the first 500 days for the
  explosions of two hot, blue massive stars with zero age main
  sequence masses of 110 \Msun \ (top) and 115 \Msun \ (bottom). The
  three solid curves show results for the same helium core exploding
  inside compact hydrogenic envelope with masses of 5 \Msun \ (green),
  10 \Msun \ (blue), and 20 \Msun \ (red). Time is normalized to zero
  at shock break out. During the first few days the explosion emits
  chiefly in the ultraviolet and would be optically faint. The
  emission during the first faint peak is the recombination of the BSG
  envelope ejected in the first pulse. All subsequent emission is
  collisionally powered. For comparison, the light curves of RSG
  models T110 and T115A are shown as dashed lines. These are about 20
  times brighter on the during the early ``plateau'' stage, but
  similar afterwards. Rapid time variations in all light curves are
  artifacts of the 1D nature of the calculation and all curves would
  be much smoother in reality.  \lFig{lbvliteb}}.
\end{figure}

While technically RSGs, many of the ``T'' series models with high mass
loss rates, actually had an envelope structure that, except for a
relatively small amount of mass near the surface, was
``BSG-like''. Presupernova Model T120, for example, had a hydrogen
envelope of 11.98 \Msun and a photospheric radius of $9.5 \times
10^{13}$ cm, but only the outer 0.83 \Msun \ of that envelope was
convectively unstable and had a low density. The radius at the base of
this surface convective shell was $4.5 \times 10^{12}$ cm.  Most of
the hydrogen envelope was thus structurally like a BSG. This was the
reason why the light curve in \Fig{t120lite} declined rapidly on the
plateau after shock break out.  In other models with mass loss ``1''
in \Tab{tmodels}, the low density convective shell was also a 
small fraction of the mass of the hydrogen envelope. Those with lower
mass loss rates had more extended convective shells, but rarely more
than half the envelope was involved.  The time spent as a RSG by those
high mass loss models models was also quite short. Most reached helium
core depletion (X($^4$He) = 0.01) while still ``yellow'' supergiants
(T$_{\rm eff}$ = 5500 - 7000 K). This was particularly true for the
models over 90 \Msun. The final expansion to the red sometimes took as
little as 10,000 years.

Still, observations hint that luminous blue stars, LBVs in particular,
may be the immediate precursors of some unusual Type IIn supernovae
\cite{Gal09,Mau13,Smi07,Tad13}. Because of the smaller initial radius,
the early light curves of these stars would be distinctively
different. To explore this possibility, a set of blue models, the
``B-series'' (\Tab{bmodels}), was artificially constructed from a
subset of the T-series. The stars chosen had main sequence masses of
80 to 120 \Msun.  It was assumed that mass loss, perhaps episodic by
processes not considered here, had removed most of envelope and left
behind a helium-rich composition.  All of the matter outside of the
desired fiducial envelope was removed from a RSG with greater mass at
the time of helium depletion (X$_{\rm cen}{^4}$He = 0.01). Reinflation
was inhibited by applying a surface boundary pressure of 10$^6$ -
10$^7$ dyne cm$^{-2}$. Most of the resulting models had, by design,
hydrogen envelopes of approximately 10 \Msun. In two cases, B110 and
B115, the effect of varying the envelope mass to $\sim$5 \Msun \ or
$\sim$20 \Msun \ was examined.  Prescribing the envelope mass
precisely proved difficult because of a small amount of dredge up that
occurred after helium depletion. The resulting star was allowed to
relax, both hydrodynamically and thermally, to its new
structure. Well before the new star ignited carbon burning, the energy
generated by helium core and hydrogen shell burning was once again in
steady state with the surface luminosity and the star was in tight
hydrostatic equilibrium. Typical photospheric radii at carbon ignition
were 2 - $7 \times 10^{12}$ cm and luminosities were $6 \times
10^{39}$ erg s$^{-1}$ to $1.1 \times 10^{40}$ erg s$^{-1}$
\Tab{bmodels}. Effective temperatures were thus 25,000 - 35,000
K. These properties overlap with common definitions for both BSGs and
LBVs. The composition in the envelope was helium enriched with only
about 20 - 30\% of hydrogen remaining by mass fraction.

The stars so generated were then allowed to evolve through their PPI
and core collapse, as before, and their light curves,
calculated. Significant differences were noted. Not only was the
initial light curve fainter because of the small initial radius, but
pulsational mass loss was weaker in the lighter stars because less
envelope mass was situated at a large radius where the binding energy
was low. Below about 80 \Msun, no matter was ejected. Even at 80
\Msun, only a small fraction of the outer envelope, 1.4 \Msun \ out of
10 \Msun, was ejected in a weak explosion with a faint light curve
(\Fig{lbvlitea}). Typical speeds were around 500 km s$^{-1}$.  After
shock breakout, the temperature declined to near 6500 K where it
remained on the plateau. Thus an observer would see an unusually faint
Type IIn supernova.

\begin{deluxetable*}{cccccccc} 
\tablecaption{BLUE STAR MODELS} 
\tablehead{ Mass & M$_{\rm He}$ & M$_{\rm H}$ & L$_{\rm preSN}$ & T$_{\rm eff}$ & Duration & M$_{\rm final}$ & KE$_{eject}$ \\
(\Msun) & (\Msun)& (\Msun) & (10$^{40}$ erg s$^{-1}$) & (10$^3$ K)  & (10$^7$ s) & (\Msun) & (10$^{50}$ erg) }
\startdata
B80    &  35.1 &  10.5  & 0.62  &  23.9  &  0.0016 & 44.1 &   0.042 \\
B90    &  39.2 &  10.6  & 0.71  &  25.0  &  0.102  & 36.4 &   4.1   \\  
B95    &  43.0 &  10.8  & 0.80  &  26.1  &  0.215  & 38.8 &   5.6   \\
B100   &  44.8 &  10.8  &  0.83 &  26.6  &  1.11   & 39.2 &   8.1   \\
B105   &  47.4 &  10.8  &  0.89 &  27.5  &  6.32   & 43.5 &   6.3   \\
B110-5 &  49.8 &   6.0  &  0.92 &  45.8  &  33.2   & 44.9 &   7.9   \\
B110   &  49.8 &  10.9  &  1.01 &  31.1  &  13.9   & 45.1 &   8.7   \\
B110-20&  49.9 &  21.0  &  1.06 &  26.3  &  16.4   & 45.5 &   7.6   \\
B115-5 &  51.8 &   4.9  &  0.97 &  48.9  &  17.7   & 45.9 &   7.3   \\
B115   &  51.9 &  11.0  &  1.01 &  29.4  &  14.6   & 45.3 &  11.5   \\
B115-20&  52.0 &  19.3  &  1.11 &  31.8  &  45.4   & 46.3 &   8.6   \\
B120   &  55.1 &  10.8  &  1.06 &  30.7  &  199    & 50.9 &  11.8   \\
\enddata
\lTab{bmodels}
\end{deluxetable*}

By 90 \Msun, the situation had changed appreciably
(\Fig{lbvlitea}). The first few pulses ejected most of the hydrogen
envelope with a kinetic energy near 10$^{50}$ erg producing a brief,
faint outburst. After the ejected envelope expanded to $5 \times
10^{13}$ cm (2 days), subsequent pulses launched strong shock into a
more extended stellar structure. The new shocks took about a day to
traverse the expanded star before breaking out and initiating a much
bright transient that lasted about 80 days. Typical velocities on the
plateau were 1500 - 2500 km s$^{-1}$ and the temperature, again near
7,000 K.  Keeping in mind that the LBV may have had considerable low
velocity mass loss just prior to the supernova that could contribute
to both the spectrum and a ``tail'' on the light curve, this explosion
might resemble a ``normal'' Type IIn, but with some ``precursor
activity'' lasting a few days. Except for this faint precursor, Model
B90 is similar to Model T90 (\Fig{t90lite}), for an appropriately
small envelope mass ($\sim$10 \Msun).

The same trend continued for the 95 and 100 \Msun \ models. The
initial transient was faint because of the small presupernova radius,
but after a month or so of expansion, the light curves were similar
e.g., for B100 (\Fig{lbvlitea}) and T100 (\Fig{t100lite}). Typical
velocities were 2000 - 4000 km s$^{-1}$. The spike in emission in Model
B100 at $\sim300$ d was not the result of a new pulse. In fact, the
iron core already collapsed in this model at day 130. The spike was
instead due to matter ejected by the last pulse colliding with a
dense, thin shell at $5 \times 10^{15}$ cm that came from the
snowplowing of earlier pulses into the original envelope after it was
ejected.  This shell would have been broadened by mixing in a more
realistic 2D or 3D study and not so prominent. A similar caveat
applies to the spike at day 390 for Model B105.

The early light curves of the 110 \Msun \ and 115 \Msun \ models are
particularly interesting (\Fig{lbvliteb}). For this mass range, the
interval between the first strong pulse that ejects most of the
hydrogen envelope and subsequent pulses that slam into it is of order
months, i.e., the duration of a typical Type IIp supernova. The light
curve thus exhibits a characteristic ``double peak'' structure. Given
the large bolometric correction near shock break out, the first optical
peak would be a faint plateau resulting from the expansion and recombination
of the BSG envelope. During this time, the supernova, though less
energetic, would resemble SN 1987A before radioactivity became
important. The second and subsequent pulses result in a dramatic
brightening due to collision with the previously ejected envelope,
which now has a large radius. For the models examined, this
brightening amounted to a factor of about 30 to 100. The duration of
the first peak was shorter for smaller hydrogen envelope masses and
lower mass helium cores. Velocities in the ejected envelope were 500 -
2000 km s$^{-1}$ for the higher mass envelopes, but extended beyond
4000 km s$^{-1}$ for the lower mass ones.

RSG T-models with the same mass helium cores (the dashed lines in
\Fig{lbvliteb}) showed a similar, but less dramatic brightening. The
hydrogen envelopes for Models T110 (13 \Msun) and T115A (20 \Msun) had
similar masses to the B-models and the core structures were
identical. The different light curves thus reflected chiefly the different
radii for the presupernova stars and, to a lesser extent, the
different binding energies of those envelopes.

These light curves, with an initial faint ``plateau'' dramatically
brightening on a time scale of weeks, resemble what has been reported
for a number of Type IIn supernovae. SN 1961v \citep{Smi11c,Koc11}, an
enigmatic event, that may have resulted from the explosion of a low
metallicity star over 80 \Msun, showed a similar light curve
morphology and peak brightness. So did the 2012 outburst of SN 2009ip
\citep{Fra15}. So did SN 2010mc \citep{Ofe13}.  At least two of these
events are thought to have come from LBV stars.

Interestingly though, and perhaps difficult to hide, the surviving
stars in the B110 and B115 models all experienced a second major
outburst 1.2 to 14 years after the initial display shown in
\Fig{lbvliteb}. The lighter models had shorter delay times. Some of
these secondary explosions \Fig{lbvlitec} also show a characteristic
``double hump'' structure with a substantial brightening after several
weeks. It might be easier to miss a previous supernova than a
subsequent one.

Three factors might act to mitigate the high predicted luminosities of
these second events. First, the collisions with the longest delay
times happened in a medium that was becoming thin to electron
scattering. Perhaps a significant fraction of the radiation would be
in non-optical wavelengths. Second, the peak brightness of the display
is possibly exaggerated by the pile up of matter from previous mass
ejections in a thin shell in the 1D study. Finally, the emission from
interaction with any circumstellar wind emitted prior to the onset of
the PPI is ignored here and might help obscure faint outbursts. Still,
the total amount of energy radiated should be close to correct and is
large ($5 \times 10^{49}$ erg for the brighter B115 models, $1.3
\times 10^{50}$ for the most luminous B110 model), suggesting that the
location of events like these should be revisited for at least several
years after the initial outburst. \citet{Chu04} have inferred
  the ejection of a massive circumstellar shell 1.5 years before the
  explosion of Type IIn SN 1994W. The unshocked gas is inferred to
  have a velocity $\sim1000$ km s$^{-1}$ and shocked gas, $\sim4000$
  km s$^{-1}$ \citep{Kie12}, consistent with the models here. Similar
  conditions could also be created, however, by the ejection of the
  envelope of a $\sim10$ \Msun \ star due to a silicon flash
  \citep{Smi13,Des16,Woo15b}.

% fig 22 - 2nd pulse 110 and 115
\begin{figure}
\includegraphics[width=0.48\textwidth]{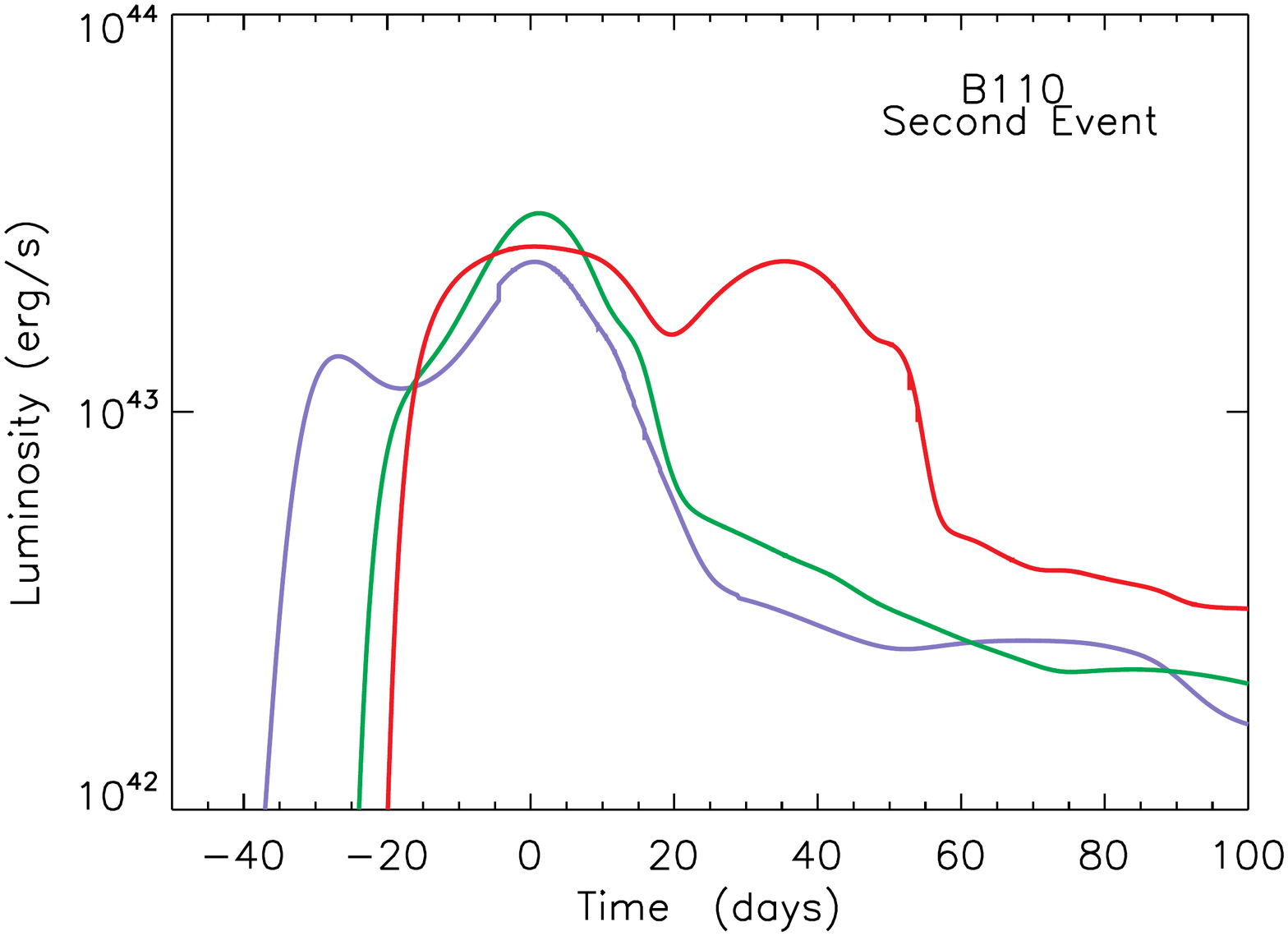}
\includegraphics[width=0.48\textwidth]{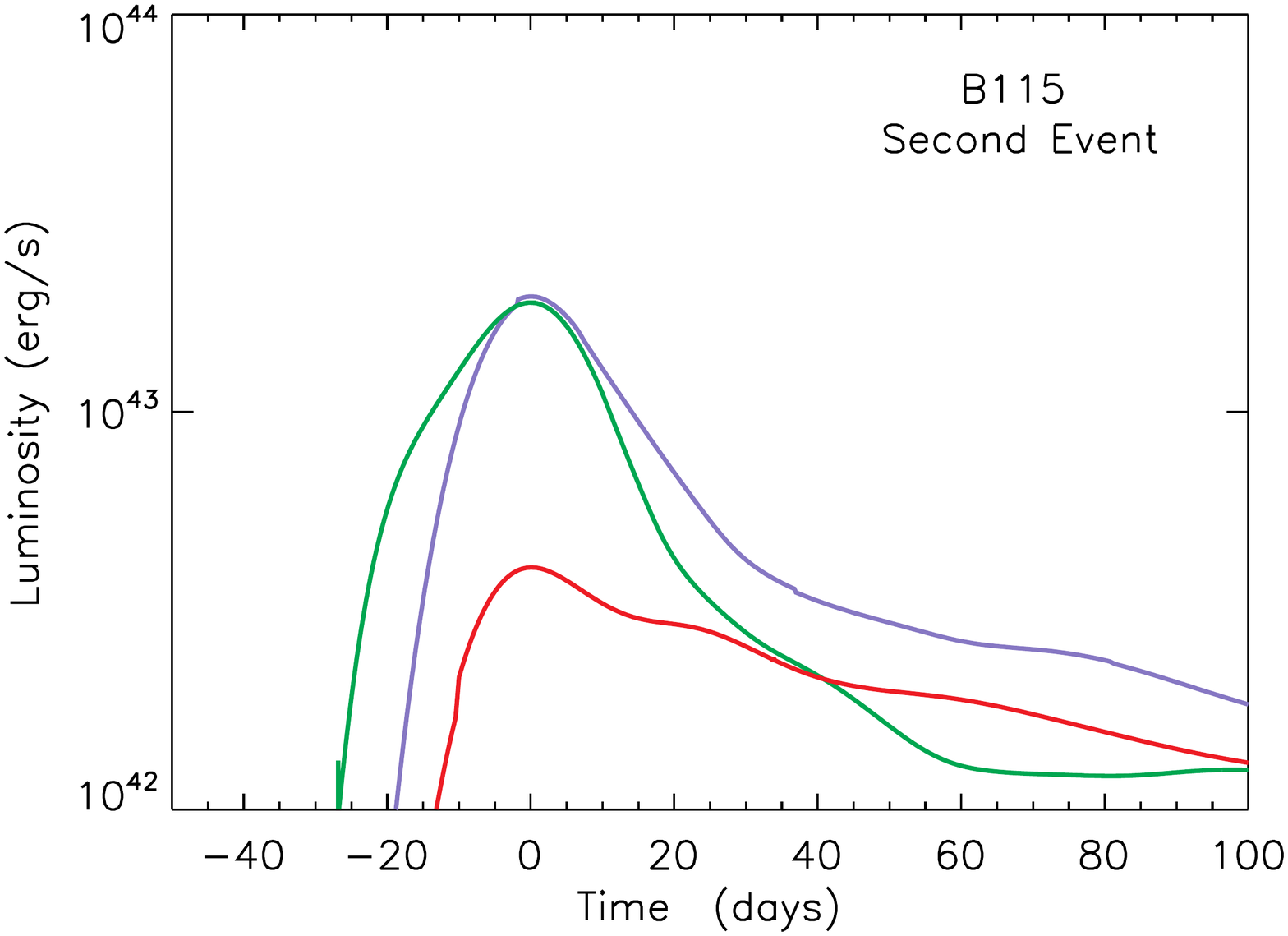}
\caption{Bolometric light curves for the second (and final) major
  peaks arising from the 110 \Msun \ and 115 \Msun \ progenitors (see
  also \Fig{lbvliteb}). The green curves are for the models with the 5
  \Msun \ envelope; the blue curves, for 10 \Msun; and the red curves,
  for 20 \Msun. Time is approximately zero at peak. For the 110 \Msun
  \ models, these peaks correspond to a time 2.4, 2.3 and 1.2 years
  after the first supernova for models with envelopes of approximately
  5, 10 and 20 \Msun \ respectively. For the 115 \Msun \ models the
  corresponding times are 5.7, 4.7, and 14.3 years after the first
  outburst. The interaction radii vary from about 10$^{15}$ cm to a
  few times 10$^{16}$ cm, and the optical depth of the shock for the
  120 \Msun \ models is small and the fraction of the emission that is
  in optical wavelengths is uncertain.  \lFig{lbvlitec}}.
\end{figure}

By 120 \Msun \ the rebrightening from the second pulse occurs well
after the initial plateau is already over.  Collectively, the blue
models are similar to the red ones, but with a fainter initial
display. The hydrodynamics of interacting with a compact envelope is
somewhat different from interacting with an extended one though, e.g.,
a stronger reverse shock in the red star \citep{Her94}, and this complicates
a direct comparison of the late time evolution of models with the
same mass. For example, the second explosion of Model B120
makes a light curve very much like Model T120 in
\Fig{t120lite}, but the interval between the first and second event is
20 years in Model B120 and 18 months in T120.

\section{Stars With Rotation}
\lSect{rotate}

Rotation induces chemical mixing that increases the helium core mass
for a given main sequence mass, reducing the threshold for the PPI
\citep{Cha12}. If sufficiently rapid, rotation can also dramatically
affect the outcome of iron core collapse, perhaps driving a final
explosion. In order to examine the effects of rotation, a smaller grid
of rotating stars was calculated. The nuclear and stellar physics was
the same, save for the addition of rotational mixing and the transport
of angular momentum as a tracer quantity. A centrifugal term was not
included in the force equation, but the ratio of centrifugal force to
gravity was small at all times. The same metallicity (10\% solar) was
employed and magnetic torques were included in all models
\citep{Heg05}.

% all models in table used a mass loss rate of 0.5
\begin{deluxetable*}{ccccccccccccc} 
\tablecaption{10\% \Zsun \ MODELS WITH ROTATION} 
\tablehead{ Model & $\dot M$ & v$_{\rm rot}$ & M$_{\rm preSN}$ & M$_{\rm He}$ & M$_{\rm Si}$ & M$_{\rm Fe}$ & J$_{\rm He}$ & J$_{\rm rem}$ & J$_{\rm Fe}$ & Duration & M$_{\rm rem}$ & Kin. Energy \\
(\Msun) & (mult.) & (km s$^{-1}$) & (\Msun) & (\Msun) & (\Msun)  & (\Msun) & (10$^{50}$ erg s) & (10$^{50}$ erg s) &(10$^{48}$ erg s) & (10$^7$ sec) & (\Msun) & (10$^{50}$ erg) }
\startdata
%use r80mb1 lower rot
R60A & 0.5 & 160 & 46.58 & 30.85 & 8.18 & 2.64 & 6.7 & 29  & 5.2 & 0.0047 & 46.6 & -   \\    
R70A & 0.5 & 175 & 54.41 & 41.68 & 6.35 & 2.92 & 11  & 8.8 & 6.4 &  0.52  & 37.0 & 8.6 \\
R80A & 0.5 & 175 & 62.20 & 47.78 & 4.06 & 2.00 & 15  & 12  & 3.3 &  26    & 43.6 & 8.6 \\
R80Ar& 0.5 & 195 & 62.47 & 55.96 & 4.89 & 2.74 & 24  & 14  & 7.2 & 7600   & 47.8
 & 22  \\
R90A & 0.5 & 180 & 68.84 & 56.04 & 5.21 & 1.83 & 21  & 13  & 3.0 & 7400   & 48.1 & 24  \\
R100A& 0.5 & 185 & 75.32 & 62.37 & 4.67 & 2.40 & 28  & 8.1 & 5.4 & 17000  & 44.8 & 38  \\
R110A& 0.5 & 180 & 80.91 & 65.68 &  -   &  -   & 26  & -   &  -  &   -    &  0   & 62  \\
C60A & 0.5 & 260 & 26.30 &   -   & 5.54 & 2.09 &  -  & 6.5 & 3.9 &   -    & 26.3 &  -  \\
C60B & 0.25& 270 & 35.40 &   -   & 8.30 & 2.49 &  -  & 37  & 15  & 0.0047 & 35.3 & 0.0086 \\
C60C & 0.1 & 275 & 46.45 &   -   & 7.42 & 2.35 &  -  & 105 & 37  & 0.76   & 41.2 & 4.9 \\
C70A & 0.5 & 250 & 28.35 &   -   & 6.21 & 2.22 &  -  & 6.6 & 3.9 &   -    & 28.4 &  -  \\
C70B & 0.25& 260 & 40.72 &   -   & 8.80 & 2.88 &  -  & 24  & 14  & 0.061  & 38.1 & 1.7 \\
C70C & 0.1 & 260 & 53.24 &   -   & 6.02 & 2.31 &  -  & 16  & 7.5 & 8900   & 41.7 & 8.8 \\
C80A & 0.5 & 240 & 30.46 &   -   & 7.00 & 2.35 &  -  & 6.9 & 4.1 &   -    & 30.5 &  -  \\
C80B & 0.25& 250 & 44.88 &   -   & 7.35 & 2.67 &  -  & 21  & 11  & 0.39   & 40.4 & 4.0 \\
C80C & 0.1 & 250 & 59.69 &   -   & 5.53 & 2.27 &  -  & 17  & 4.6 & 12900  & 46.3 & 14  \\
C90A & 0.5 & 235 & 31.43 &   -   & 7.27 & 2.39 &  -  & 7.1 & 4.1 & 0.0013 & 31.4 & 0.0072 \\
C90B & 0.25& 245 & 49.39 &   -   & 4.12 & 2.60 &  -  &17.5 & 8.1 &  41    & 43.4 & 4.0 \\
C90C & 0.1 & 250 & 65.81 &   -   &  -   &  -   &  -  & -   &  -  &   -    &  0   & 76  \\
\enddata
\lTab{rmodels}
\end{deluxetable*}

Two sets of models were calculated to illustrate the effect of
rotation in ``ordinary'' slowly rotating stars (the ``R-series'') that
make red giants, and in more rapidly rotating stars that experience
CHE (the ``C-series'') and remain compact throughout their
evolution. For stars of 60, 70, 80, 90, 100, and 110 \Msun, the
R-series had initial angular momenta on the main sequence of 1.1, 1.6,
2.0, 2.5, 3.0, and $3.4 \times 10^{53}$ erg s respectively. One model,
R80Ar, rotated a bit faster, $J_{\rm init} = 2.2 \times 10^{53}$, and
bordered on CHE. The surface rotational speeds on the main sequence,
when the central hydrogen mass fraction had declined to 0.4, are given
in \Tab{rmodels} and cluster around 180 km s$^{-1}$. Increasing these
speeds by about 50\% led to CHE. The C-series models had initial
masses 50, 60, 70, 80, and 90 \Msun \ and angular momenta of 1.5, 1.9,
2.3, 2.7 and $3.2 \times 10^{53}$ erg s, respectively. The rotational
mixing parameters employed were those of \citet{Heg00}, that is $f_c$
= 0.0333 and $f_{\mu}$ = 0.05. If the more recent calibration of
\citet{Bro11} is employed, $f_c$ = 0.0228 and $f_{\mu}$ = 0.1, which
implies more inhibition to mixing. Qualitatively similar results are
then obtained for the CHE models for an equatorial speed about 20\%
larger than in \Tab{rmodels}, i.e., 300 - 320 km s$^{-1}$.

For the CHE models, all hydrogen was burned or lost before any
pulsations began, and a different mass loss prescription was
required. For surface mass fractions of hydrogen in excess of 0.4, the
mass loss rate of \citet{Nie90} continued to be used (\Sect{mdot}),
appropriately scaled for metallicity. The mass lost during this stage
was relatively unimportant, however. For surface hydrogen mass
fractions less than 0.4, the treatment was the same as \citet{Woo06},
including a metallicity scaling of Z$^{0.66}$.  The mass loss rate
was
\begin{equation}
\begin{split}
{\rm log}_{10}\, \dot M \ &= \ -12.43 +1.5 \, {\rm log}_{10}\, \left(\frac{L}{10^6
  \Lsun}\right) \\ & - 2.85 X_H +0.66 \, {\rm log}_{10}\left(\frac{Z}{\Zsun}\right).
\end{split}
\end{equation}
This implies a loss rate at 10$^6$ \Lsun, X$_{\rm H}$ = 0.15,
and solar metallicity, of $1.4 \times 10^{-4}$ \Msun \ y$^{-1}$. This
is large compared with modern estimates \citep[see e.g., Fig. 1
  of][]{Yoo05}, so this rate was multiplied by factors of 0.5 (Models
CxxB), 0.25 (Models CxxC), and 0.1 (Models CxxD), with 0.5 and 0.25
perhaps being most appropriate, but 0.1 within the realm of
possibility.

\Tab{rmodels} gives the major results and shows a strong dependence of
the supernova progenitor mass of the CHE models on the uncertain mass
loss rate. No helium core mass is given since the presupernova mass
was less than the maximum helium core mass. All C-series stars ended
their lives as Wolf-Rayet stars with surfaces containing mostly
carbon, and oxygen with some helium, but devoid of hydrogen. Their
explosions would produce supernovae of Type I. Typical presupernova
radii were $\sim$50 \Rsun \ and effective temperatures, $\sim$10$^5$
K.  In contrast, the R-series models ended their lives as RSGs with
similar luminosities (within a factor of two of 10$^{40}$ erg
s$^{-1}$), but with extended atmospheres that still contained
appreciable hydrogen and had radii near $1.5 \times 10^{14}$ cm.

Also given in the table are $J_{\rm He}$, $J_{\rm rem}$, and $J_{\rm
  Fe}$, the angular momentum of the helium core at carbon depletion,
the angular momentum of the final star, after any pulsational activity
is over (and hence the angular momentum of the black hole formed if no
matter is ejected), and the angular momentum of the iron core at the
time it collapses. Some possible implications for the explosion and for black
hole properties are discussed in \Sect{superl} and \Sect{gwave}.

Even a moderate amount of rotation substantially decreases the initial
masses necessary to form helium and CO cores of a given mass
\citep{Cha12}, thus lowering the threshold for encountering the
PPI. Model R90A shows that a 56 \Msun \ helium core is now made in a
star with a main sequence mass of only 90\Msun \ whereas without
rotation it took 120 \Msun \ (\Tab{tmodels}). For CHE models, the
threshold for making PPISN is, in principle, even lower.  Once the PPI
is encountered, the cores of these rotating stars evolve similarly to
their non-rotating counterparts with the same CO mass.  Because of
rotationally-induced mixing the helium and nitrogen abundances in the
winds and presupernova atmosphere of these stars are increased. For
example, the helium mass fraction in the envelope of R100A is
90\%. 

The Kerr parameters of the black holes formed from the collapse of the
remaining cores can be calculated from the quantities in \Tab{rmodels}
using $a = J_{\rm rem} c/ G M_{\rm rem}^2$ and are typically of order
0.01 - 0.1, though larger values are possible for the CHE models with
small mass loss.  If no further mass ejection occurred during the
collapse, Model C60D would leave a 41 \Msun \ black hole with a =
0.7. These large values of the Kerr parameter might be an observable
signature of CHE.

% mass loss rates  c60
% Xh = 0.12  L = 5e39  Mdot full = 5.3e-5
% Xh = 0.0   L = 5e39              1.2e-4
%                  c90
% Xh = 0.11  L = 8e39             1.3e-4
% Xh = 0         7e39             2.1e-4     note that these are for Z = 10% solar
%                                            and would be 4.6 tims greater for solar
% Mdot = 0.33*10.0**(-11.95+1.5*xl-2.85*xh)*(Z**.66)
% This is certainly a factor of 2 too high see Figure 1 Yoon and Langer
% Could easily be 4 x too big and possibly 10 times
% E.g., Z = solar, L = 10**6, Xh = 0.15  Mdot = 1.38e-4  log = -3.86

\section{Superluminous Supernovae}
\lSect{superl}

While some of the models in \Sect{tmods} were unusually bright
(\Fig{t120lite} and \Fig{t130lite}), none of them emitted as much
light as the brightest SLSN - events like SN 2003ma \citep{Res11}, SN
2006gy {Smi10}, SN 2005ap \citep{Qui11}, and SN 2008es
\citep{Mil09}. If stars in this mass range are to explain such events,
it seems likely that something beyond purely thermonuclear explosions
- PISN and PPISN - is necessary.  The natural time for any additional
energy input is when the iron core collapses, or shortly thereafter,
but the large iron core masses and binding energies outside those
cores, $\sim 5 \times 10^{51}$ erg, preclude neutrinos acting alone
from powering an explosion
\citep[\Fig{fecore}][]{Wil86,Fry99,Fry01,Ugl12,Pej15}. This leaves
rotation as the likely alternative.

The rotation could be so extreme as to form a disk around the black
hole, a collapsar \citep{Woo93}, though none of the models in
\Tab{rmodels} rotated that fast, or a ``millisecond magnetar''
\citep{Uso92,Met11}. In either case, different conditions from those
required to make a GRB might produce a more isotropic, but still
very energetic supernova. In the case where a neutron star remained,
the magnetar could power a prompt explosion with an energy as great as
$2 \times 10^{52}$ erg \citep{Maz14}, and might even contribute later
to the light curve itself \citep{Kas10,Woo10,Cha16,Suk16b}. But would
the magnetar survive the explosion? \citet{Mun06} reported the
discovery of an x-ray pulsar with magnetar-like properties in a region
where only stars with main sequence masses greater than about 40 \Msun
\ had died. If one includes the mass loss appropriate to solar
metallicity stars, however, the mass at death of such initially
massive stars was probably substantially less than 12 \Msun
\ \citep{Eks12}. This is far lighter than the 32 \Msun \ threshold for
helium cores that encounter the PPI, suggesting that the vast majority
of magnetars are probably made in lighter, more abundant stars.

% fig 23 - fe core 115
\begin{figure}
\begin{center}
\leavevmode
\includegraphics[width=0.49\textwidth]{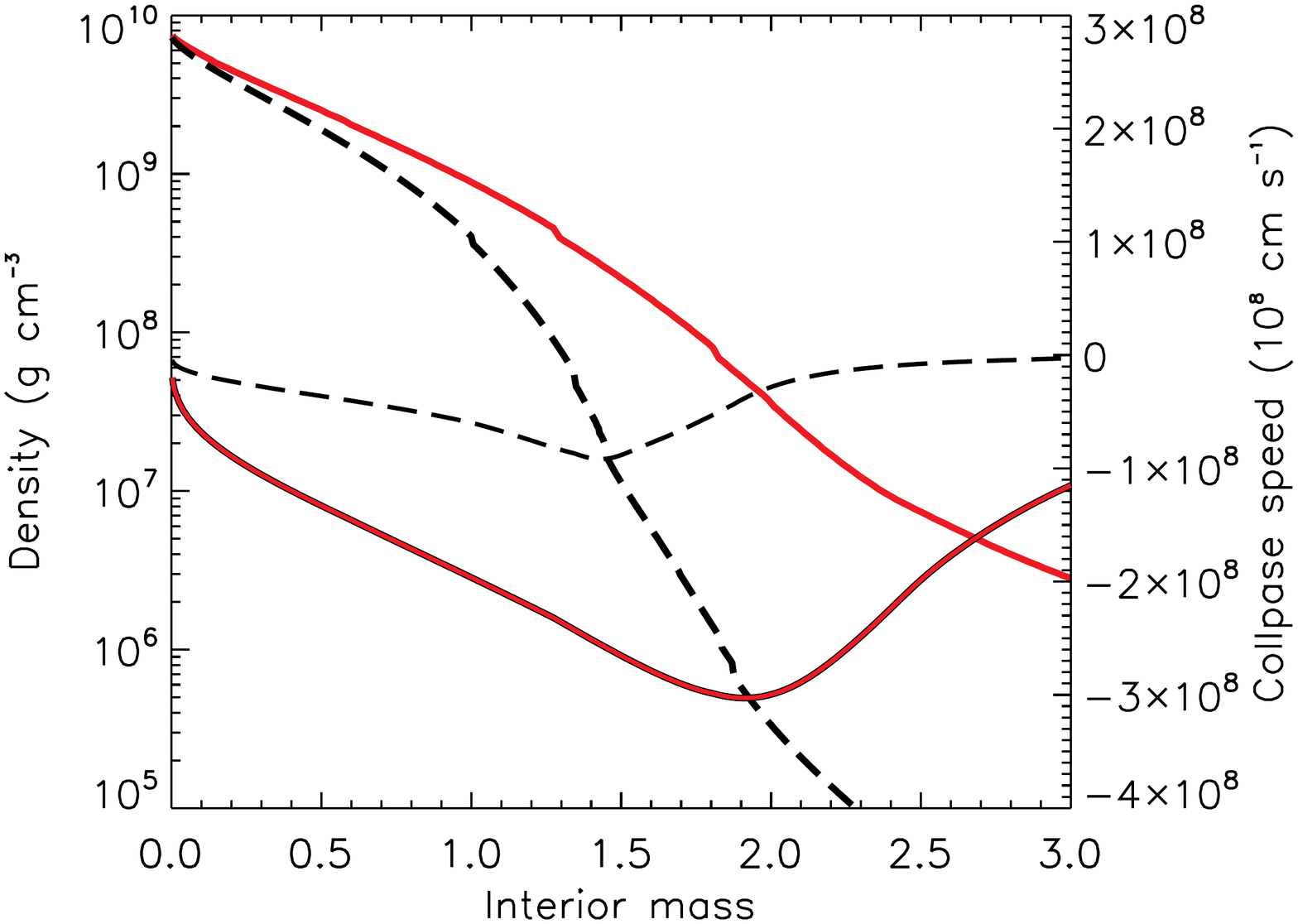}
\caption{Density and velocity at the time of final collapse for Model
  T115A (solid red lines) and a typical 15 \Msun \ presupernova star
  (dashed black lines) \citep{Woo07a}. Both are evaluated at a central
  density of $7.5 \times 10^9$ g cm$^{-3}$ at which time the maximum
  collapse speed is 1000 km s$^{-1}$ core the 15 \Msun \ model but $3
  \times 10^8$ cm s$^{-1}$ for Model T115A.  The iron core masses are
  1.48 \Msun\ and 2.38 \Msun \ for the 15 and 115 \Msun \ models
  respectively, but more important is the lack of any appreciable
  density decline at the edge of the iron core in the 115 \Msun
  \ model. The radius enclosing 2.5 \Msun \ is $1.4 \times 10^{9}$ cm
  for the 15 \Msun \ model and $3.3 \times 10^8$ cm for the 115 \Msun
  \ model implying a compactness parameter of 0.76 \citep{Oco11} for
  the latter. The core of Model T115A, which is typical of the stars
  in this study, will be very difficult to explode using neutrinos
  alone.  The net binding energy external to the iron core in Model
  T115A is $4.1 \times 10^{51}$ erg. \lFig{fecore}}
\end{center}
\end{figure}

On the positive side, even with magnetic torques and mass loss
included, the short lifetimes of the stars considered here result in
considerable angular momentum remaining trapped in the presupernova
core. Typical angular momenta for the iron cores inside the giant star
models are $3$ to $7 \times 10^{48}$ erg s (\Tab{rmodels}), and some
of the CHE models rotate even faster.  For a variety of equations of
state, \citet{Lat07} suggest a moment of inertia, I, for a neutron
star of I/M$^{3/2}$ = 35 - 45 km$^2$ \Msun$^{-1/2}$, where M is the
gravitational mass. For a fiducial gravitational mass of 2.0 \Msun,
which may be near the maximum in nature, this implies a moment of
inertia near $2 \times 10^{45}$ cm$^2$ gm. This, in turn, implies an
angular velocity for the cold pulsar of 1500 to 3000 rad s$^{-1}$, or
a period of 2 to 3 ms and a rotational energy $2 - 4 \times 10^{51}$
erg. Still more energy is available from some of the CHE
models. Models C60C and C60D would produce sub-millisecond pulsars
(magnetars?), hence substantial deformation and gravitationa;
radiation would be expected in their collapse.

This energy could be difficult to extract, however, since the final
rotational energy of the neutron star is only available once its
binding energy has been radiated as neutrinos. This takes of order
seconds which, given the expected high accretion rate from the dense
silicon shell, may not be available before an event horizon forms.
The black hole masses inferred from the recent detection of
gravitational radiation in the event GW 150914 also suggest that black
hole formation from these sorts of stars is a common event
(\Sect{gwave}).

% fig 24 - SLSN
\begin{figure*}
\begin{center}
\leavevmode
\includegraphics[width=0.45\textwidth]{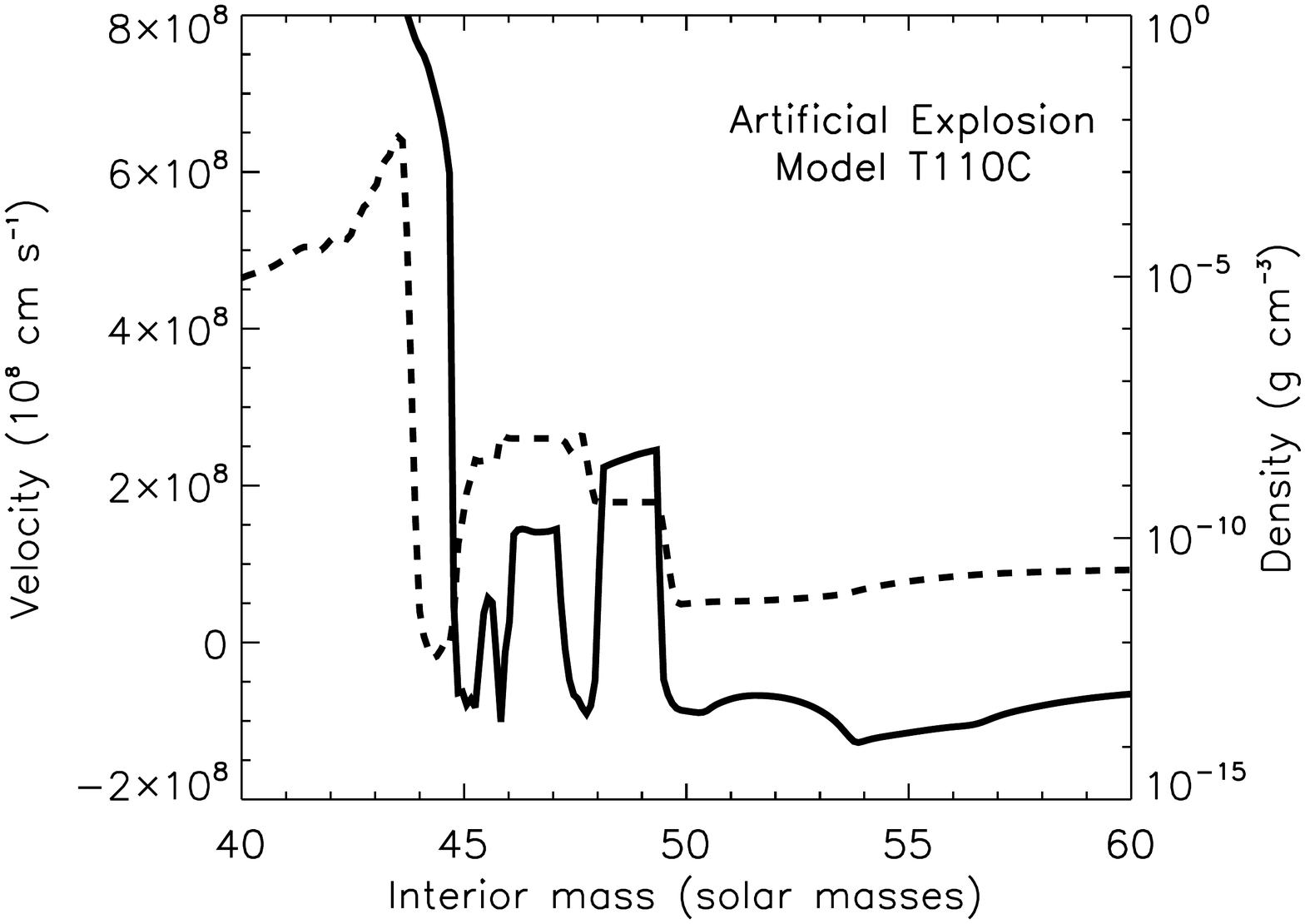}
\includegraphics[width=0.45\textwidth]{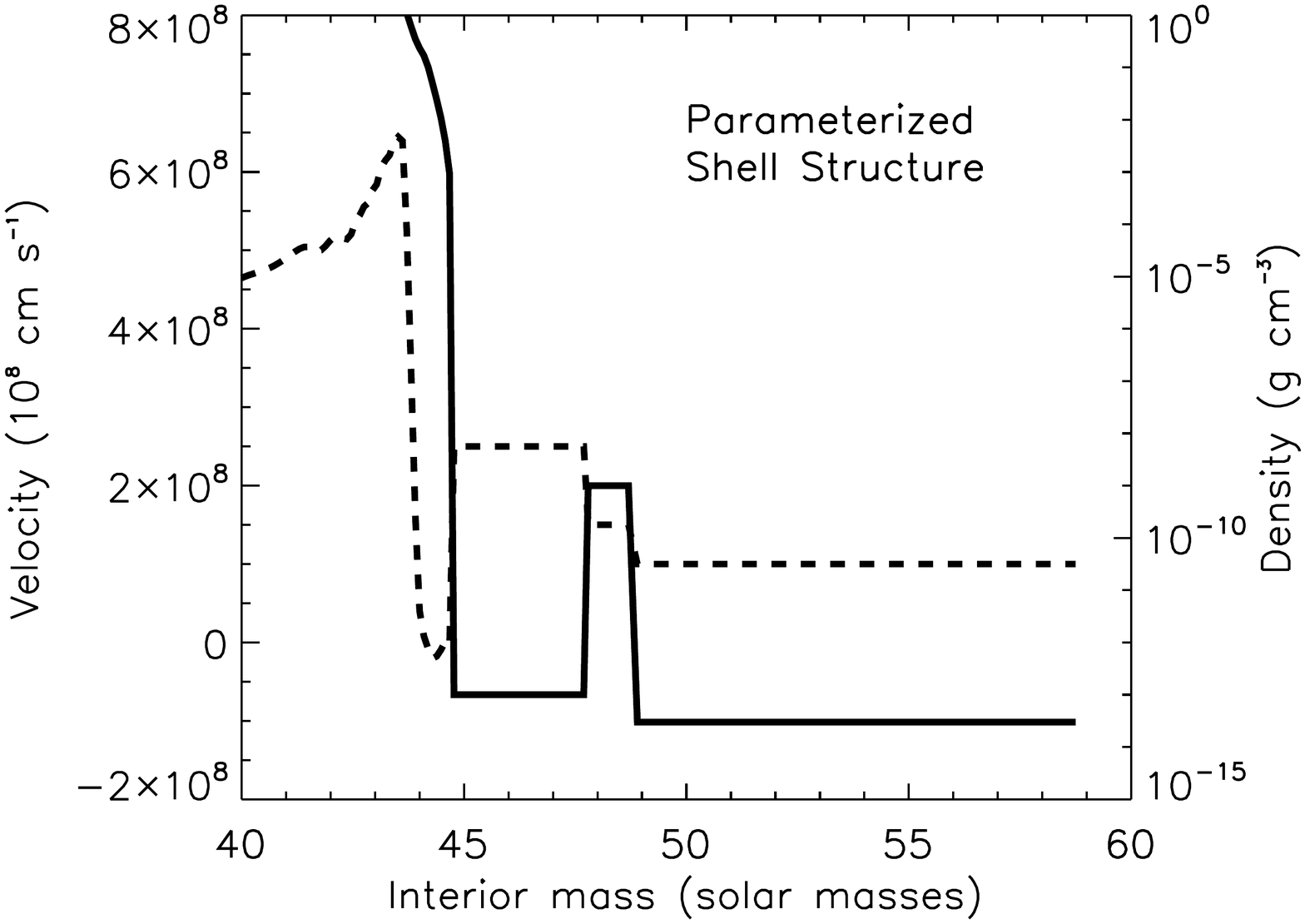}
\vskip 24pt
\includegraphics[width=0.45\textwidth]{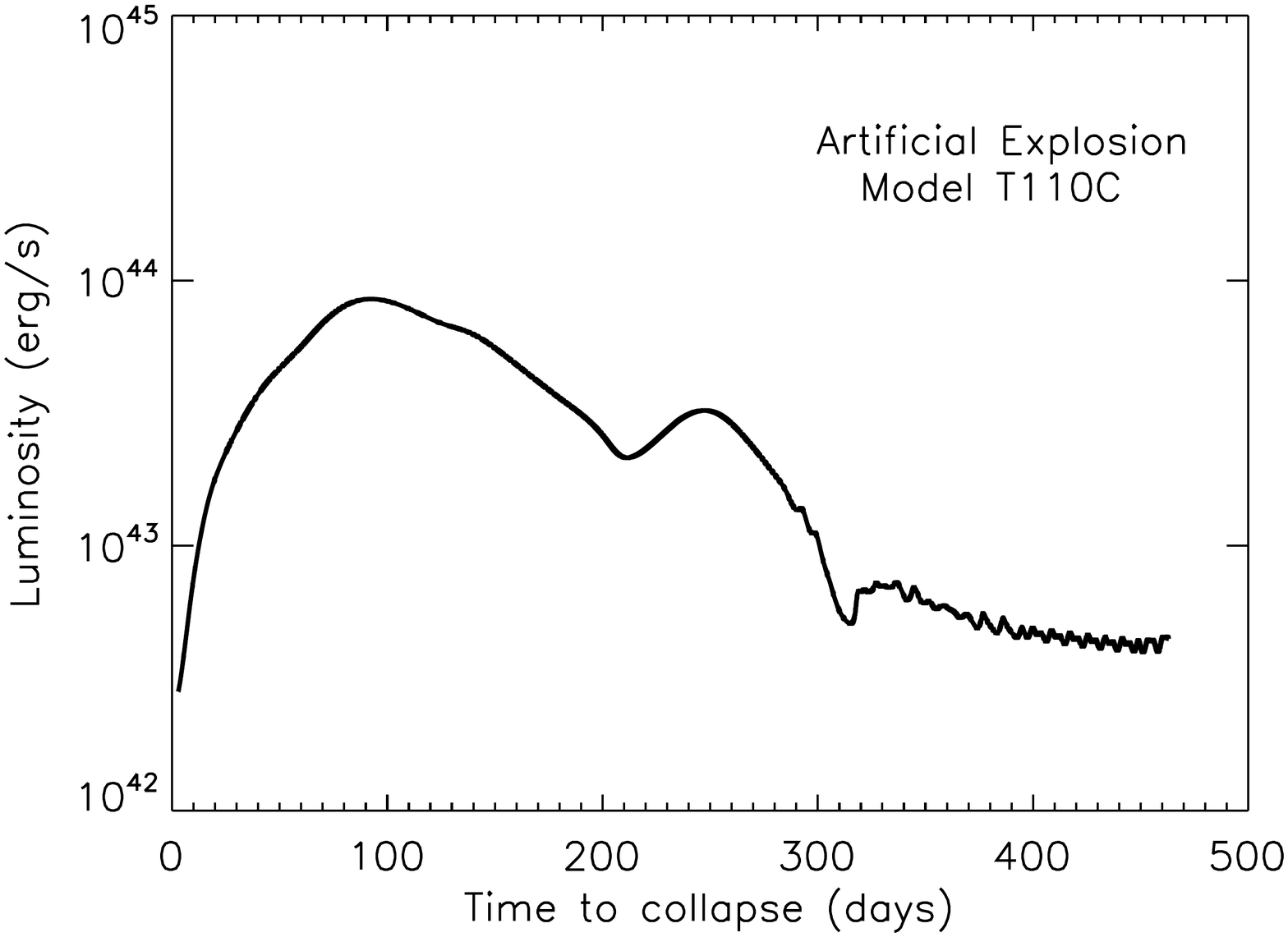}
\includegraphics[width=0.45\textwidth]{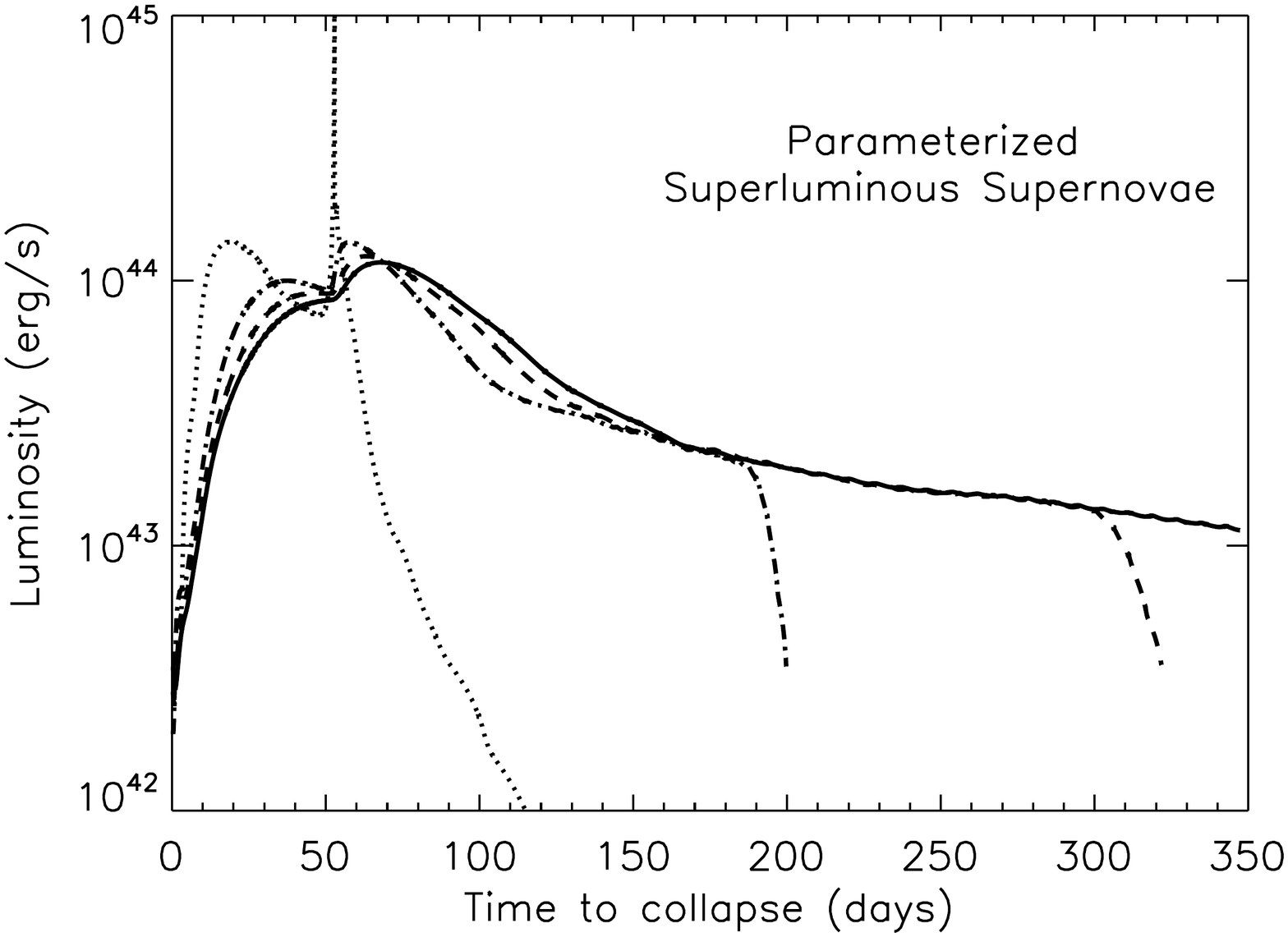}
\caption{Artificial explosions in Model T110C. (Top left:) Density
  (solid line) and velocity (dashed line) of
  Model T110C at the time the shock wave from an artificially induced
  explosion arrives at the edge of the bound remnant at 44.7 \Msun
  \ 100 s after core collapse.  Outside the core are several shells of
  matter ejected by previous pulses. The boundary of the large density
  spike at 49 \Msun \ is at $2.07 \times 10^{15}$ cm. The ejected
  matter actually extends to 96 \Msun \ though only the inner 60
  \Msun \ is shown. 
%The matter outside the core is composed of almost pure helium out to
%  47.7 \Msun, but material further out contains a small amount of
%  hydrogen, 5 - 20\% by mass.
(Bottom left:) Bolometric light curve resulting from the evolution of
  the velocity and density structure shown in the top left panel. At
  the end of the curve shown, the external shock had reached 76.7
  \Msun \ and $1.1 \times 10^{16}$ cm and had a speed of 1900 km
  s$^{-1}$. 
%The supernova would have continued to be bright for a longer time.
(Top right:) Parametrized density and velocity profiles are adopted
  outside of the core, including one major shell at 47.7 - 48.7 \Msun,
  radius, $2.4 \times 10^{15}$ cm, and speed 1500 km s$^{-1}$. The
  profile qualitatively resembles that for Model T110C shown on the
  left, but with only one major shell. The densities, speeds, and
  masses are now user adjustable parameters. (Bottom right:) The
  bolometric light curve resulting from the configuration shown in the
  upper right panel. Several light curves were calculated in which the
  mass of the ejected envelope exterior to 48.7 \Msun \ was 0, 10, 20,
  and 30 \Msun. The results are shown as the dotted, dash-dotted,
  dashed, and solid line respectively. The dip at about 50 days is
  artificial and reflects the arrival of the main shock at the inner
  edge of the dense shell at 47.7 \Msun. The energy emitted in light
  for the four models is 0.54, 0.92, 1.09, and 1.12 $\times 10^{51}$
  erg. \lFig{slsnlite}}
\end{center}
\end{figure*}

An intermediate possibility is that the star only partly explodes,
with strong bipolar outflows accompanied by appreciable fall back and
accretion in the equatorial plane. The final product would still be a
black hole, but its birth need not be quiet.  Evolution as a
proto-neutron star always precedes the formation of an event horizon
for stars that develop iron cores in hydrostatic
equilibrium. Multi-dimensional studies of MHD core collapse
\citep{LeB70,Mei76,Mue79,Aki03,Ard05,Bur07,Des08,Tak11,Mos14,Mos15}
universally show jets or strong bipolar outflows developing during the
proto-neutron star stage.  So far, these studies have been for
lighter, less tightly bound stars, and are not directly applicable
here. They suggest, however, that, even if most of the helium and
heavy element core does collapse to a black hole, a mildly-collimated
polar outflow might emerge. This outflow could have dramatic
consequences when interacting with the shells previously ejected by
the PPI. The objects considered would be intermediary between those
that make the powerful, tightly focused, relativistic jets seen in
long soft GRBs from massive stars, and the roughly spherical
explosions of ordinary supernovae.

To illustrate the possible consequences, consider Model T110C. When
its iron core collapses, its mass is 2.59 \Msun, external to
which the net binding energy is $4.6 \times 10^{51}$ erg. A bipolar
outflow focused into a solid angle of $\pi$ steradians ($\pi/2$ in
each hemisphere) would only need an energy slightly greater than
10$^{51}$ erg to eject, or to push aside the matter in its path. A
more energetic explosion at larger angles would require more rapid
rotation than calculated for the giant star models in \Tab{rmodels},
but is not ruled out.

Using a piston, an explosion was launched at the edge of the iron core
of T110C with sufficient energy to provide the still bound material
with a final kinetic energy of $2.2 \times 10^{51}$ erg at
infinity. Since the star's binding energy must also be provided, this
amounts to the central engine doing about $7 \times 10^{51}$ erg
of work, more than the total rotational energy of even a 2 ms cold
neutron star.  Even with this large assumed energy input, about half
of the core eventually reimploded, leaving a black hole mass of about
22 \Msun. For an asymmetric explosion, fallback and accretion would
probably be greater and the remnant mass larger.

The matter that was ejected interacted with the existing circumstellar
shells, producing a very luminous supernova that lasted hundreds of
days (\Fig{slsnlite}). The initial rise to peak was given by the
interaction with the more recently ejected shell of helium, but a long
``tail'' resulted from interaction with the previously ejected
hydrogen envelope with structure imposed by different shells. The
abundance of hydrogen in this envelope was low, ranging from 5\% by
mass at its base to 20\% farther out. Most of the rest was helium and
nitrogen. As the light curve developed, the effective temperature
declined from 7000 K near peak to 4000 K out on the tail.

\Fig{slsnlite} also shows the results of using a parametric
representation of the shell structure which allows more control of the
shell masses, densities, speeds and radii. Varying the
hydrogen envelope mass in this model affects the light curve duration,
but not so much the rise to peak. The slope during the decay phase and
the abrupt termination of light when the shock reaches the edge of the
ejected envelope are sensitive to the assumed density structure which
was assumed to be constant.

Pulling out all the stops, the results of forcing a much more
energetic isotropic explosion of $\sim2 \times 10^{52}$ erg in three
models are shown in \Fig{hyper}. These models were selected on the
basis of having experienced a major mass ejection roughly a year prior
to iron core collapse. A longer wait and the ejected matter would have
moved to such a large radius (well beyond 10$^{16}$ cm) that the
interaction would be too faint, though longer lasting, and perhaps not
an optical supernova. A shorter wait, and the ejected matter would
still have been very optically thick and the energy from the collision
subject to adiabatic degradation. Model He50 ejected a total of 6.3
\Msun \ during 6 pulses spanning the last 0.35 years before its iron
core collapsed (\Fig{hepulses}); Model R80Ar ejected 8.2 \Msun \ 1.0
year earlier; and Model T105C, 47.7 \Msun, 1.4 years earlier. The
kinetic energies of these ejected shells were 0.86, 0.55, and 0.70
$\times 10^{51}$ erg, respectively. Typical shell velocities were
3500, 2500, and 1200 km s$^{-1}$ for Models He50, R80Ar, and T105C
respectively. For R80Ar, this final shell ejection came after losing
its hydrogen envelope to pulses 2400 years earlier.  Models He50 and
R80Ar would thus be of Type I, while T105C would be Type II. Because
these are a small subset of all PPISN, which itself is already a rare
class (\Sect{conclude}), they would be exceedingly infrequent events,
less that 1\% of core collapses, even in metal-poor regions.

Explosions in these three stars were simulated by removing the iron
core and placing a piston at the inner boundary that imparted a large
explosion energy to the external matter. For He50, the kinetic energy
before radiative losses was $2.1 \times 10^{52}$ erg. Since the
binding of the matter external to the core was $4.6 \times 10^{51}$
erg, this implies a total energy delivered by the central engine of
about $2.6 \times 10^{52}$ erg, close to the upper bound on rotational
energies for pulsars. For Models R80Ar and T105C the kinetic energies,
before radiative losses, were $1.4 \times 10^{52}$ (plus $4.7 \times
10^{51}$ erg for the binding energy), and $2.0 \times 10^{52}$ (plus
$4.6 \times 10^{51}$ erg for the binding energy). No matter fell
back. A large amount of $^{56}$Ni was synthesised, 2.7 \Msun, 1.8
\Msun, and 2.5 \Msun \ respectively for He50, R80Ar, and T105C, but
this had little effect on the light curve. Typical expansion speeds
for the interacting matter behind the shock were 10,000 km s$^{-1}$,
declining with time due to shock interaction to about 4000 km
s$^{-1}$.

The resulting light curves approached $10^{45}$ erg s$^{-1}$ at
maximum in all three cases, but the duration and hence the total
emitted power was very sensitive to the speed, mass, and radius of the
shell that was being impacted (\Fig{hyper}). Integrated luminous
powers were 1.2, 2.5 and $6.6 \times 10^{51}$ erg. The latter, from
Model T105C, is probably close to the maximum that can be attained in
any PPI plus magnetar-powered explosion since the mass of the shell
was approximately equal to the mass of the exploding core and the
shell speed was low. 

When two masses, $m_1$ and m$_2$ with speeds $v_1$ and $v_2$,
experience an inelastic collision and radiate all dissipated energy,
conservation of momentum and energy implies that the fraction of the
initial kinetic energy radiated is $f (1+f)^{-1}(1-g)^2$ where $f$ is
the ratio of the two masses, $f = m_2/m_1$, and $g$ is the ratio of
the initial speeds $g = v_2/v_1$. Here $m_1$ is the that part of the
core mass that collides with the shell of mass $m_2$ during the time
the light curve is mostly generated. The most efficient production of
light then occurs when $f < 1$, i.e, the shell is more massive than
that part of the core with which it interacts, and $g = 0$, i.e., the
shell is initially stationary.  For models with envelopes comparable
to the core mass (it can hardly be much bigger) $f \approx 1$ and g is
at best, 1/10, so no more than 40\% of the kinetic energy is
radiated. For models without envelopes, the efficiency is smaller. For
typical shell masses of, at most, 5 - 10 \Msun, a core mass of 40
\Msun, and shell speed 1/3 of the edge of the exploding core, the
maximum efficiency is closer to 10\%.  These estimates of upper
bounds, $2 \times 10^{51}$ for Type I and $8 \times 10^{51}$ agree
reasonably well with the results for Models He50, R80Ar, and
T105C. They are also consistent with observtions of the brightest SLSN
- e.g., SN 2003ma, SN 2006gy, SN 2005ap, and SN 2008es.

% fig 25 -  hypernovae
\begin{figure}
\includegraphics[width=0.49\textwidth]{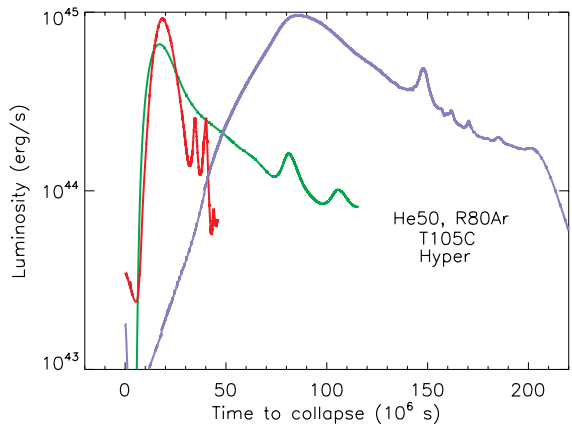}
\caption{Results of artificial ``hyper-energetic'' explosions, at the
  time of iron core collapse, of Models He50 (red), R80Ar (green), and
  T105C (blue). The three models emitted $1.2 \times 10^{51}$ erg
  (He50), $2.1 \times 10^{51}$ erg (R80Ar), and $6.6 \times 10^{51}$
  erg (T105C) of light.  Time is measured in days since the simulate
  explosion.  The post-peak variabilty is artificially exaggerated in
  this 1D calculation.  \lFig{hyper}}.
\end{figure}

Still more luminous supernovae are possible, in principle, if a
magnetar contributes directly to the light curve
\citep{Kas10,Woo10,Cha16,Suk16b}. While it seems increasingly likely
that many SLSN are indeed illuminated by magnetars \citep[][and
  references therein]{Ins13}, that possibility is not explored here
because of the additional complexity and uncertain parameters required
for a model that provides {\sl both} a prompt hyper-energetic
explosion ($\sim10^{52}$ erg) and a large amount of electromagnetic
energy at late times.

\section{Eta Carinae}
\lSect{etacar}

One of the most enigmatic of astronomical icons, Eta Carinae is also
one of the most massive stars in our galaxy \cite{Dav97}.  Depending
upon its mass-loss history, Eta Carinae seems likely to encounter, or
to have encountered, the PPI at the end of its life. Indeed, the
generally accepted mass, $\sim120$ \Msun \ \citep{Hil01}, places it
squarely in the range treated in this paper. But at what stage in its
life are we viewing the star? Will mass loss ultimately remove the
hydrogen envelope and shrink the core so much that the PPI is avoided?

Eta Carinae's last millennium has been complicated. There is evidence
for unusual mass ejections starting at least 700 years ago
\citep{Wal78,Dav97,Kim16}.  Beginning around 1837, the star underwent
a major structural change known as the ``Great Eruption'' that lasted
roughly 20 years with frequent large variations in brightness during
that period \citep{Smi11a} that are sometimes counted as separate
eruptions.  In 1843, the star's apparent magnitude briefly increased
to approximately -1, making it the second brightest extrasolar
object. A lesser eruption occurred in the 1890's. The current
luminosity of Eta Carinae is about $1.9 \times 10^{40}$ erg s$^{-1}$
\citep{Hil01}, which is being emitted mostly in the
infrared. \citet{Dek05}. \citet{Smi03} estimates a mass for the
material that was ejected in the major outburst, a portion of the
``Homunculus'', of more than 10 - 15 \Msun. \citet{Smi03} further
estimates that this matter carries a kinetic energy of 10$^{49.6}$ -
10$^{50}$ erg. \citet{Smi08} and \citet{Smi13} have made a compelling
case that the production of the Great Eruption required an explosive
event, not just a strong wind.  The duration of the major mass
ejection was less than 5 years \cite{Smi06} and the velocity and
energy in the ejecta is quite asymmetric \citep{Smi06} with $\sim$90\%
of the explosion energy concentrated at latitudes above 45
degrees. Velocities as high as 3500 - 6000 km s$^{-1}$ have been
reported, though a more typical speed is 650 km s$^{-1}$
\citep{Smi08}.

% fig 26 -  density and velocity in t125a0 at 180 years
% despite the figure name the plot is for T125B
% file t125a1#31752
\begin{figure}
\includegraphics[width=0.49\textwidth]{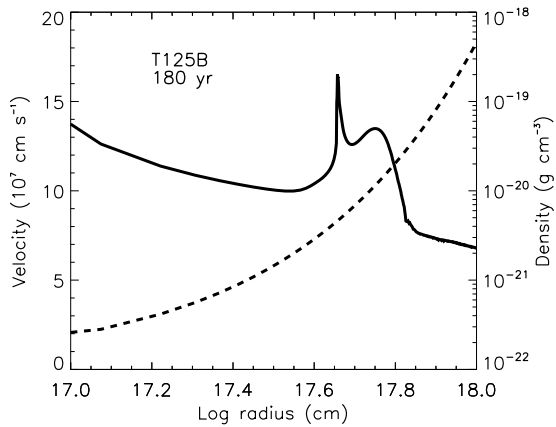}
\caption{Density (solid line) and velocity (dashed line) 180 years
  after the first pulse (i.e., at present epoch) in the ejecta of
  Model T125B. Velocity is in 100 km s$^{-1}$ and the density ranges
  from 10$^{-22}$ g cm$^{-3}$ to 10$^{-18}$ g cm$^{-3}$. Roughly 13
  \Msun \ of the 34 \Msun \ ejected is currently in a thin shell
  between 4 and 6 $\times 10^{17}$ cm. The dense concentration is a
  result of the reverse shock operating as the forward shock plowed
  through an envelope of nearly constant density during the
  explosion. \lFig{t125a0un}}
\end{figure}

% fig 27 -  light curve t125a0 and t125a1 second pulse
\begin{figure}
\includegraphics[width=0.49\textwidth]{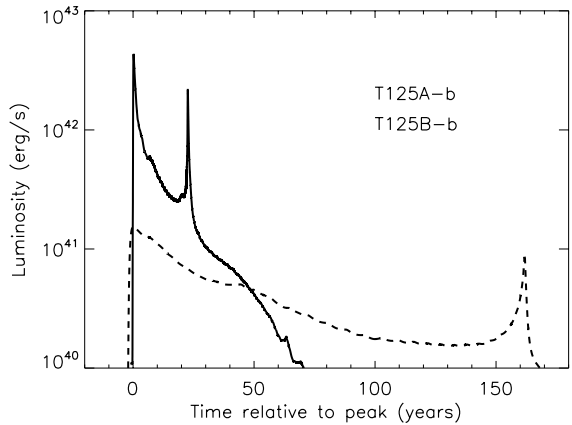}
\caption{Bolometric light curve resulting from the collision of the
  second mass ejection in Models T125A (solid line) and T125B (dashed
  line) with the first. Relative to zero here, the first pulse
  happened 70 years earlier (T125A) and 470 years earlier (T125B).  If
  these models were taken to represent Eta Carina, the solid line
  would be the expected light curve for the second eruption in the
  1890's and the dashed line would still lie two centuries in the
  future. The sharp spikes in luminosity are artificial and would be
  smoothed by mixing in a 2D simulation or by additional opacity. The
  second outburst in both models happens as the second pulse sweeps
  over the density enhancement left by the reverse shock in the first
  pulse (see \Fig {t125a0un}). The collision giving the light curves
  shown happens between roughly 5 and $10 \times 10^{17}$ cm and and
  the radiation might be emitted in wavelengths other than
  optical.\lFig{homua0}}
\end{figure}

These energies, masses, and a star that survives for centuries after
the first explosive outburst with a luminosity $\sim10^{40}$ erg
s$^{-1}$ are just what one would might expect for a PPISN
\citep{Woo07,Smi08} with a helium core mass near 55 \Msun
\ (\Tab{tmodels}). This would be derived from a main sequence star of
$\sim$125 \Msun, if rotation is not included, and about 90 \Msun \ if
it is. Though there is no reason to assume that Eta Carinae has low
metallicity, but it is the helium core and envelope mass that matter
most, so the ``T'' models can be a useful guide. Making Eta Carina as
described here in a solar metallicity star would require a significant
reduction in currently favored mass loss rates though.

Consider the two models, T125A and T125B. Some relevant properties are
given in \Tab{eta}. Here M$_{ej1}$ and M$_{ej2}$ are are the masses in
solar masses ejected in pulses 1 and 2, and E$_1$ and E$_2$ are their
kinetic energies in units of 10$^{50}$ erg. There are only two
pulses. t$_{1-2}$ is the time between the first pulse (nominally t =
0) and the second, and t$_{\rm PreSN}$ is the time between the second
pulse and the collapse of the iron core. Both are measured in
years. $M_{\rm now}$ is the mass of the primary star in Eta Carinae
today, which should be in the range 50 - 60 \Msun.  Most of the mass
ejected in the first pulse is envelope, although helium is its
dominant constituent. Most of the mass ejected in the second pulse is
helium and carbon. Nitrogen is overabundant in both pulses because of
extensive CNO processing. Both models leave Wolf-Rayet stars in the
present day remnant. These WR stars have a luminosity near 10$^{40}$
erg s$^{-1}$ as is observed.  Both models explosively eject a mass
comfortably above the lower limit for the observed mass of the
Homunculus.

\begin{deluxetable}{cccccccc} 
\tablecaption{125 \Msun \ Models for Eta Carina} \tablehead{ Model &
  M$_{ej1}$ & E$_1$ & t$_{1-2}$ & M$_{ej2}$ & E$_2$ & t$_{\rm PreSN}$
  & M$_{\rm now}$ }
% & (\Msun) & (10$^{50}$ erg) & (yr) & (\Msun)  & (10$^{50}$ erg) & (yr) & (\Msun) } 
\startdata
T125A  & 22.5 & 8.3 & 70  & 7.1 & 8.0  &  2650 & 51.8 \\    
T125B  & 34.0 & 9.6 & 470 & 7.4 & 5.8  &  1100 & 58.2 \\
\enddata
\lTab{eta}
\end{deluxetable}

% half the ejected mass in model b is below 1100. Speeds range up tp 3000

Assuming that the first pulse and the ejection of the envelope
occurred around 1837, Model T125A has a second pulse that, within
the generous error bars of the models, might coincide with the 1890
outburst. This model is too energetic however.  By now the ejecta
of the two pulses would have merged and most of the matter would have
a speed near 2000 km s$^{-1}$, well above the observed average
650 km s$^{-1}$ \citep{Smi08}. Model T125B fares somewhat better if
only one pulse has happened so far. Half of the ejected mass is
moving slower than 1100 km s$^{-1}$ (\Fig{t125a0un}). In this case
though, the pulsing is not over and another Great Eruption is due in
the next few centuries.  It could of course be that the actual
pulses were somewhat weaker than in Model T125A, e.g., because of a
lower mass helium core or larger hydrogen envelope mass.  Two models
do not fully explore the range of possibilities.

This hypothesis has two major difficulties though.  One is the
expected brightness of the first mass ejection which, if the
progenitor was a RSG, would have exceeded $4 \times 10^{42}$ erg
s$^{-1}$ for roughly 100 days, i.e., resembled an ordinary Type IIp
supernova, not an ``impostor''. The other is the gross asymmetry of
the observed ejecta. The latter might relate to the well-established
presence of a binary companion with a current mass $\sim30$ \Msun
\ \citep{Mad12}, period 5.54 years, semi-major axis 16.64 AU, and
eccentricity 0.9 \citep{Hil01,Dam08,Par11}. The large eccentricity
implies that at closest approach the stars are separated by only 1.5
AU. Given that the radius of all the RSG studied here, both solar and
low metallicity, are over 10 AU, the companion has spent a lot of the
time inside the primary. Some sort of direct interaction would have
been unavoidable \citep{Smi11b}. This might be avoided or lessened if
the primary were a BSG or LBV. A dramatic expansion of the solar
metallicity stars frequently occurs at the end of helium burning
(\Sect{solar}) when the core is contracting to ignite carbon, a
process that takes about 10,000 years. If so, a substantial fraction
of the primary's envelope might be ejected in the plane of the orbit
\citep[e.g.][]{Nor06,Mor09,Smi11b} just before the pulses begin.

A bigger problem may be how to hide the 100 day light curve from the
first pulse. Someone probably would have noticed the sudden appearance
of a magnitude -6 star, even in the southern hemisphere in 1830. One
possibility is that the progenitor was a blue star not a RSG (see
\Sect{lbv}). Another is that the Great Outburst was heavily extincted
by dust. The latter seems unlikely, however, since historical
observations of Eta Carinae itself, well before the Great Eruption
\citep{Smi11a} do not indicate a large amount of extinction.

Another possibility is that the first pulse in Model T125B and its
bright supernova happened in the distant past, centuries {\sl before}
the Great Outburst (\Tab{eta}), and that the bright episode in the
1830's was the {\sl second} pulse running into the first. The dashed
line in \Fig{homua0} is the resulting light curve. A second delayed
brightening occurs about 150 years later as the shock wave encounters
a density spike left behind by the reverse shock in the first
eruption. The sharpness of the spikes in the light curve in
\Fig{homua0} are an artifact of the 1D calculation and would be
smoothed out in 2D. Again though, the velocity of the average ejecta
170 years after second pulse produced the Great Eruption (i.e., today)
is about 2000 km s$^{-1}$, larger than what is presently seen.

The future will tell, though not right away, if either of these
scenarios is correct. Eta Carinae could experience another great
eruption in the next few centuries (Model T125B) or disappear in 1000
years. Eta Carinae itself is not a star in the regular sense. It is
shining by gravitational contraction on its way to a final episode of
core silicon burning. Eventually the primary ``star'' in Eta Carinae
will collapse, probably to a black hole. If rotation and magnetic
fields generate a strong bipolar explosion, it could become a
superluminous supernovae, but at least as likely, it will not.  It
will become a black hole of about 50 \Msun. Given the persistence of
the close binary companion, Eta Carinae would then possibly become a
very luminous x-ray source.

If not Eta Carina itself, the class of PPISN with its broad range of
luminosities and durations with the possibility of recurrent
supernovae probably relates to some other supernova
``impostor''\citep[e.g.][]{Smi11c}.

\section{GW 150914}
\lSect{gwave}

The detection of gravitational radiation from two merging black holes
in GW 150914 \citep{Abb16a} offers new insights into the evolution of
stars in the mass range that might make PPISN. The inferred masses,
$36_{-4}^{+5}$ and $29_{-4}^{+4}$ \Msun, are what one would expect
from the evolution of low metallicity, non-rotating stars with masses
near 70 and 90 \Msun \ \citep[\Tab{tmodels}][]{Woo16}. If rotation is
included, the inferred main sequence masses are closer to 60 and 70
\Msun \ (\Tab{rmodels}). Model R60A is a special case with a residual
hydrogen envelope of about 16 \Msun \ that would have been lost in a
close binary capable of merging in a Hubble time. The estimated black
hole masses for Models R60A and R70A in a close binary are thus 30.9 and
41.7 \Msun.

Models from CHE can also, given a freely adjustable mass loss rate,
produce the observed black hole masses for any progenitor mass above
the masses of the black holes themselves, e.g., Model C90B makes a
black hole of 31.4 \Msun \ (all masses might be reduced by a few
tenths \Msun \ to account for neutrino losses during the protoneutron
star stage). A possible discriminant is the rotation rate of the black
hole. More mass loss means greater braking and a slower spin for the
black hole. Model C90B has a Kerr parameter of 0.08; Model C60C which
makes a similar 35.3 \Msun \ black hole gives a Kerr parameter of
0.34.  Given the small radii of the CHE models throughout their
evolution, it might be possible that a close binary could merge in a
Hubble time without experiencing a common envelope phase
\citep{Man16}.

An important inference from GW 150914 is that stars in the mass range
that makes PPISN must at least occasionally, and probably frequently,
collapse to black holes. Though low mass jet-like outflows cannot
excluded (\Sect{superl}), the cores cannot always explode completely
in rotationally-powered supernovae that leave neutron star remnants.
The calculations presented here also have implications for the kind of
black holes that might be discovered in the future. It has long been
known that there should be a gap in black hole production between 64
and 133 \Msun \citep{Heg02}. The hydrogen envelope is loosely bound in
stars that have helium cores in this mass range and such stars
robustly explode as PISN leaving nothing behind. This paper extends
that range downwards from 64 \Msun \ to 52 \Msun. No matter what the
star's mass when it encounters the PPI, it will pulse until sufficient
mass is lost for the star to complete its silicon burning evolution in
hydrostatic equilibrium. The heaviest such core in \Tab{tmodels} is 52
\Msun. There can be lighter black holes, but none heavier until a
helium core mass of 133 \Msun \ is reached. Single stars like Model
T70B could, in principle, make a black hole of over 52 \Msun \ if the
envelope as well as the core participated in the collapse. For zero
mass loss, this mass could be as large as 70 \Msun. Given how loosely
bound the envelope is, getting it to collapse might prove difficult
\citep{Wei08,Lov13}, but this cannot be ruled out. The close binaries
that make x-ray sources or merge in a Hubble time are a different
issue, however. There the envelope will be lost in a common envelope
or by mass exchange. There should be no black holes in close binary
systems with masses between 52 and 133 \Msun. CHE models will also
lack any loosely bound envelope and not produce black holes in this
mass range, even as single stars.

As with all rules, there are exceptions. Two black holes might merge
in a triple system, to make a black hole of up to 100 \Msun \ that
later merged with a third, or the black hole binary might form in a
dense cluster by dynamical processes \citep{Abb16b}.  Which is to say
the discovery of a black hole of say 60 or 70 \Msun \ would have
profound implications.

\section{Nucleosynthesis}
\lSect{nucleo}

\begin{deluxetable*}{cccccccccc} 
\tablecaption{NUCLEOSYNTHESIS IN LOW METALLICITY MODELS} 
\tablehead{ Species & 0.1 \Zsun & T80A & T90A & T100A & T110A & T120A &  T130A
& T140A & T150A \\
      & X & (\Msun) & (\Msun)  & (\Msun)  & (\Msun) & (\Msun) & (\Msun) & (\Msun) & (\Msun)}
\startdata
$^1$H   & 0.721   &  -9.2    &  -15.3   &  -17.7   &  -18.9   &  -19.2   &  -23.7   &  -62.4   & -70.8   \\
$^4$He  & 0.278   &   9.2    &   15.2   &   16.6   &   18.3   &   19.2   &   20.2   &   6.7    &   5.8   \\
$^{12}$C & 2.5(-4) & -3.5(-4) &  2.0(-2) &  1.3(-1) &  9.0(-2) &  5.0(-4) &  2.9(-1) &    1.4   &   1.4   \\
$^{13}$C & 3.0(-6) &  8.2(-5) &  7.4(-5) &  5.6(-5) &  4.9(-5) &  4.6(-5) &  3.6(-5) & -8.2(-5) & -1.1(-4)\\
$^{14}$N & 8.1(-5) &  4.1(-3) &  3.3(-3) &  2.1(-3) &  2.3(-3) &  3.2(-3) & -3.2(-4) & -4.9(-2) & -5.7(-2)\\
$^{15}$N & 3.2(-7) & -1.2(-6) & -7.3(-7) & -4.8(-7) & -1.2(-7) &  2.8(-7) &  5.8(-7) &  2.9(-4) &  3.6(-4)\\
$^{16}$O & 6.7(-4) & -4.3(-3) &  6.0(-2) &  9.1(-1) &  4.7(-1) & -5.1(-3) &   3.0    &   44.8   &   45.6  \\
$^{17}$O & 2.7(-7) & -5.0(-6) & -6.6(-6) & -6.9(-6) & -6.9(-6) & -6.8(-6) & -7.4(-6) & -1.3(-5) & -1.4(-5)\\
$^{18}$O & 1.5(-6) & -2.2(-6) &  6.8(-4) &  7.8(-4) &  8.9(-4) &  9.4(-4) &  1.1(-3) &  1.9(-3) &  2.3(-3)\\
$^{19}$F & 4.7(-8) & -6.1(-8) &  2.7(-7) &  2.4(-7) &  3.7(-7) &  1.9(-7) &  4.0(-7) &  1.7(-5) &  1.2(-5)\\
$^{20}$Ne& 1.2(-4) & -7.5(-4) & -8.3(-4) &  2.9(-2) &  1.1(-2) & -1.7(-3) &  2.2(-1) &    2.2   &   2.3   \\
$^{21}$Ne& 3.0(-7) & -2.5(-6) & -2.6(-6) &  8.4(-6) &  2.5(-6) & -3.0(-6) &  1.4(-5) &  1.6(-4) &  1.3(-4)\\
$^{22}$Ne& 9.6(-6) & -8.5(-6) &  1.1(-3) &  1.3(-3) &  1.4(-3) &  5.2(-4) &  1.3(-3) &  1.3(-4) &  1.2(-5)\\
$^{23}$Na& 3.9(-6) &  1.4(-4) &  7.8(-5) &  6.5(-5) &  8.2(-5) &  1.2(-4) &  9.2(-5) &  4.6(-3) &  4.8(-3)\\
$^{24}$Mg& 5.7(-5) &  7.6(-4) &  1.3(-3) &  4.3(-3) &  2.6(-3) &  1.8(-3) &  3.6(-2) &   1.7    &    1.6  \\
$^{25}$Mg& 7.6(-6) & -8.9(-5) & -1.1(-5) &  8.0(-4) &  3.9(-4) & -2.1(-5) &  3.0(-3) &  2.0(-2) &  2.0(-2)\\
$^{26}$Mg& 8.7(-6) & -2.1(-4) & -2.8(-4) &  5.9(-4) &  5.2(-5) & -4.5(-4) &  2.9(-3) &  2.3(-2) &  2.3(-2)\\
$^{27}$Al& 6.7(-6) &  3.2(-4) &  5.0(-4) &  5.8(-4) &  6.4(-4) &  6.7(-4) &  8.0(-4) &  3.9(-2) &  3.5(-2)\\
$^{28}$Si& 7.7(-5) &  7.0(-6) &  1.1(-5) & -7.0(-6) &  6.0(-6) &  1.8(-5) & -4.2(-5) &   4.0    &    8.5  \\
$^{29}$Si& 4.0(-6) & -5.9(-7) &  1.3(-6) &  1.8(-5) &  9.4(-6) & -4.0(-7) &  6.2(-5) &  2.6(-2) &  2.2(-2)\\
$^{30}$Si& 2.7(-6) &    -     &  2.6(-6) &  2.9(-5) &  1.5(-5) &  9.0(-7) &  9.3(-5) &  1.5(-2) &  1.0(-2)\\
 Total  &  1.0    &   41.0   &   54.1   &   60.7   &   65.5   &   69.4   &   78.7    & 135.6   &   150   \\
\enddata
\lTab{nucleo}
\end{deluxetable*}

Nucleosynthesis for a representative series of models is given in
\Tab{nucleo} and \Fig{nucleo110}. The models given are the ``A''
series of low metallicity, non-rotating stars for which a standard
mass loss prescription was used. ``Total'' in columns 3 through 10 is
the total mass ejected in winds and in pulses by the given model. The
quantity 0.1 \Zsun \ times ``Total'' in each column is the starting
composition for that material. Column 2 is the starting composition
for all the low metallicity, full star models, given by mass
fraction. The other numbers in other columns give the changes from
these initial values. While the network in all cases extended to above
germanium, only the species below titanium are tabulated. There was
little contribution to heavier elements except for a mild s-process
(\Fig{nucleo110}). Except for the last two cases, in which nearly
(T140A) or all (T150A) the star exploded, the nucleosynthesis is
evaluated at the time of carbon depletion just before pulses
started. In those cases subsequent nuclear burning further in did not
affect the ejecta and presumably all ended up in black holes. In those
cases where nuclear processing after helium burning is negligible, the
pulses, nucleosynthetically, merely act to augment mass loss.

The main products of PPISN are thus helium, carbon, nitrogen and
oxygen. Small amounts of fluorine through silicon are also ejected and
some s-process up to mass A = 80. No appreciable primary iron-group
synthesis occurs in any model. Even in T150A, a full up PISN, only
0.045 \Msun \ of iron is ejected and most of that is $^{54}$Fe, not
$^{56}$Ni. In the same model 8.5 \Msun \ of $^{28}$Si and 45.6 \Msun
\ of $^{16}$O is produced, so this small iron production is nearly
negligible.

The nucleosynthesis of Models 120A and 140A are anomalous compared with
the rest because of the large remnant mass in the former and the small
remnant mass in the latter. T140A in fact closely resemble Model
T150A. If the initial mass function for some reason resulted in the
production of no stars above 130 \Msun, then CNO would be the
principal nucleosynthetic contribution of this mass range. However,
just a few stars of 150 \Msun \ and more would quickly come to
dominate. Given the relatively small amount of iron made all the way
up to mass 200 \Msun, one would expect from a generation of such
stars (truncated below 200 \Msun) a composition rich in the elements
carbon through magnesium with an increasing concentration of silicon
through calcium at higher masses. It would be quite deficient in iron
group species.

% fig 28 -  nucleosynthesis in Model n110a0
\begin{figure}
\includegraphics[width=0.49\textwidth]{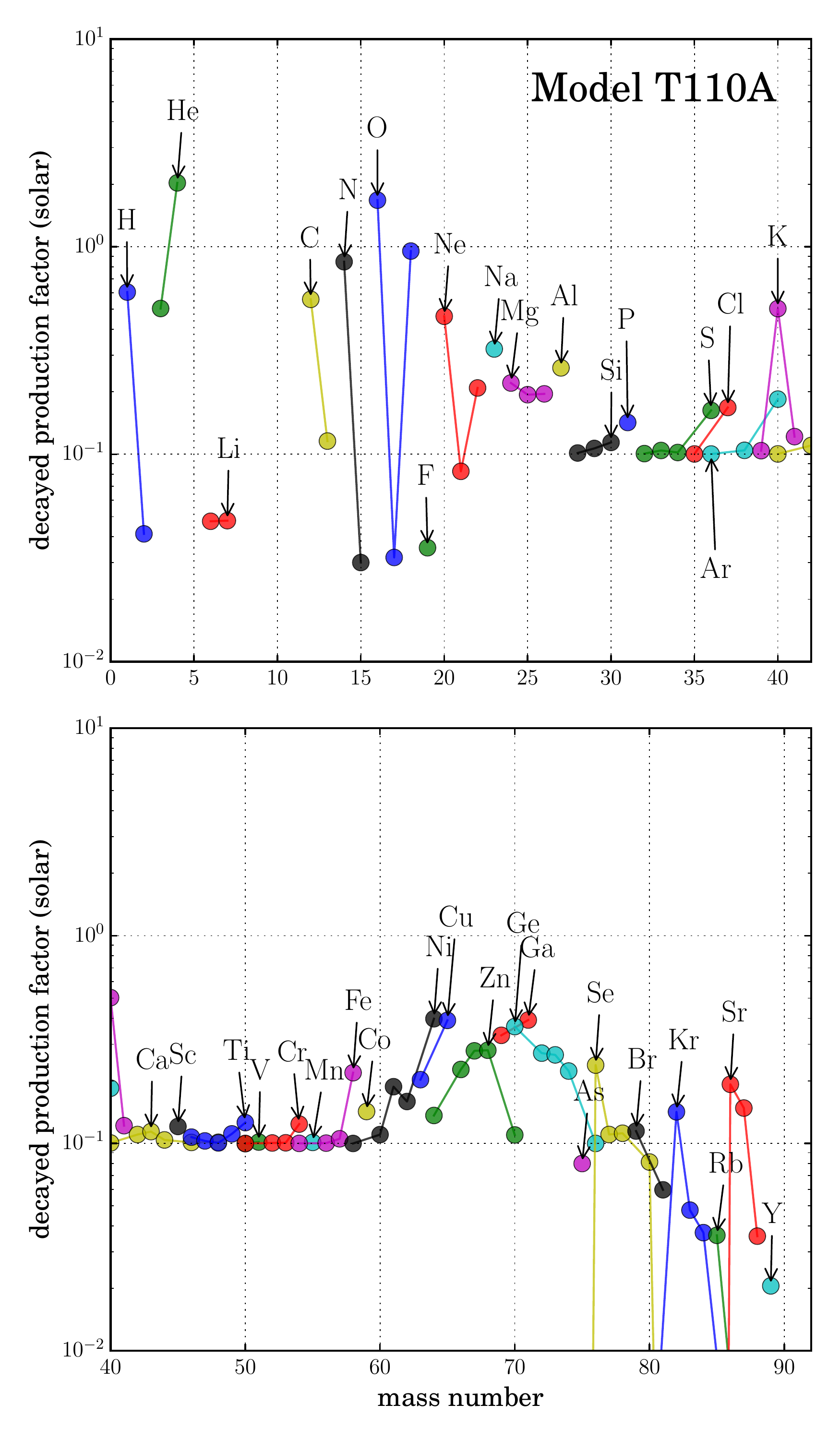}
\caption{Nucleosynthesis in Model T110A. Production factors for the
  isotopes of a given element are connected by lines. A value of 1 in
  the ejecta would correspond to a solar mass fraction. The
  concentration of points around a production factor of 0.1 between A
  = 28 and 58 reflects unchanged abundances in a star that had 0.1
  solar mass fractions of these species to begin with, i.e., the
  metallicity of the initial star was 0.1 solar. Species above 0.1
  thus have a net creation in the event and those below are at least
  partly destroyed.  Species with production 0.1, like $^{56}$Fe,
  would have a much lower production factor in a lower metallicity
  star. The abundances of $^{40}$K and species above A = 58 show the
  operation of a limited s-process in the helium shell prior to the
  PPI. $^{14}$N is produced by the CNO cycle and would be smaller in a
  star of lower metallicity. $^{12}$C and $^{16}$O are mostly primary,
  produced by helium shell burning.  \lFig{nucleo110}}
\end{figure}

While not presented in this paper, zero metallicity models have also
been calculated and have similar nucleosynthesis, with the notable
exception of large quantities of primary nitrogen made in the very low
metallicity stars.

\section{Conclusions}
\lSect{conclude}

The PPI, operating in stars with various final helium core masses,
envelope masses, and radii, gives rise to a broad range of observable
phenomena. These include single supernovae ranging from very faint to
very bright; supernovae with complex, distinctive light curves and
multiple peaks; recurrent supernovae; bright, enduring radio and x-ray
sources; and supernova remnants that contain luminous WR stars. This
paper has been a first attempt to characterize these diverse
possibilities and describe their observable properties.

PPISN occur when mass loss does not shrink the final helium core below
30 \Msun \ (40 \Msun \ for the more luminous events). They are thus
favored by low metallicity, and a threshold value of about one-third
\Zsun \ is estimated here (\Sect{mdot}). Uncertain mass loss rates
make this limit very approximate, however, and PPISN are not excluded
even at solar metallicity. For bare helium cores above 62 \Msun, the
pair instability is so violent that the entire star is disrupted in a
single pulse, i.e., a PISN. Slightly larger cores, up to 65 \Msun, can
still be PPISN if that core is embedded in a massive hydrogen envelope
(\Tab{tmodels}).

For the stellar physics used, this range of helium core mass is produced
by single, non-rotating stars of 10\% solar metallicity with main
sequence masses between 70 and 140 \Msun. This same mass range would
yield PPISN for other non-rotating stars in which mass loss failed to
uncover the helium core before the star died. In particular, the same
limit would apply to stars with less metallicity, including Pop III
stars, and to solar metallicity stars with unusually low mass loss
rates (\Tab{smodels}).  For a moderate amount of rotation, the
threshold main-sequence mass for the PPI is reduced to 60 \Msun, or
even less in the case of CHE (\Tab{rmodels}).  On the other hand,
ending life as a non-rotating blue supergiant can raise the lower
bound for a faint optical display to 80 \Msun. Regardless of
radius, the presupernova stars will have bolometric luminosities in
the range 0.5 to 1.3 $\times 10^{39}$ erg s$^{-1}$. These progenitors
may be RSG, BSG, LBV, or WR-stars, and all four possibilities were
explored here.

The fraction, by number, of core-collapse supernovae (all stars above
8 \Msun) in the range 70 \Msun \ to 140 \Msun \ for a Salpeter IMF
with $\Gamma= -1.35$ is small. Taking an upper limit on the stellar
mass of e.g. 150 \Msun \ (the answer is not sensitive to this limit),
\begin{equation}
f_{\rm PPISN} \ = \ \frac{70^\Gamma - 140^\Gamma}{8^\Gamma - 150^\Gamma} \ = 0.033.
\end{equation}
The fraction in the more restricted mass range, 90 \Msun \ to 140
\Msun, that makes optically bright events is smaller still, about
1.7\%. These estimates might be multiplied by two or so since a
substantial fraction of core collapses below 70 \Msun \ go directly to
black holes with no bright display and the IMF is not precisely
Salpeter-like, but the fraction of bright supernovae from PPISN is
probably no more than a few percent. On top of that, PPISN probably
only happen in metal poor regions, If, as appears likely, PPISN often
end up looking like Type Ibcn and IIn supernovae, they can only
explain a small fraction of the observed events. Type IIn supernovae
alone are estimated to be 2\% to 9\% of core-collapse events
\citep{Kie12}. PPISN might, however, account for some of the more
unusual cases.
 
Provided a helium core in the unstable mass range survives to the
presupernova stage, the PPI and its consequences are unavoidable and
its qualitative features, robust and simple to calculate
(\Sect{taukh}). For low mass cores the PPI is weak, lasts a short
time, and is characterized by many low energy pulses that, in total,
release only a small amount of energy. For more massive cores, the
converse is true (\Tab{hecore}, \Sect{hecoreexp}). Fewer pulses,
separated by longer intervals, eject more mass with greater energy.
The maximum duration of any PPI episode is close to the
Kelvin-Helmholtz time for the typical helium-come core mass starting
from a loosely bound state, about 10,000 years, and the maximum
explosion energy is a fraction of the binding energy of that core,
$\sim4 \times 10^{51}$ erg.  Full stellar models
(\Tab{hecore},\Tab{tmodels}) demonstrate these characteristics.

Once the pulsational episode is ended, which typically requires the
exhaustion of oxygen in the inner 6 \Msun \ of the more massive stars,
silicon burning ignites, either centrally or in a shell, and the star
forms an iron core in hydrostatic equilibrium that collapses in the
usual way. Unlike supernovae in lighter stars, however, the iron core
is very massive, and the density outside declines slowly with radius
(\Fig{fecore}). Such a star is virtually impossible to explode with
neutrinos and difficult to completely explode even with rotation. It
may thus be that the pulsations and the colliding shells they produce
are the sole optical and chemical manifestations of star death in the
mass range that makes PPISN (though see \Sect{superl}).

Some PPISN progenitors will have lost most or all of their hydrogenic
envelopes, or will have experienced CHE, and will be Wolf-Rayet
stars. Their surface abundances will reflect the extent of that mass
loss and rotational mixing. Here, all of these compact progenitors were
approximated by pure helium stars of constant mass. The structure and
explosive characteristics of the PPI will not be greatly modified by
charging the surface composition of the star, or by adding a low mass
extended envelope. Pulses in the lightest of these helium cores
produced short, faint, blue transients (\Sect{helite},\Fig{helight})
that ejected a small amount of mass and lasted only a week or
so. Slightly heavier stars made Type Ibn or Icn supernovae as bright
as a typical Type Ia, but with more structured, irregular, longer
lasting light curves. These light curves, produced by colliding shells
could have ``tails'' of a sort, especially if they interacted with
pre-pulsational mass loss (not included here), but the explosions
ejected no radioactivity. The colliding shells were usually more massive
than a SN Ia and had less kinetic energy, so the velocities were
slower. The colors were bluer, and the initial display from the the
matter ejected by the first pulse was faint.

Particularly intriguing are the helium cores between 52 and 62 \Msun
\ that give supernovae separated by long delays. Following an initial
mass ejection which, if circumstellar interaction is ignored, is faint
(\Fig{helight2}), the star becomes a ``dormant supernova'', a compact
star shining with approximately the Eddington luminosity (near
10$^{40}$ erg s$^{-1}$), embedded in a supernova remnant. If there was
pre-pulsational mass loss, the remnant could also be a bright radio or
x-ray source. This dormant phase lasts from several years to several
thousand years. Light from the central star is provided by its
Kelvin-Helmholtz contraction and the accretion of matter that falls
back from the first mass ejection. The latter may occur at an
irregular rate, since the shells are not perfect spheres, and produce
some variability in the emission. Finally, just before the star truly
dies, several pulses in rapid succession produce a bright optical
display (\Fig{helight} and \Fig{helight1}). These secondary light
curves often exhibit a characteristic ``double peak'', a rise to a
ledge or first peak lasting 10 - 20 days, followed by a dramatic
brightening to $\sim10^{43}$ erg s$^{-1}$.  This light curve
morphology, as well as the color and multiple velocity components,
resembles what was seen in SN 2005bf \citep{Fol06}, and further study
of this event as the possible explosion of a WN star of around 55
\Msun \ is warranted (\Sect{helite}).  The matter ejected by these
final pulses will also interact for a long time afterwards with the
matter ejected by the first pulse powering, again, a bright radio and
x-ray source with, perhaps, some optical emission. The physical
conditions and appearance might be similar to what is happening now in
the rings of SN 1987A \citep{Hel13,Man05,Zan14}.

Circumstellar interaction with a presupernova wind can complicate and
enrich the possibile outcomes of PPISN in all the models. Consider,
for example, the case of SN 2009ip, nominally an ``supernova
impostor'' that only became a ``real supernova'' (of Type IIn) in
2012.  The spectrum of the 2009 outburst showed hydrogen lines with a
characteristic speed of 550 km s$^{-1}$, but with evidence for a high
velocity component up to 3000 - 5000 km s$^{-1}$ \citep{Smi10b}, or
even 7000 km s$^{-1}$ \citep{Fol11}. The peak luminosity in 2009 was
$\sim10^{41}$ erg s$^{-1}$, but in 2012, a second brighter outburst
occured.  The second light curve had a ledge at 10$^{41.5}$ erg
s$^{-1}$ that lasted for roughly a month, followed by a rapid rise to
10$^{43}$ erg s$^{-1}$ \citep{Fra15}. Interaction continues today in
SN 2009ip \citep{Kim16}. Qualitatively at least, this history
resembles what would be expected for a 52 \Msun \ helium core blowing
up inside a shell of pre-explosive, hydrogen-rich mass loss with a
mass loss rate about $5 \times 10^{-4}$ \Msun \ y$^{-1}$ and
charateristic wind speed 500 km s$^{-1}$ (\Fig{helight} and
\Fig{lbvliteb}). The first mass ejection in Model He52 ejects 1.0
\Msun \ with an energy of $7 \times 10^{49}$ erg and a velocity that
peaks at about 8000 km s$^{-1}$ (average about 2500 km s$^{-1}$, but
the highest speed would collide with the wind first). Interacting with
the wind would give a peak luminosity $\sim10^{41}$ erg s$^{-1}$.
Time structure could be added if the wind were clumpy or unsteady. 4.6
years later in Model He52, a second eruption makes the light curve in
\Fig{helight} as two more shells collide. While He52 was a pure helium
star and incapable of making a Type II supernova, similar dynamics
would result for a WN star or a compact LBV with the same helium core
mass. A slight change in core parameters might make a fainter first
peak like for Model He58 in \Fig{lbvliteb}. More study is warranted.

The study of bare helium cores also provides insight into the
energetics and luminosity of the brightest Type I PPISN. In no case
did the total kinetic energy in the pulses exceed $2.3 \times 10^{51}$
erg, and this was for a rare case on the verge of becoming a PISN. A
more common limit was 10$^{51}$ erg, and that energy was shared among
several pulses.  Because of the low energies in individual pulses and
the large masses ejected in the more energetic models, typical
velocities are less than 4000 km s$^{-1}$, except in a small amount
of material near the outer edge. These events might thus be classified
as Type Ibn or Icn \citep{Fol07,Smi12,Pas08a,Pas08b}. Given that only
a fraction of this energy can be converted to light, PPISN from
compact progenitors should not exceed a few $\times 10^{50}$
erg. The brightest helium core explosions here (e.g., Models He48 and
He50) radiated a total energy close to $1 \times 10^{50}$ erg. This
omits the considerable energy radiated by any star during the dormant
stage and any interaction with preexplosive mass loss.

{\black Stars that retain an appreciable hydrogen envelope, i.e., a mass
greater than the mass ejected by the first pulse in a bare helium core
explosion, have different dynamics and light curves. Because the
envelope tamps the expansion of the helium core, recurrence times can
be shorter and a greater fraction of kinetic energy is turned into
light.  The explosions can be more luminous. Two possibilities were
explored: red supergiants (\Sect{tmods}) and blue supergiants or
luminous blue variables (\Sect{lbv}). For the physics assumed,
hydrogenic stars with 10\% solar metallicity most naturally ended
their lives as RSGs. Only a fraction of their envelope was convective
though, and throughout most of their mass and lifetimes, the stars
with appreciable mass loss resembled BSGs. The final helium core mass
for a given ZAMS mass did not vary greatly for a large range of mass
loss rates, though the size of the CO core for a given helium core was
slightly larger in these stars that retained envelopes. This affected
the comparison between the results of helium cores evolved at constant
mass and full star models.}

{\black 
That part of the envelope contained in the surface convection zone of
a RSG is very loosely bound and easily ejected.  Even a weak shock can
eject a mass typical of a Type II supernova. The velocity of the
ejected matter, however, is very low, $\sim$100 km s$^{-1}$ in the
lightest cases, so the transients for lowest mass PPISN in stars with
envelopes are long and faint and have low characteristic speeds
(\Fig{t70lite} and \Fig{t75lite}).  They would be classified as Type
IIn. Discovering such events will be challenging though. Even at peak,
they are only about 10 times brighter than the star that made
them. They occur only at low metallicity, so perhaps are far away, and
only in star-forming regions where other bright stars might be present.}

{\black PPISN happening in RSGs over 90 \Msun \ are brighter and easier
to discover, but could easily be confused with ordinary Type IIp
supernovae (\Fig{t80lite} and \Fig{t90lite}). The light curves from
the heavier ones are distinctly overluminous and structured
(\Fig{t100lite}), but the duration, energetics, color, and luminosity
are not out of bounds for what is observationally, a diverse class
anyway. In cases where the envelope mass is large, the light curves
also last longer than is typical for Type IIp.} Another discriminant
might be the lack of a radioactive tail on the light curves. It may
be difficult to distinguish a radioactivity-powered tail from
circumstellar interaction though.  For a presupernova mass loss rate
of 10$^{-4}$ \Msun \ y$^{-1}$, a shock speed of 4000 km s$^{-1}$ and a
wind speed of 100 km s$^{-1}$, the contribution of circumstellar
interaction to the luminosity would be $L_{\rm CSM} \approx 2 \pi \dot
M v_{\rm shock}^3/v_{\rm wind} = 2.5 \times 10^{41}$ erg
s$^{-1}$. Given the uncertainty in late time mass loss rates
and wind speeds for presupernova stars, this could easily be an
underestimate by 10 or more. The ``tail'' could dominate the display!
Circumstellar interaction with presupernova mass loss was omitted
here because of the uncertain parameters, and a desire to highlight
what the PPI, acting alone, would do.

Observationally, there is no clear evidence for a Type IIp supernova
with no tail. There are cases of faint supernovae with very faint
tails, requiring as little as 0.005 \Msun \ of $^{56}$Ni for their
explanation \citep[see Table 6 of][]{Pej15b}, but these are probably
the neutrino-powered explosions of lower mass stars.  An interesting
case is supernova LSQ13fn \citep{Pol16}. This unusual event resembles
Model T90A (\Fig{t90lite}). It had about the same luminosity, a
duration longer than most Type IIp supernovae (though somewhat less
than T90), an unusually slow velocity, exhibited a dramatic drop to an
unresolved tail, and was inferred, spectroscopically, to have low
metallicity. Further modeling of this specific event might be
desirable.

For main sequence masses above about 100 \Msun, the interval between
pulses becomes longer than the duration of any single event and there
can be multiple supernovae and dormant supernovae. If the star is a
RSG, the first event resembles a normal Type IIp as the envelope is
ejected, but the later displays powered by colliding shells, can be
especially bright, long lasting, and have multiple maxima. The
brightest supernovae from PPISN are produced by secondary pulses in
the 100 - 130 \Msun \ mass range. Activity there can continue for
years (\Sect{t110}, \Fig{t110lite}, and \Fig{t120lite}), centuries
(\Fig{homua0}), or even millennia (\Fig{t130lite}). Typically, these
explosions have only one or two violent episodes of pulsing activity
after the first pulse ejects the envelope. Structure is added to the
light curves by the collision of thin high density shells resulting
from reverse shocks. These shells are artificially thin in 1D
calculations and, in addition to causing unrealistic short excursions
to very high luminosity, pose computational difficulties. Further
two-dimensional studies that include radiation transport are needed to
properly simulate the mixing that goes on and the modification to the
light curve.  Until such calculations have been done, a more realistic
prediction would come from drawing the best smooth line through the
calculated light curve. The smoothing length $\Delta t/t$ is set by
the degree to which the shells are spread by mixing, and might be
$\Delta r/R \sim 10 - 20$\% \citep{Che14,Che16}. The smoothing should
preserve energy emitted, i.e., the area under the curve.

A potential observational counterpart to explosions in this 100 - 130
\Msun \ range is SN 2008iy, an exceptionally luminous Type IIn
supernova with a very long rise time of 400 d that showed evidence for
a major mass loss about 55 years before \citep{Mil10}. Compare this
with the second outburst in Model T125A in \Fig{homua0} which took
place 70 years after the first pulse ejected 23.5 \Msun. The rise time
from $5 \times 10^{41}$ erg s$^{-1}$ to a peak of $4 \times 10^{42}$
erg s$^{-1}$ in Model T125A was only about 200 d and the peak
luminosity half that of SN 2009iy, but the width of shells is very
uncertain in these 1D models and the masses and speeds are model
sensitive. Perhaps more problematic, for any collisional model, is
that the collision takes place in an optically thin medium and it is
not obvious that most of the power would come out as optical light.

Light curves and, to a lesser extent, the dynamics of PPISN will be
different if the presupernova star is a BSG or an LBV (\Sect{lbv})
rather than a RSG.  The blue progenitors here had properties
that overlapped with both BSGs and LBVs. Envelopes of approximately 5,
10 and 20 \Msun \ (\Tab{bmodels}) had radii of a few $\times 10^{12}$
cm and effective temperatures of 25,000 - 50,000 K. Because of the
lack of weakly bound matter in the convective RSG envelope, the
threshold for making faint supernovae was increased to above 80
\Msun. The smaller initial radius of blue progenitors also made the initial
display much fainter.  Apart from affecting the lighter fainter events
and the light curve resulting from the first pulse, blue progenitors
in the more massive stars resembled their RSG counterparts. Compare,
for example, Models T110B and B110 in \Fig{t110lite}, \Fig{lbvlitea},
and \Fig{lbvliteb}. Some variation in shell mass and thickness is
expected, but the duration, interval, peak brightness, and
structures are qualitatively similar, especially if the sharp time
structure in T110B is smoothed.

Some of the prompt light curves of the blue progenitors are especially
interesting though, given both the lack any compelling evidence for
RSGs at such high masses, regardless of metallicity, and observations
of supernovae that resemble the models. SN 2005gl \citep{Gal09}, for
example, had a roughly month long ``precursor'' at $\sim$10$^{41}$ erg
s$^{-1}$ before abruptly rising to a peak luminosity of $5 \times
10^{42}$ erg s$^{-1}$. Compare that with the models in
\Fig{lbvliteb}. Gal-Yam et al estimated that a mass of only $\sim0.01$
\Msun \ was necessary to power the light curve at peak, but this
estimate overlooks the velocity gradient in the ejected matter. The
mass could be several solar masses, but the maximum luminosity comes
from interacting with only the innermost, slowest moving, highest
density part of that shell. A similar dramatic rise from a faint
initial outburst was seen in SN 1961v \citep{Smi11c,Koc11}; SN 2010mc
\citep{Ofe13}; the 2012 outburst of SN 2009ip \citep{Fra15}; and SN
2015bh \citep{Eli16}. There is also some evidences that these events
came from the explosion of very massive stars \citep{Fol11} and, at
least in the case of 2009ip, that the star was a LBV. Strong radio and
x-ray emission has continued for years after the 2012 exposion of SN
2009ip \citep{Smi16}.  These light curves with ledges and second peaks
are also similar to the previously mentioned Type Icn SN 2005 bf and
one cannot help but feel that PPISN are responsible for at least some
events with this double-peaked morphology. Further detailed study of
individual events is clearly warranted. On the more negative side
though, all the blue models that produced these interesting initial
displays became bright supernovae again shortly afterwards. It seems
doubtful that these subsequent events would have been missed. Some of
the {\sl later} explosions (\Fig{lbvlitec}) also exhibit double
maxima, however, and there the converse problem arises of hiding an
{\sl earlier} supernova.

There is also substantial observational evidence for other explosions in
LBVs producing Type IIn supernovae from one to several years after a
major mass ejection. SN 2006aa, 2006jd, 2006qq, and 2008fq may be
examples \citep{Tad13}. Compare with Models B110 and B115 (\Fig
{lbvlitec}). SN 1994W, a Type IIn, also ejected a circumstellar shell
1.5 years before explosion. The interaction there produced a
luminosity of 10$^{43}$ erg s$^{-1}$ at about 10$^{15}$ cm
\citep{Chu04,Kie12}. SN 2015U lost of order a solar mass during the
last few years before exploding \citep{Shi16}. It is unlikely that all
these events were PPISN, but further individual study could be
warranted.

Hydrogenic stars above 120 \Msun \ produce some interesting transients
that may not all be particularly optically bright (\Sect{t120}).  The
star's core survives 100's to 1000's of years after the first mass
ejection, mostly in a dormant state, but occasionally experiencing
additional pulsational mass ejection. These ejections collide with
previously ejected shells at such large radii ($> 10^{16}$ cm) that
the display lasts a long time and is optically thin to electron
scattering. Radiative powers are $\sim10^{40} - 10^{41}$ erg
s$^{-1}$. These might resemble what is happening now as the supernova
collides with its ring in SN 1987A \citep{Lar11} and SN 2009ip
\citep{Smi16}. Another feature is the dormant central star itself,
shining with a a luminosity of about 10$^{40}$ ergs$^{-1}$ and a hot
spectrum.

A possible example of a dormant PPISN could be Eta Carinae
(\Sect{etacar}), though the high metallicity, gross asymmetry, and
long history of recurrent faint outbursts \citep{Kim16} argue against
such an interpretation. If a PPISN, Eta Carina is best modeled as a
star near 125 \Msun \ on the main sequence. Near death, the helium
core mass was 57 $\pm$ 2 \Msun.  A residual core of 50 to 60 \Msun
\ has survived previous outbursts (\Tab{eta}) and is currently
radiating near its Eddington luminosity. This star would resemble an
ordinary massive Wolf-Rayet star, but perhaps with an extended
atmosphere from the fall back of previous explosions and a wind. In
order that the event have a time scale of centuries, a PPISN
explanation requires that the total kinetic energy be closer to
10$^{51}$ erg than the previously claimed 10$^{50}$ erg. The unusual
asymmetry of the object is not explained in this model, but might
involve interaction with its binary companion.

A possible problem with this speculation is that the first pulse and
envelope ejection should have produced a supernova that was brighter
than seen in the 1830's. This could be alleviated, in part, if the
progenitor was a compact blue star and not a RSG. The luminosity from
the first pulse of Model B120 (not illustrated) was only a few times
10$^{41}$ erg s$^{-1}$ after the first week (and the bolometric
correction was large during that first week). Or the first pulse may
actually have happened several centuries earlier and the supernova it
produced, despite being bright for several months, was not recorded in
the southern hemisphere where it was visible. The second pulse,
happening in the 1830's would then be responsible for the Great
Eruption. Eta Carina would then have had two outbursts in the past,
but, unfortunately, not three as inferred by \citet{Kim16}. If this
speculative scenario is valid, Eta Carina should transition into a
black hole in the next few millennia (\Tab{eta}).  The PPI seems
unlikely to explain all the complex history of Eta Carina, but it
could be playing a partial role and further study is definitely
warranted.

Part of the motivation for this study was the hope that PPISN would
provide a robust explanation for SLSN \citep{Woo07}. The results here
confirm that the colliding shells made by PPISN can indeed make
supernovae that are very bright for extended periods (e.g.,
\Fig{helight}, \Fig{t100lite}, \Fig{t120lite}, \Fig{t130lite}, and
\Fig{lbvlitea}), but none approach the level of e.g., SN 2003ma, which
may have emitted $3.6 \times 10^{51}$ erg of light \citep{Res11}. The
most luminous events here, T120, (\Fig{t120lite}) and T130,
(\Fig{t130lite}) emitted less than $5 \times 10^{50}$ erg and only
briefly exceeded 10$^{44}$ erg s$^{-1}$. Most models emitted less than
10$^{50}$ erg. While not all possibilities have been explored, it
seems unlikely that thermonuclear PPISN, unassisted, can explain the
integrated light of events like SN 2006gy \citep[$2.4 \times 10^{51}$
  erg of light;][]{Smi10}, SN 2005ap \citep[$1.7 \times 10^{51}$ erg
  of light;][]{Qui11}, and SN 2008es \citep[$1.1 \times 10^{51}$ erg
  of light;][]{Mil09}. This is disappointing. SN 2006gy, was the first
PPISN identification suggested in the literature \citep{Woo07}. Later,
\citet{Smi10} showed that this event required a kinetic energy of $5
\times 10^{51}$ erg which may be just out of reach. Indeed, Woosley et
al. had to artificially enhance the collision velocity by a factor of
two (hence explosion energy by a factor of 4) in their model in order
to get a good fit to the light curve.  Further study, especially of
2006gy, is certainly needed.

A more energetic outcome is possible if, contrary to current opinion,
the helium and heavy element core of these very massive stars does not
go quietly into a black hole \citep[see also][]{Yos16}. The
observation of gravitational radiation from two merging black holes
with masses 29 and 36 \Msun \ shows that PPISN of moderate metallicity
do frequently make black holes \citep{Woo16}, but collapse to a black
hole does not necessarily exclude making a supernova.  The rotation
rates of the iron cores for the rotating models here are large
(\Sect{rotate}, \Tab{rmodels}), corresponding to a cold pulsar
rotation period of 2 - 3 ms for stars with hydrogenic envelopes and
possibly faster for CHE. These periods imply rotational energies up to
$\sim4 \times 10^{51}$ erg for stars with envelopes and $\sim2 \times
10^{52}$ erg for CHE. Much more work is needed to determine the
outcome here. Are there jets? What is their energy? Are they broad or
narrow?  Does a neutron star survive?

Lacking adequate theoretical guidance, the effect of very energetic
central explosions happening at the time of iron core collapse in
PPISN was explored in a parametric way (\Sect{superl}). The most
interesting cases had pulses that ejected many solar masses
approximately one year before collapse.  Helium cores near 50 \Msun
\ had pulsing activity that spanned $\sim1$ year and were
candidates. Some heavier stars also had late stage pulsations that
lasted months to years after previously ejecting their hydrogen
envelopes centuries or more before. The collision of the rapidly
expanding core with the massive shell at 10$^{15}$ - 10$^{16}$ cm
produced very luminous supernovae potentially capable of explaining
even the brightest SLSN (\Fig{slsnlite}, \Fig{hyper}). For
``moderate'' kinetic engies of a few $\times 10^{51}$ erg (plus the
non-trivial binding energy of the disrupted star), large amounts of
matter fell back and the final remnant was a black hole. For more
energetic explosions, $\sim2 \times 10^{52}$ erg, the light curves
were brighter, no matter fell back, and up to several solar masses of
$^{56}$Ni were ejected. Even this large amount of nickel had no effect
on the light curve.  The results of 10$^{52}$ erg explosions in stars
that retain extended hydrogenic envelopes should be treated with
special caution though. Not only is the explosion mechanism
unspecified, but retaining so much rotation in the core of a giant
star has been an enduring theoretical problem in the context of GRB
models, and may be why GRBs are associated with Type I supernovae and
not Type II. Be that as it may, the brightest explosions, and the only
ones producing over a few times 10$^{51}$ erg of light, required a
very massive shell for the impact, i.e., the hydrogen-rich envelope of
the presupernova star, and thus would be Type IIn supernovae. Somewhat
more realistic models derived from CHE that avoided ever becoming
giants, gave Type I supernovae that radiated up to $\sim2 \times
10^{51}$ erg.  If these MHD explosions in PPISN exist, they would be
close cousins to GRBs, but lacking, in their final ejecta, any strong
relativistic component.

It has been known for some time that normal PISN leave no bound
remnants. A helium core over 133 \Msun \ is required for the direct
production of a black hole, and for some range of masses below that
no black hole is made, but how far down in mass does this void extend?
This study answers that question \citep[\Sect{gwave}; see
  also][]{Woo07,Woo15}. PISN will occur down to helium core masses of
about 64 \Msun, but the PPI will eject any mass in excess of 52 \Msun
\ (\Tab{hecore}, \Tab{tmodels}, and \Tab{rmodels}). A generation of
bare helium cores or of CHE models that managed to span all masses at
death would not produce any black holes between 52 and 133 \Msun. But
what about lighter stars that retained some hydrogen envelope? There
the answer is more nuanced and depends upon the mass loss history of
the star.  Consider Model T70B (\Tab{tmodels}) for example, which has
a helium core of 30.5 \Msun \ when it dies and thus avoids the
PPI. Will only the helium core collapse to a black hole, or will the
hydrogen envelope also participate in the collapse? If the envelope
collapses and there has been no mass loss, a black hole of 70
\Msun \ could result.

If the mass loss rate (i.e., metallicity) is high enough, then the
envelope will have little or no mass and the 52 \Msun \ limit will not
be violated. If late time mass ejection removes the envelope
\citep{Lov13,Wei08}, the limit will also hold. More importantly, since
black hole mass determinations, including those from gravitational
radiation, come from interacting binaries, a companion star may have
robbed the PPI candidate of its envelope, perhaps through common
envelope evolution. Systems that produce binary black holes that must
merge in a Hubble time by emitting gravitational radiation do not
accommodate RSGs, BSGs, or even LBV's in their final stages
\citep[][de Mink, private communication]{Pet64}. Either the envelope
is removed through common envelope evolution or never was there
(CHE). So if the black hole pairs seen in gravitational radiation
experiments have been produced by the evolution of a single stellar
system and not, e.g., by dynamical merger in a dense cluster, they
will exhibit the predicted gap from 52 to 133 \Msun. In reality, since
133 \Msun \ helium cores require extremely massive stars for their
production (260 \Msun?), the ``gap'' will probably be seen as a ``cut
off'' above which no black holes are found in merging systems.

In addition to limiting the range of black hole masses that exist in
nature, the present study also limits the mass of $^{56}$Ni ejected in
a core-collapse supernova. As \Tab{hecore} and \Tab{tmodels} show
there can be no CO core mass bigger than 51 \Msun \ at the time of
iron core collapse, even in stars that avoid the PPI, i.e., the
minimum of M$_{\rm preSN}$ and M$_{\rm final}$.  Most likely these
cores collapse to black holes, but even if they are artificially
exploded with very high energy, the $^{56}$Ni production is
limited. Model He50 with a CO core of 42 \Msun \ exploded with a
central energy deposition of $2.6 \times 10^{52}$ erg (\Fig{hyper})
only made 2.7 \Msun \ of $^{56}$Ni. It thus seems unlikely that credible
models can produce more than about 4 \Msun \ of $^{56}$Ni without
encountering the PPI and shrinking in mass first \citep[though
  see][]{Ume08}.

Since the matter ejected in PPISN experiences no explosive nuclear
processing, the nucleosynthetic yields are the same as if, towards the
end of its life, the star had very rapid mass loss. Assuming the
collapse of the entire core to a black hole, most of the elements
heavier than magnesium, and all appreciable radioactivities, are
lost. The nucleosynthesis for a representative set of non-rotating
models is given in \Tab{nucleo} and shows the appreciable
nucleosynthesis of helium and CNO, with traces of neon and
magnesium. One has to go to full PISN before the ejection of
intermediate mass elements competes, in solar proportions with oxygen,
and even then very little iron-group elements are made until the
star's helium core exceeds about 90 \Msun \ (stellar mass about 190
\Msun). At very low metallicity, rotating models may also copiously
produce primary $^{14}$N \citep{Mey02,Heg10,Yoo12}. A first generation
of such stars might thus contribute a composition rich in either CO or
CNO, with a trace of neon and magnesium and very little silicon and
iron \citep{Woo15}. Stars with this sort of composition have been seen
\citep{Fre05,Kel14}. Given that a single one of the more massive stars
discussed here could produce, e.g., 1\% solar oxygen for thousands of
solar masses of second generation stars, the nucleosynthetic role of
Pop III PPISN deserves greater attention. This is also an interesting
topic for future work.

% note n110 and t110 give different nucleosynthesis. Very sensitve to mass cut.

\acknowledgements

This work, which spanned many years, has profited from conversations
with many people, especially Alex Heger, Nathan Smith, Jorick Vink,
Selma de Mink, and Thomas Janka.  Alex Heger also contributed
important parts of the KEPLER code that were necessary to this study,
e.g., the physics used in the rotating models, the adaptive network
for nuclear burning, and many other features that made the code easier
to use and the results easier to analyze. Early on, the research was
supported by the NSF (ARRA AST-0909129) and by NASA (NNX09AK36G). More
recently it has been supported solely by NASA (NNX14AH34G).

\end{document}